\documentclass[epj]{svjour}
\usepackage{graphics}
\usepackage{hyperref}
\usepackage{blindtext}
  \newcommand{\oo}{\mathrm{O}}
  \newcommand{\Qa}{\mathcal{Q}}
 \usepackage[latin1]{inputenc}
\usepackage{color}
\usepackage{url}
\usepackage{graphicx}
\hypersetup{linkcolor=red,citecolor=blue,filecolor=cyan,urlcolor=magenta}
\usepackage{float}
\usepackage{amsmath}
\usepackage{graphicx}
\usepackage{latexsym,color}
\usepackage{amssymb}
\usepackage{xcolor}
\usepackage{amsfonts,dsfont}

\usepackage[T1]{fontenc}
\usepackage{ae,aecompl,latexsym}

\definecolor{lightgray}{rgb}{0.75,0.75,0.75}

\def\be{\begin{equation}}
\def\ee{\end{equation}}
\def\bea{\begin{eqnarray}}
\newcommand{\cc}{\mathrm{C}}

\newcommand{\apjs}{APJS}
\newcommand{\apj}{APJ}
\newcommand{\prd}{Phys. Rev. D}
\newcommand{\prl}{Phys. Rev. Letter}
\newcommand{\mnras}{MNRAS}
\newcommand{\jj}{\mathrm{J}}
\newcommand{\aap}{A\&A}
\newcommand{\nat}{Nature}
\def\eea{\end{eqnarray}}

\newcommand{\pp}{\textbf{()}}

\newcommand{\mso}{\mathrm{mso}}
\newcommand{\mbo}{\mathrm{mbo}}
\newcommand{\il}{~}

\usepackage[varg]{txfonts}
\begin{document}
\title{\textbf{RADs} energetics and constraints on emerging tori collisions  around super-massive Kerr Black Holes}
\author{D. Pugliese\inst{1} \and Z. Stuchl\'{\i}k\inst{2}
}                     
\offprints{}          
\institute{Institute of Physics and Research Centre of Theoretical Physics and Astrophysics, Faculty of Philosophy\&Science,\\
 Bezru\v{c}ovo n\'{a}m\v{e}st\'{i} 13, CZ-74601 Opava, Czech Republic
}
\date{Received: date / Revised version: date}
%
\abstract{
We provide constraints on  possible  configurations and interactions  of  two coplanar tori  orbiting  a   central   Kerr black hole (BH),   in dependence on   its  dimensionless spin. The two-tori configurations can be directly linked to the current models featuring the obscuration of galactic BH X-ray emission.  The emergence of   each torus instability phases  is discussed and  tori  collision has been  also  investigated.  The first  simple evaluation of the     center-of-mass energy proves  that  collision-energy-efficiency   increases with the   dimensionless BH spin. We explore the  phenomenological aspects of the corotating and counterrotating  tori  by analyzing   properties of the orbiting toroidal configurations related to   the fluid enthalpy flux, the mass-flux, the  mass-accretion-rates, and the  cusp luminosity in the two cases of corotating and counterrotating fluids in dependence on the SMBH spin. The analysis resulted ultimately in a comparative investigation of the properties of corotating  versus counterrotating tori, demonstrating that two accretion tori can orbit around the central Kerr attractor only under very specific conditions.  Our results also demonstrate that the dynamics of the unstable phases of these  double tori systems  is significant   for the high energy phenomena  which could be observable in the  X-ray  emission and  extremely energetic phenomena in active galactic nuclei   and quasar.
%
} 
\maketitle

\section{Introduction}\label{Sec:I}
Observations  in the  X-ray emission are on the verge of the  astrophysical development in the high energy  sector and in many senses we can safely say that  thay contribute  significantly to make   a challenging period for astrophysics. Surely, the planned   new satellite  observatories represent   expected breakthrough mission  in this sector\footnote{For example from \textbf{XMM-Newton}: \url{https://www.cosmos.esa.int/web/xmm-newton}, to  NASA's \textbf{Chandra} X-ray Observator: \url{http://cxc.harvard.edu/index.html}, \textbf{NuSTAR}  (Nuclear Spectroscopic Telescope Array) \url{http://www.nustar.caltech.edu/page/about} and \textbf{Swift BAT} (Burst Alert Telescope) \url{https://swift.gsfc.nasa.gov/about_swift/bat_desc.html}.}
 \cite{WW-AAVV}.
Consequently, theoretical models are  continuously  challenged by new data that require new emerging hypotheses   supported by   computational simulations by     appropriately  adapted  numerical integration  codes.
There is then   a remarkable area  which  is potentially capable to unify  very different issues  of high energy  astrophysics   connecting X-ray emission, black hole \textbf{(BH)} physics related to accretion processes, and galaxy mergers see, e.g.,  \cite{Ricci.:2017wmr} and
 \cite{Tadhunter:2017qji,Regan:2017vre}.
The origin of X-ray emission in \textbf{BH} environments is especially  viewed  as accretion related, and possibly a bridge between galaxy and  a galactic \textbf{BH}  activity keeping  fingerprints of   galaxy mergers\cite{edge-1,Tucci:2016tyc,Xie:2017jbz,Collinson:2016rmg,Parker:2017wnh,Regan:2017vre,Zhang:2015eka}.
There are various  studies    of the X-ray emission profile,  and more generally jet emission,  using more or less the general scheme of the
emission ``screened'' by some  ``bubbles'' of matter configurations in equilibrium, assumed to be located in the inner region between the  \textbf{BH} and an  outer  disk  orbiting the central attractor. We mention also  obscuring torus in Galaxy X-ray emission \cite{Marchesi,Gilli:2006zi,Marchesi:2017did,Masini:2016yhl} and \cite{DeGraf:2014hna}, for a discussion on  the role of the inner and outer part of an accretion disk \cite{Storchi-Bergmann}.
In fact,
active galactic nuclei \textbf{(AGN)} ``obscuration'' is generally linked to some toroidal configurations  possibly
surrounding the \textbf{AGN} accretion disk \cite{2011PhRvD..84h4002K,2013CQGra..30b5010K,2014PhRvD..90d4029K,2016PhRvD..93l4055K}. The results we found in the present paper should be  directly comparable with  investigation of  \textbf{AGN}-\textbf{BH} X-ray spectra. We set  opaque geometrically thick toroidal surfaces   that are  likely  produced  during \textbf{BH} growth due to  ``clumpy'' episodic accretion  processes, implying potentially interesting    consequences on the X-ray emission and oscillations in the black hole accretion systems{--see also \cite{Slany:2013rml,Kovar:2011uh,Schroven:2018agz,Trova:2018bsf} for different charged models of structured disks}.

Motivated by these facts, in the present work we  consider   the possibility to provide reliable general constraints   for any ``adjoint'' orbiting structures which   may serve as a guide to  model developments of the accreting systems.
We  investigate this situation by carefully studying  coplanar structured toroidal  disks, so called \emph{ringed accretion disks} (\textbf{RADs}),  first introduced in \cite{pugtot} and then  detailed in  \cite{ringed,open,multy,Proto.Jets}. The set-up for the ringed accretion disk model was drown in \cite{ringed}, where  constraints and discussion on perturbations were provided.
Then in \cite{open} sequences of unstable configurations were discussed, the investigation was focused on the unstable phases of  accreting multi toroidal structures.
The paper \cite{dsystem} focused on the case of  two tori as ``seed'' for larger configurations. The \textbf{RAD} configurations were then discussed  in more detailed form in two works\cite{long,multy}.
The fourth fundamental study of the \textbf{RAD} is related to  their energetics and  it is subject of   the present paper.
We provide indication on  the   doubled toroidal structures that can be detected   independently on the attractor characteristic spin and,  as a sideline result,  we also   explicitly present  first evaluation of the efficiency of the  tori collisions and energy release in \textbf{RADs} colliding tori.
Our results show that even in the simplest but effective set-up considered here,     the presence of such ``screening bubbles'' cannot  be considered as  a general universal property of any spinning attractor,  but  on the contrary, the systems composed by  orbiting multi structures
(within the   symmetry conditions fixed here) must undergo strict constraints on the relative tori rotation and  they  strongly depend on the dimensionless spin of the central Kerr attractor.
This result remarkably sets, for the first time to our knowledge, a strong unifying framework for studying of an  orbiting multiple configuration and  its  \textbf{BH} attractor \cite{multy}. Specifically, we show that only four     \textbf{BHs} classes, distinguished  according to their dimensionless spin, can  host  specific  \textbf{RAD} configurations, in   dependence  on periods   of the life of the attractor and the \textbf{RAD}. This finding thus  represents a  precise  guidance and  gives directions to orient any studies featuring  possible   screening effects in this framework. The constraints provided here   are very restrictive--
    we argue that,  by considering  different processes leading to   \textbf{RAD}  formation,  some \textbf{RAD} configurations are likely to be formed  around some  specific attractors in some periods of the \textbf{RAD-BH} life\cite{multy,ringed,open,long,Pugliese:2018zlx}.

From the observational point of view, the   phenomenology associated with these toroidal complex structures  can    be   very wide. Our studies  could  open up a new field of investigation in astrophysics,  leading us to reinterpret  various phenomena analyzed  so far in the  single torus  framework. In  the new   framework, requiring the possibility  of  multi-tori  systems, we can simultaneously  revisit the actual analysis of screened X-ray emission by considering constraints provided here.
The related observational evidence may be gained by   the spectral  features
of \textbf{AGNs} X-ray emission line shape, due to the  X-ray obscuration and absorption
by one of the tori. In  \cite{S11etal,KS10} and \cite{Schee:2008fc,Schee:2013asiposs}, for instance, it was proposed that the  \textbf{AGNs} X-ray spectra  should provide  a fingerprint of the tori:
relatively indistinct excesses
 of the relativistically broadened  emission-line components were predicted, arising in a well-confined
radial distance in the accretion structure
  originating by a series of  episodic accretion events.
Thus, hints of formation processes are also briefly addressed.

  The \textbf{RADs} feature  systems made up of several axis-symmetrical matter configurations orbiting in the equatorial plane of a single central Kerr \textbf{BH}. In the model development, and especially in the choice for \textbf{RAD} toroidal components, we have taken into account that  tori  may be formed   during several accretion regimes occurred in the lifetime of non-isolated  Kerr \textbf{BHs}.
Indeed, formation of several  accretion tori  orbiting one central super-massive  Kerr black hole (\textbf{BH}) has been conjectured   specifically    as a  peculiar  feature of {\textbf{AGNs}} and   Quasars-\cite{pugtot,ringed,open,NixonKing(2012b),S11etal,KS10,Schee:2008fc,Schee:2013asiposs,AKM03}. Therefore,  evidences of these special configurations are  expected to be found in  the associated  X-ray spectra emission in \textbf{AGNs}.
In these environments,  tori  might be   formed as remnants of several accretion regimes  occurred in various phases of the \textbf{BH} life  and  could eventually be   reanimated in non-isolated systems where the central attractor is interacting with the  environment, or in some kinds of binary  \textbf{BH} systems{\cite{Aligetal(2013),Lovelace:1996kx,Gafton:2015jja}}.  Some additional matter  could be  supplied into the vicinity of the
central \textbf{BH} due to  tidal distortion of a star, or if some cloud of interstellar matter is captured by the strong gravity \cite{natures,Blanchard:2017zfe}.
During  evolution of   \textbf{BHs} in these environments both  corotating  and counterrotating accretion stages are
mixed   in  various accretion periods  of the attractor  life \cite{Carmona-Loaiza:2015fqa,Dyda:2014pia,Volonteri:2002vz}. For the construction of a reliable model, this makes it necessary to consider, for a rotating central attractor, the possibility of different orientation of the spin for several aggregations of matter orbiting the attractor, and to consider this possibility  in the  associated phenomena. We will show that remarkably for some attractors only counterrotating tori could be considered as inner screening object.

From methodological view-point,
it is clear that  \textbf{RAD} investigation   is affected by  a series of challenging issues.
This rather complex scenario envisages  different aspects of black hole and accretion torus life subject to numerous and articulated studies.
  The existence of  different evolution periods of the  \textbf{BH}-\textbf{RAD}   systems should be carefully considered in any model construction   requiring   several   assumptions on the  \textbf{BH}-\textbf{RAD}  history  according to the different scenarios  of the force balance of  the torus in each period.   To fix our idea we may conveniently
    introduce the following three \textbf{RAD}-periods: $\mathbf{I}$- period features tori formation, $\mathbf{II}$-period  consists in the torus  accretion    onto the central Kerr  attractor  and finally a  $\mathbf{III}$-period  deals with  tori interaction   and  tori collision  emergence.  Here we mainly focus on  period   \textbf{(II)}, discussing the unstable configurations as defined in  this analysis, and  period  \textbf{(III)}, by considering  conditions for the emergence of tori collisions, providing very simple approximative evaluation  for the collision energy release.

The morphology of each  component of the \textbf{RAD}  is a fat torus in equilibrium, centered on the attractor. This is an example of
opaque (large optical depth) and super-Eddington (high  matter accretion rates)
model: a radiation pressure supported accretion torus, cooled by advection with low viscosity.
 The development of this model was  firstly drawn up by Abramowicz and his collaborators in a
series of works
\cite{Jaroszynski(1980),Pac-Wii,cc,Koz-Jar-Abr:1978:ASTRA:,abrafra} and then adopted in different contexts: particularly this is   today widely adopted    as the initial conditions in the set up for simulations of the GRMHD (magnetohydrodynamic)  accretion structures \cite{Shafee,Fragile:2007dk,DeVilliers,arXiv:0910.3184,Porth:2016rfi}. Consequently, the adaptability and common use of this model in much more complex dynamical scenarios has driven our choice towards the  application for each  \textbf{RADs} component,   taking care to guide our investigation into results on the ranges of variation of the relevant quantities.
The individual toroidal  (thick disk) configurations are prescribed by  barotropic models, for which the time scale of the dynamical processes $\tau_{dyn}$ (regulated by the gravitational and inertial forces) is much lower than the time scale of the thermal ones $\tau_{therm}$   (heating and cooling processes, radiation) that is lower than the time scale of the viscous processes $\tau_{\nu}$; thus  $\tau_{dyn}\ll\tau_{therm}\ll\tau_{\nu}$ and the effects of strong gravitational fields are dominant with respect to the  dissipative ones and predominant to determine  the unstable phases of the systems \cite{F-D-02,abrafra,pugtot,Pac-Wii},
see also \cite{Hawley1990,Fragile:2007dk,DeVilliers,Hawley1987,Hawley1991,Hawley1984,Fon03,Lei:2008ui}.
As a consequence, during the  dynamical processes, the functional form of the angular
momentum and entropy distribution depends on the initial conditions of the system and on
the details of the dissipative processes.
Paczy\'nski realized that it is physically
reasonable to assume  ad hoc distributions. This feature  constitutes a great advantage of these models  and  render their  adoption   extremely useful and predictive.
The tori are governed by ``Boyer's condition'' of the analytic theory of equilibrium configurations of rotating perfect fluids \cite{Boy:1965:PCPS:}. The toroidal structures of orbiting barotropic perfect fluid are determined by an {effective potential} reflecting the spacetime geometry, and the centrifugal force component through the  distribution of the specific angular momentum $\ell(r)$ of the orbiting fluid -- \cite{pugtot,Raine,PuMonBe12,Lei:2008ui,PuMon13,2011,Rez-Zan-Fon:2003:ASTRA:,Stuchlik:2012zza,Sla-Stu:2005:CLAQG:,astro-ph/0605094,Stu-Sla-Hle:2000:ASTRA:,arXiv:0910.3184,Stu-Kov:2008:INTJMD:}.
The entropy is constant along the flow and, according to the von Zeipel condition, the surfaces of constant angular velocity $\Omega$ and of constant specific angular momentum $\ell$ coincide \cite{M.A.Abramowicz,Chakrabarti0,Chakrabarti} (this implies in particular that the rotation law $\ell=\ell(\Omega)$ is independent of the equation of state \cite{Lei:2008ui,Abramowicz:2008bk}).
Consequently the perfect fluid equilibrium tori are regulated  in a given spacetime by the specific angular momentum distribution function $\ell(r)$, and a \emph{constant} $K$, related to the fluid effective potential, determining the matter content of the tori.
The equipressure surfaces, $K=$constant,   could be closed, determining equilibrium configurations, or open (proto-jet configurations  related to  jets \cite{open,Proto.Jets}). The special case of cusped equipotential surfaces allows for  accretion onto the central black hole \cite{Koz-Jar-Abr:1978:ASTRA:,Jaroszynski(1980),Pac-Wii,abrafra}. The outflow of matter through the cusp occurs due to an instability in the balance of the gravitational and inertial forces and the pressure gradients in the fluid, i.e., by the so called Paczynski mechanism of violation of mechanical equilibrium of the tori \cite{Jaroszynski(1980)}.

The \textbf{RADs} model is based on several  symmetry assumptions on  the  tori-central attractor system.
 In fact, in our approach  we take full advantage of the symmetry of the Kerr geometry, considering  a stationary and axisymmetric, full general relativistic (GR) model for a single  thick accretion disk with a toroidal shape;
   a generalization to different symmetry misaligned disk and even tilted disk may be convened in future work as a modification of the general case considered here.
 Thus
 we address directly the question, if   two aligned   axi-symmetric  accretion tori  can orbit a central coplanar Kerr attractor.
The response to this issue  provided in this work is   positive: a double tori  may orbit and even accrete  around a central  Kerr \textbf{BH}.  However,  this may occur only in four special cases,    constrained   by  the dimensionless spin of the central attractor, the tori  specific angular momentum and the relative rotation of the two  tori. The  tori evolution may lead  to the  interaction and eventual destruction of the tori system. We  briefly discuss   conditions for    these critical  phenomena      emergence, providing an evaluation of the center-of-mass energy for  colliding fluid particles originating in both tori.

Tori  may  collide and merge   or eventually turn  to origin some \emph{feeding--drying} processes with some  possible interesting phenomenological consequences on the periodical radiation emission: the accreting matter from the outer torus can  impact on the  inner torus  located between the outer one  and the central \textbf{BH}, or    the outer torus may be inactive  with  an active inner torus accreting  onto the \textbf{BH}, or both tori  may be accreting \cite{ringed,open}; for tidal destruction see \cite{Tadhunter:2017qji}. Radially oscillating tori  could be also  related to the high-frequency quasi periodic oscillations observed in non-thermal X-ray emission from compact objects \cite{Stuchlik:2013esa}.
We expect  therefore that  future   analysis      should  throw new light on those aspects of the dynamical processes not covered by our first analysis.

Following the three previous papers related to various aspects of the \textbf{RAD} model, we present here a fourth  paper of this series, devoted to the energetics of the \textbf{RAD} and constraints
of collisions of tori constituting the \textbf{RADs}.

\medskip

{The plan of this article is as follows: we summarize the basic properties of stationary toroidal
fluid configurations in  Kerr  spacetime  in Sec.\il(\ref{Sec:Kerr-2-Disk}), where  the system formed by the pair of tori is also introduced.{
The construction of the \textbf{RAD} tori aggregate led to the introduction of new notations adapted to the new  concepts, most of which are suited to represent the agglomerate as a whole accreting   disk.
In Sec.\il(\ref{Sec:Kerr-2-Disk})    a general review  of the  model is provided,  the  main notation related to the new concepts   are listed in
Table\il(\ref{Table:pol-cy-multi}).}
Our analysis leads to  certain number of sideline results,  considering  especially the possibility of counterrotating accreting tori orbiting  a central Kerr \textbf{SMBH}. While this subject has been developed in literature,   we had to specify in the \textbf{RAD} framework some more details of the tori features performing a comparative analysis of the main features  of the counterrotating tori with respect to the corotating tori.
We report in the Sec.\il(\ref{Sec:theory-weell}) results concerning the tori morphology (tori edges and  elongation on the equatorial plane) for these two classes of tori, considering very carefully the different behavior in relation to the \textbf{BH} spin   according to different values of the model parameters,  in  particular the magnitude of the  fluid specific angular momentum. Throughout  this  analysis we also clarify   some  aspects  constraining  the emerging of the tori collisions.
In Sec.\il(\ref{SeC:coll}) we discuss the occurrence of tori collisions:
constraints are provided  in Sec.\il(\ref{Sec:const}) while in Sec.\il(\ref{Eq:CM-ene}) we provide an  evaluation of
 {center-of-mass   energy }      for two colliding particles  from the two interacting tori.
  In Sec.\il(\ref{Sec:more-l}) we focus our investigation  directly on the phenomenological aspects of the corotating and counterrotating  tori.  We analyze  quantities defining the fluid enthalpy flux, the mass-flux, the  mass accretion rates, and the  cusp luminosity in the two cases of corotating and counterrotating fluids in dependence on the \textbf{SMBH} spin, the fluid specific angular momentum, and further parameters  related to the fluid density. Sec.\il(\ref{SeC:poly}) includes some  considerations on the polytropic fluids considered  in our framework.
  Finally we close this article in Sec.\il(\ref{Sec:grosss})   presenting a  general  discuss on  collisional phenomena in the \textbf{RAD} from a more global perspective,  considering the case of tori in accretion.
 Concluding remarks follow in Sec.\il(\ref{Sec:conl}), where  future perspective of our investigations is also provided.
  Appendix\il(\ref{Sec:coroterf}) specifies  details on the role of the frame-dragging in the ergoregion of the  Kerr geometry, in relation to the accreting tori of the \textbf{RAD}.
\section{Axi-symmetric tori in a Kerr spacetime}\label{Sec:Kerr-2-Disk}
We consider two  axially symmetric  tori with symmetry plane coinciding with   the equatorial plane of the central Kerr   \textbf{BH} of mass parameter $M$  and dimensionless spin  $a/M\in[0,1]$ \cite{ringed,open,pugtot,abrafra,Raine}.
The Kerr  metric tensor can be
written in the Boyer-Lindquist (BL)  coordinates
\( \{t,r,\theta ,\phi \}\)
as follows
\bea\nonumber &&ds^2=-\frac{\Delta-a^2 \sin ^2\theta}{\rho^2}dt^2+\frac{\rho^2}{\Delta}dr^2+\rho^2
d\theta^2+
\\
&&\nonumber \frac{\sin^2\theta\left(\left(a^2+r^2\right)^2-a^2 \Delta \sin^2\theta\right)}{\rho^2}d\phi^2\quad-
\\
&&\nonumber 2\frac{a  \sin^2(\theta ) \left(a^2-\Delta+r^2\right)}{\rho^2}d\phi dt\ ,
\\
&&\label{alai}
 \rho^2\equiv r^2+a^2\cos\theta^2, \quad \Delta\equiv r^2-2 M r+a^2.
\eea
The horizons $r_-<r_+$ and the outer static limit $r_{\epsilon}^+$ are respectively given by:
\bea
&&
r_{\pm}\equiv M\pm\sqrt{M^2-a^2};
\\
&& r_{\epsilon}^{+}\equiv M+\sqrt{M^2- a^2 \cos\theta^2};
\eea
there is $r_+<r_{\epsilon}^+$ on   $\theta\neq0$  and  $r\_{\epsilon}^+=2M$  in the equatorial plane, $\theta=\pi/2$--\cite{Pugliese:2011aa}.  The extreme Kerr black hole  has spin-mass ratio $a/M=1$, while  the non-rotating  limiting case $a=0$ is the   Schwarzschild metric.
As the line element (\ref{alai}) is independent of $\phi$ and $t$,  the covariant
components $p_{\phi}$ and $p_{t}$ of a particle four--momentum are
conserved along the   geodesics, therefore\footnote{We adopt the
geometrical  units $c=1=G$ and  the $(-,+,+,+)$ signature, Greek indices run in $\{0,1,2,3\}$.  The   four-velocity  satisfy $u^a u_a=-1$. The radius $r$ has unit of
mass $[M]$, and the angular momentum  units of $[M]^2$, the velocities  $[u^t]=[u^r]=1$
and $[u^{\varphi}]=[u^{\theta}]=[M]^{-1}$ with $[u^{\varphi}/u^{t}]=[M]^{-1}$ and
$[u_{\varphi}/u_{t}]=[M]$. For the seek of convenience, we always consider the
dimensionless  energy and effective potential $[V_{eff}]=1$ and an angular momentum per
unit of mass $[L]/[M]=[M]$.}
the quantities
\be\label{Eq:after}
{E} \equiv -g_{\alpha \beta}\xi_{t}^{\alpha} p^{\beta},\quad L \equiv
g_{\alpha \beta}\xi_{\phi}^{\alpha}p^{\beta}\ ,
\ee
are  constants of motion, where  $\xi_{t}=\partial_{t} $  is
the Killing field representing the stationarity of the Kerr geometry and  $\xi_{\phi}=\partial_{\phi} $
is the
rotational Killing field.

The constant $E$ for
timelike geodesics represents the total energy of the test particle
 coming from radial infinity, as measured  by  a static observer at infinity, while  $L$  is  the axial component of the angular momentum  of the particle.
 The
Kerr metric (\ref{alai}) is  invariant under the application of any two different transformations: $x^\alpha\rightarrow-x^\alpha$
  for one of the coordinates $(t,\phi)$, or the metric parameter $a$, and   the    test particle dynamics is invariant under the mutual transformation of the parameters
$(a,L)\rightarrow(-a,-L)$. This makes possible to   limit the  analysis of the test particle circular motion to the case of  positive values of $a$
for corotating  $(L>0)$ and counterrotating   $(L<0)$ orbits with respect to the black hole \cite{PQN,PQK,PQKN}.

We focus  here on the case of
 a one-species particle perfect  fluid (simple fluid),  described by  the  energy momentum tensor
\be\label{E:Tm}
T_{\alpha \beta}=(\varrho +p) u_{\alpha} u_{\beta}+\  p g_{\alpha \beta},
\ee
where $\varrho$ and $p$ are  the {total fluid density} and
pressure, respectively, as measured by an observer moving with the fluid whose four-velocity $u^{\alpha}$  is
a timelike flow vector field.  Due to
symmetries of the problem, we always assume $\partial_t \mathbf{Q}=0$ and
$\partial_{\varphi} \mathbf{Q}=0$, $\mathbf{Q}$  being a generic spacetime tensor.
The  fluid dynamics  is described by the \emph{continuity  equation} and the \emph{Euler equation} respectively:
\bea\nonumber
&&
u^\alpha\nabla_\alpha\varrho+(p+\varrho)\nabla^\alpha u_\alpha=0\,
\\\label{E:1a0}
&&
(p+\varrho)u^\alpha \nabla_\alpha u^\gamma+ \ h^{\beta\gamma}\nabla_\beta p=0,
\eea
where the projection tensor $h_{\alpha \beta}=g_{\alpha \beta}+ u_\alpha u_\beta$ and $\nabla_\alpha g_{\beta\gamma}=0$ \cite{pugtot,Pugliese:2011aa}.
 We investigate  the  fluid toroidal  configurations centered on  the  plane $\theta=\pi/2$, and  defined by the constraint
$u^r=0$. No
motion is assumed also in the $\theta$ angular direction ($u^{\theta}=0$).
Considering a  barotropic equation of state $p=p(\varrho)$, the  continuity equation
is  identically satisfied as consequence of the conditions, while  from
 the Euler  equation  in (\ref{E:1a0})   we find
\bea\label{Eq:scond-d}
&&
\frac{\partial_{\mu}p}{\varrho+p}=-{\partial_{\mu }W}+\frac{\Omega \partial_{\mu}\ell}{1-\Omega \ell};\quad W\equiv\ln V_{eff}(\ell)
,\\\nonumber
 &&\mbox{where}\quad V_{eff}(\ell)=u_t= \pm\sqrt{\frac{g_{\phi t}^2-g_{tt} g_{\phi \phi}}{g_{\phi \phi}+2 \ell g_{\phi t} +\ell^2g_{tt}}},
\eea
{$\Omega$ is the relativistic angular frequency of the fluid relative to the distant static observers, and  $V_{eff}(\ell)$ provides an \emph{effective potential} for the fluid, assumed here  to be  characterized by a  conserved and constant specific angular momentum $\ell$  (see also \cite{Lei:2008ui,Abramowicz:2008bk}).}

Similarly to the case of the test particle dynamics,
the  function  $V_{eff}(\ell)$  in Eq.\il(\ref{Eq:scond-d})  is invariant under the mutual transformation of  the parameters
$(a,\ell)\rightarrow(-a,-\ell)$, therefore we can limit the analysis to  positive values of $a>0$,
for \emph{corotating}  $(\ell>0)$ and \emph{counterrotating}   $(\ell<0)$ fluids and    we adopt the notation $(\pm)$  for  counterrotating or corotating matter  respectively.
Therefore,
the accretion tori  corotate $(-)$ or counterrotate $(+)$ with respect to the  Kerr  \textbf{BH}, for $\ell_{\mp} a\gtrless0$ respectively.
As a consequence of this,
considering the case of two orbiting tori, $(i)$ and $(o)$  respectively, we need to introduce   the concept  of
 \emph{$\ell$corotating}  tori,  $\ell_{i}\ell_{o}>0$, and \emph{$\ell$counterrotating}  tori,   $\ell_{i}\ell_{o}<0$.  The  $\ell$corotating tori can be both corotating, $\ell a>0$, or counterrotating,  $\ell a<0$, with respect to the central Kerr attractor \cite{ringed}.

{On the other hand, the specific  angular momentum distribution of circular geodesics governs  the  ringed accretion disk
\bea\label{Eq:continuus}
\ell_{\pm}(r;a)\equiv \frac{L}{E} =\\\nonumber
&&
=\frac{a^3M +aMr(3r-4M)\pm\sqrt{Mr^3 \left[a^2+(r-2M)r\right]^2}}{[Ma^2-(r-2M)^2r]M},
\eea
namely centers (and cusps) of the individual tori.}

Note that the constant specific angular momentum $\ell$ in Eq.\il(\ref{Eq:scond-d}), characterizing each toroid of the   \textbf{RAD}, is related to the ``geodesic angular momentum distribution'' $\ell_{\pm}(r;a)$ of
Eq.\il(\ref{Eq:continuus}), due to the fact that for a fixed $r=\bar{r}{}^{\pm}>r_{mso}^{\pm}$,
the value $\ell_{\pm}(\bar{r}{}^{\pm};a)$ provides the specific  angular momentum $\ell$ of the torus centered in  $\bar{r}{}^{\pm}$, see also \cite{abrafra}.

The tori are regulated by  the balance of the    hydrostatic  and   centrifugal  factors due to the fluid  rotation and by the curvature  effects  of the  Kerr background, encoded in the effective potential function $V_{eff}$.

\begin{figure}[h!]
\includegraphics[width=1\columnwidth]{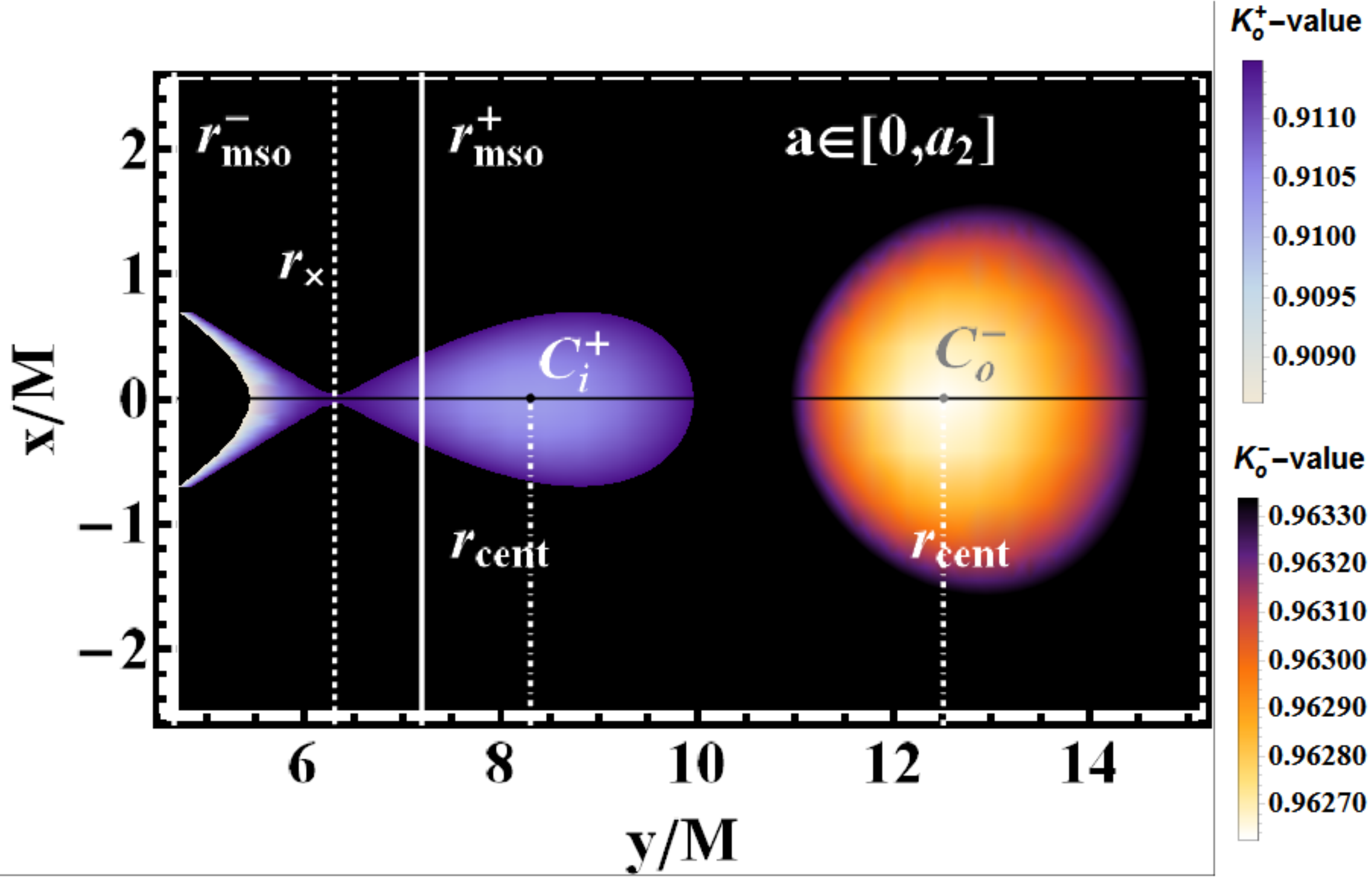}\\
\includegraphics[width=1\columnwidth]{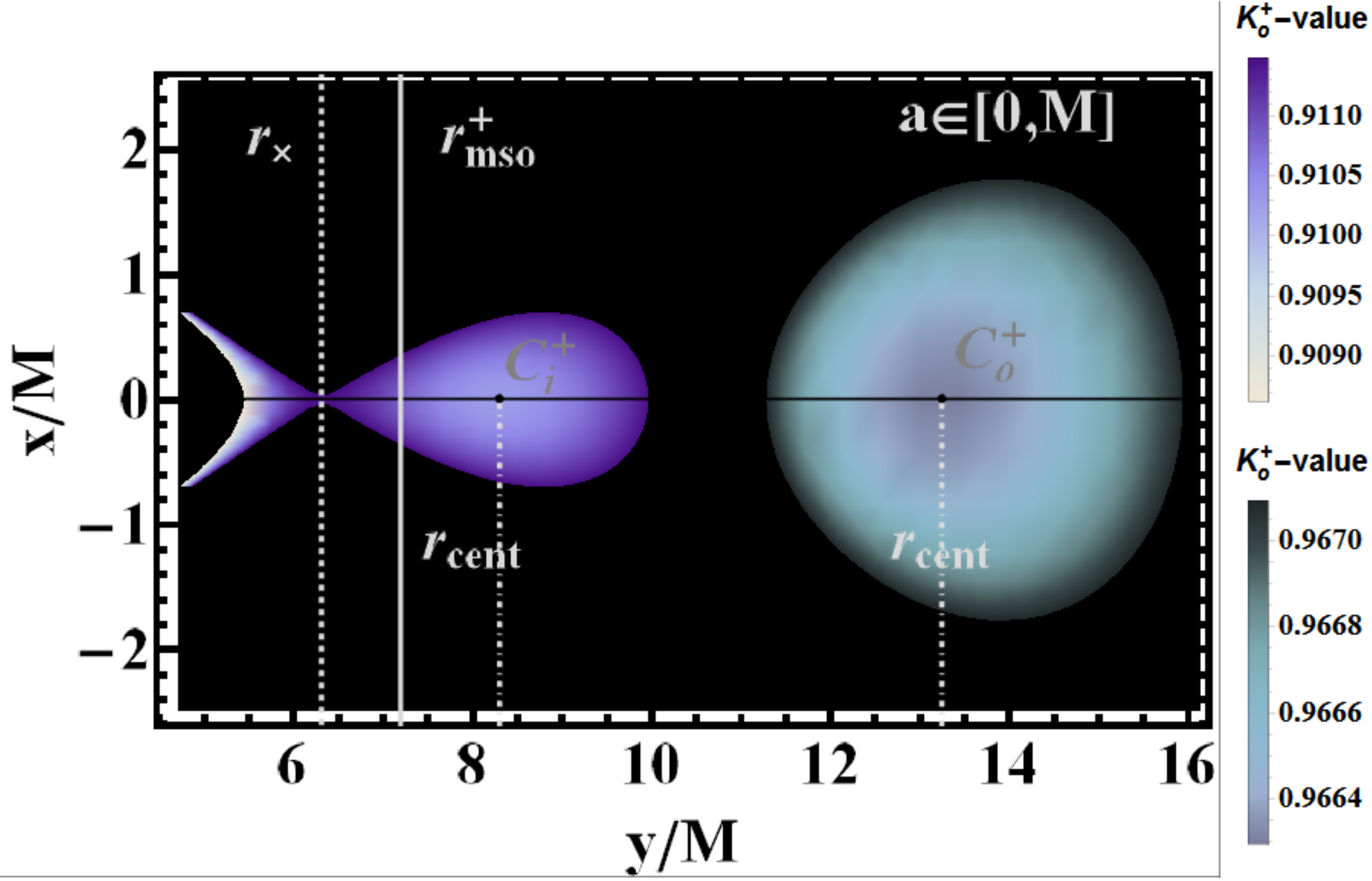}\\
\includegraphics[width=1\columnwidth]{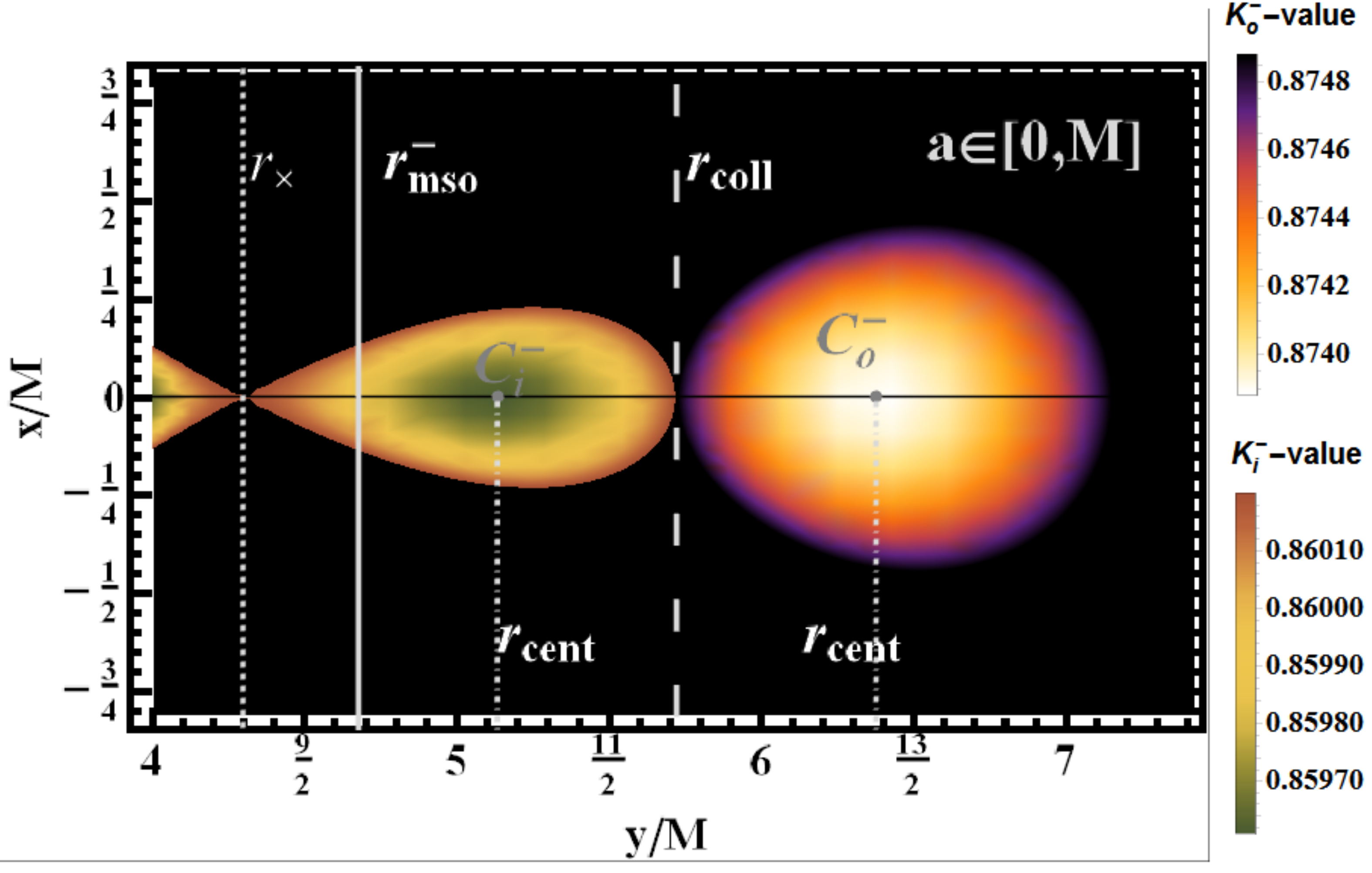}%
\caption{\textbf{GRHD} numerical 2D  integration  {of Eq.\il(\ref{Eq:scond-d}) for the }    couples  $\cc_{\times}^+<\cc^-$ (upper panel),     $\cc_{\times}^{\pm}<\cc^{\pm}$ (center and bottom panel), orbiting a central Kerr \textbf{BH}  attractor with spin $a=0.382 M$, $(x, y)$ are Cartesian coordinates. Tori centers $r_{cent}$, tori collision point $r_{coll}$, and inner edge of accreting tori are also shown. See also Fig.\il(\ref{Fig:AliceB}).
\label{Fig:teachplo} }
\end{figure}
The procedure adopted  in the present article
borrows from the  Boyer theory on the equipressure surfaces applied to a   torus \cite{Boy:1965:PCPS:,arXiv:0910.3184},
where   the Boyer surfaces are given by the surfaces of constant pressure  or\footnote{{More generally $\Sigma_{\mathbf{Q}}$ is the  surface $\mathbf{Q}=$constant for any quantity or set of quantities $\mathbf{Q}$.}}  $\Sigma_{i}=$constant for \(i\in(p,\varrho, \ell, \Omega) \), \cite{Boy:1965:PCPS:,Raine}, where the angular frequency  is indeed $\Omega=\Omega(\ell)$ and $\Sigma_i=\Sigma_{j}$ for \({i, j}\in(p,\varrho, \ell, \Omega) \).  Many features of the tori dynamics and morphology like their thickness, their stretching in the equatorial plane, and the location of the tori are predominantly determined by the geometric properties of spacetime via the effective potential $V_{eff}$.  The boundary of any stationary, barotropic, perfect fluid body is determined by an equipotential surface,  i.e., the surface of constant pressure (the Boyer surface) that is orthogonal to the gradient of the effective potential.
 The toroidal surfaces  are the equipotential surfaces of the effective potential  $V_{eff}(\ell) $, considered  as function of $r$,  solutions $ \ln(V_{eff})=\rm{c}=\rm{constant}$ or $V_{eff}=K=$constant.
The  couple of parameters $(\ell,K)$ uniquely identifies each Boyer surface--Figs\il(\ref{Fig:teachploeff}).  %
\begin{figure}
\includegraphics[width=7.1cm]{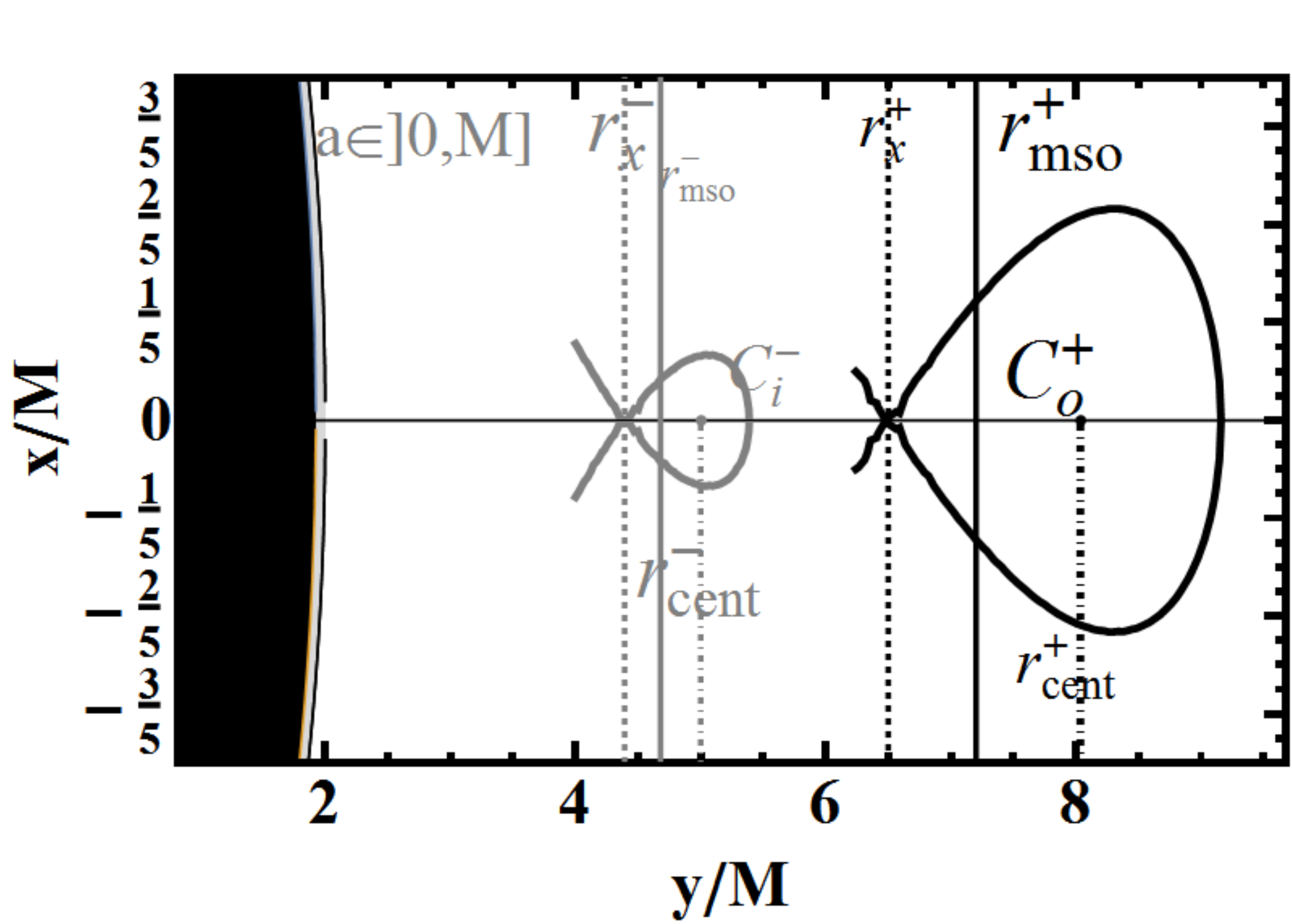}
\includegraphics[width=7.4cm]{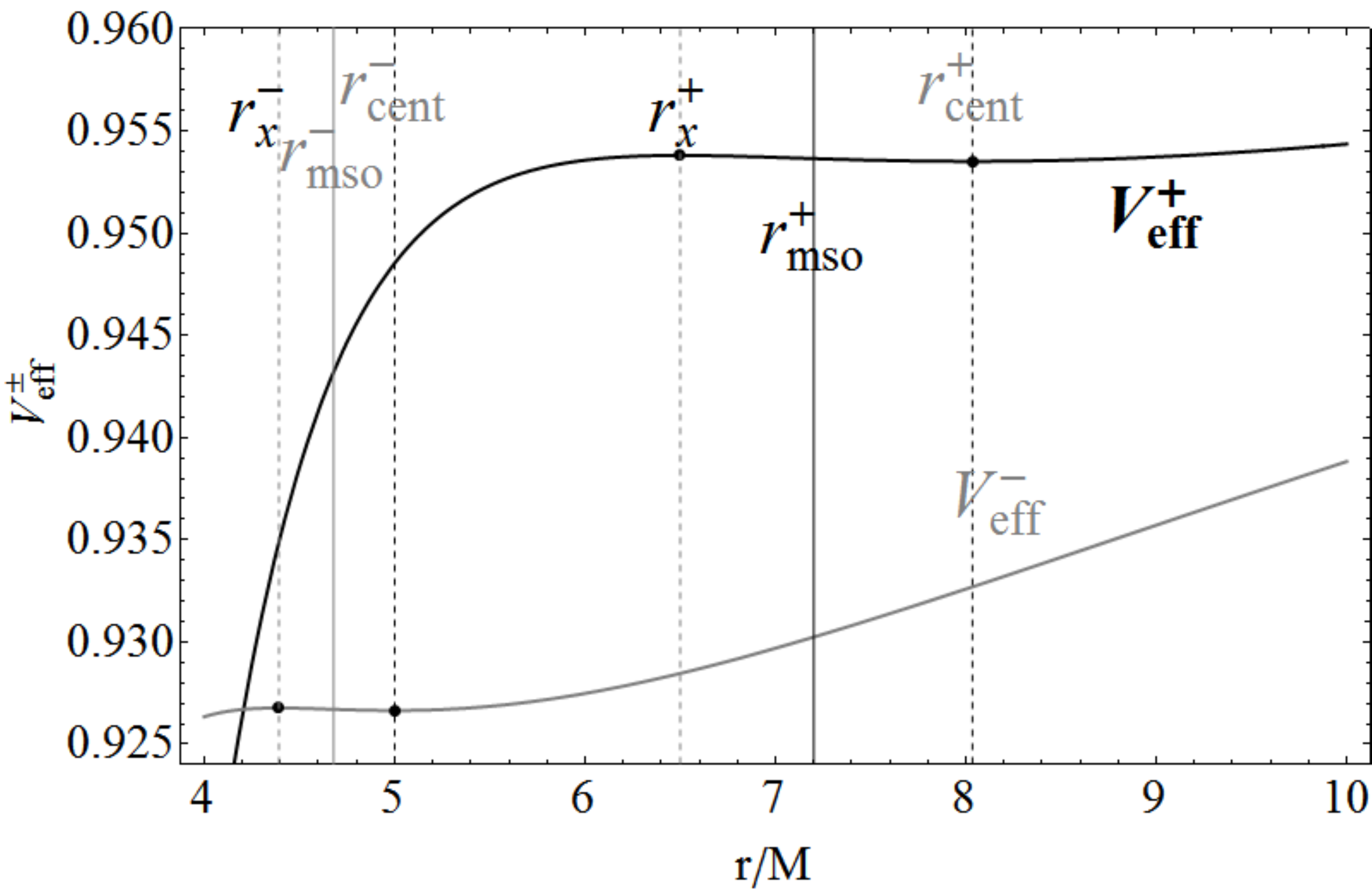}
\caption{\emph{Upper panel}: Cross sections on the equatorial plane of the outer Boyer surfaces   Roche lobes   for  $\ell$counterrotating tori $\cc_{\times}^-<\cc_{\times}^+$ (as there is $r_{cent}^-<r_{cent}^+$, and  there is $\cc_{\times}^-<_{\times}\cc_{\times}^+$ for it is $r_{\times}^-<r_{\times}^+$)  orbiting a central Kerr \textbf{BH}  attractor with spin $a=0.382 M$, $(x, y)$ are Cartesian coordinates and fluid specific angular momentum $\ell^-=3.31$, $\ell^+=-3.99$, parameters $K^-=0.927516$ and $K^+=0.953141$.
\emph{Bottom panel:} Effective potential $V_{eff}$ as function of $r/M$. This special tori couple is investigated also in Figs\il(\ref{Fig:Fig3D}) (\ref{Fig:Stopping}), (\ref{Fig:AliceB})-bottom   and (\ref{Fig:Hologra}).
\label{Fig:teachploeff} }
\end{figure}
It should also be noted that, according to  Eq.\il(\ref{Eq:scond-d}), the maximum  of the hydrostatic pressure corresponds to the minimum of the  effective potential $V_{eff}$, and it is the torus center $r_{cent}$.   The instability points of the tori, as envisaged by the  P-W mechanics, are located at the minima of the pressure and therefore maximum of $V_{eff}$---Figs\il(\ref{Fig:teachploeff}). To identify these points,  we  therefore need to compute the  critical points of  $V_{eff}(r)$  as function of the radius $r$. Equation  $\partial_r V_{eff}$  can be solved for the specific angular momentum of the fluid  $\ell(r)$--Fig.\il(\ref{PlotdisolMsMb}).
\begin{figure}
\includegraphics[width=1.\columnwidth]{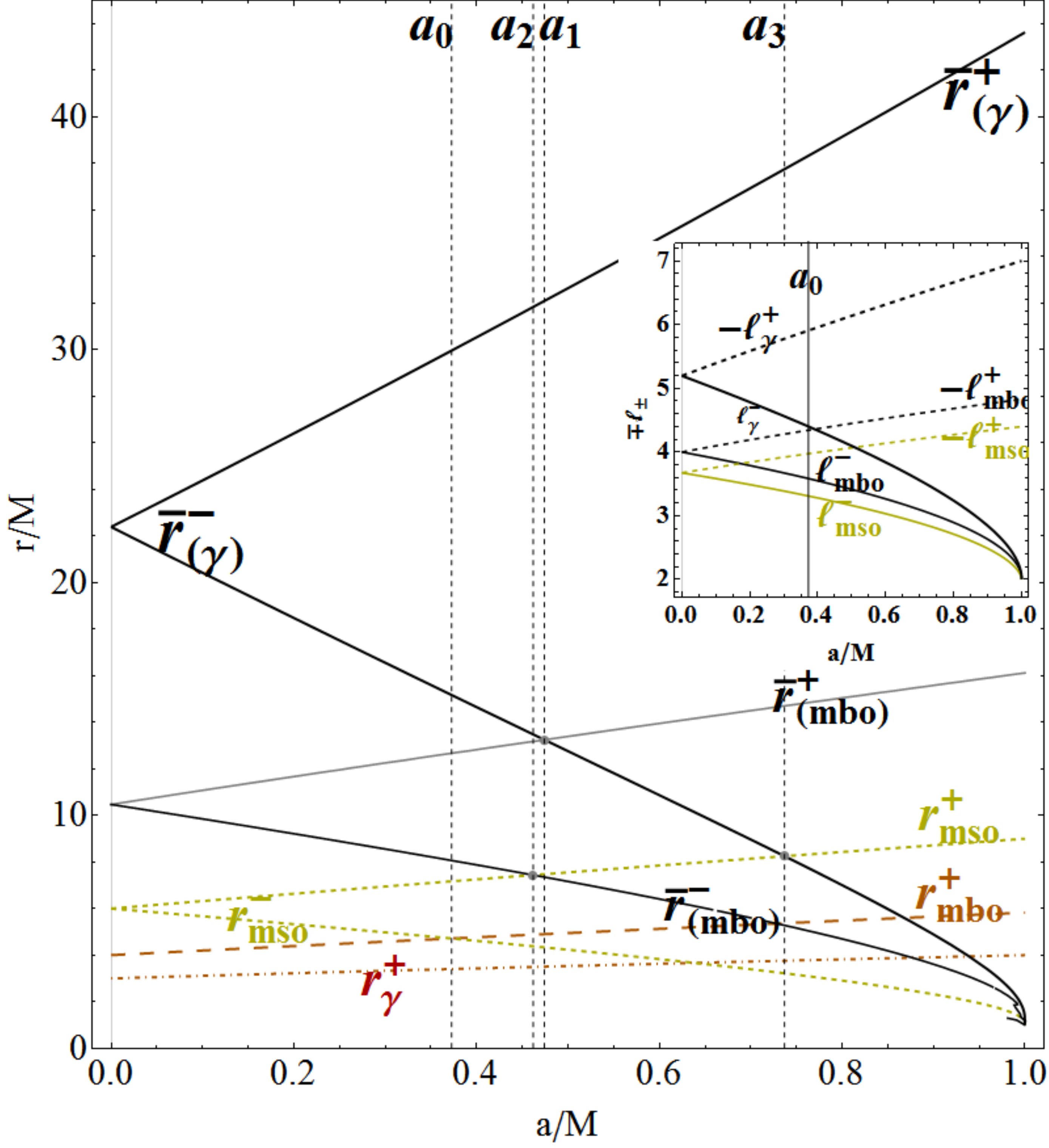}
\caption{\label{PlotdisolMsMb}{Radii  $\{\bar{r}_{(\gamma)}^{\pm}, \bar{r}_{(\mbo)}^{\pm},\bar{r}_{(\mso)}^{\pm}\}$ versus $a/M\in[0,1]$.
 }    Spins $\{a_1, a_2, a_3\}$, introduced in Eq.\il(\ref{Eq:spins}), and $a_0$ introduced in Sec.\il(\ref{Eq:CM-ene}) are also shown.  The {marginally stable circular orbit} $r_{mso}^{\pm}$, the {marginally bounded circular orbit} $r_{mbo}^{\pm}$ and finally the {marginal (photon) circular orbit} $r_{\gamma}^{+}$, for  counterrotating (+) and corotating (-) fluids are also shown. \emph{Inside  panel}: Fluid specific  angular momentum
 $\{\ell_{\gamma}^{\pm},\ell_{mbo}^{\pm},\ell_{mso}^{\pm}\}$ as function of the dimensionless spin $a/M\in[0,1]$.   }
\end{figure}
\begin{figure}[h!]
\centering
\includegraphics[width=.571\columnwidth,angle=90]{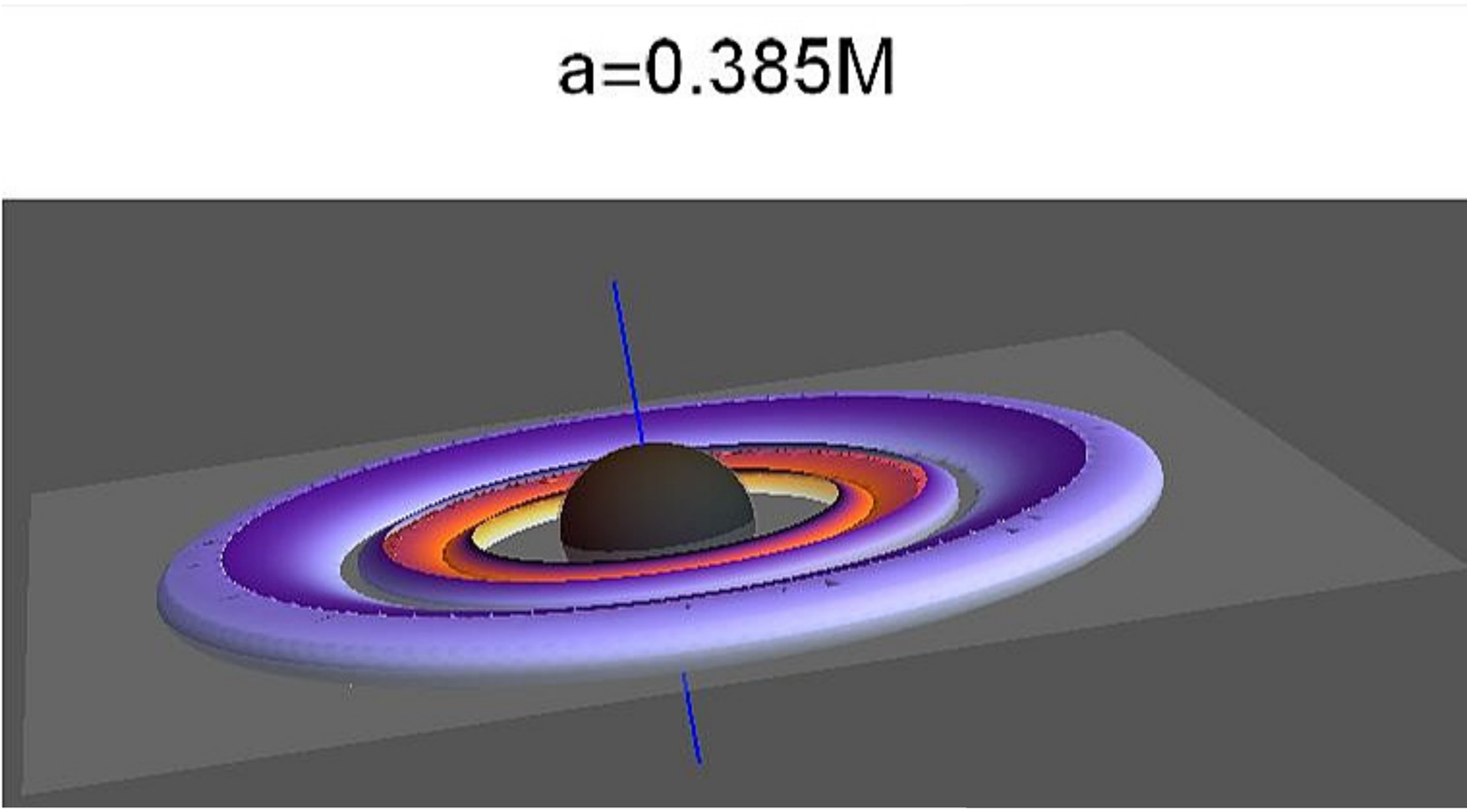}\\
\includegraphics[width=.571\columnwidth,angle=90]{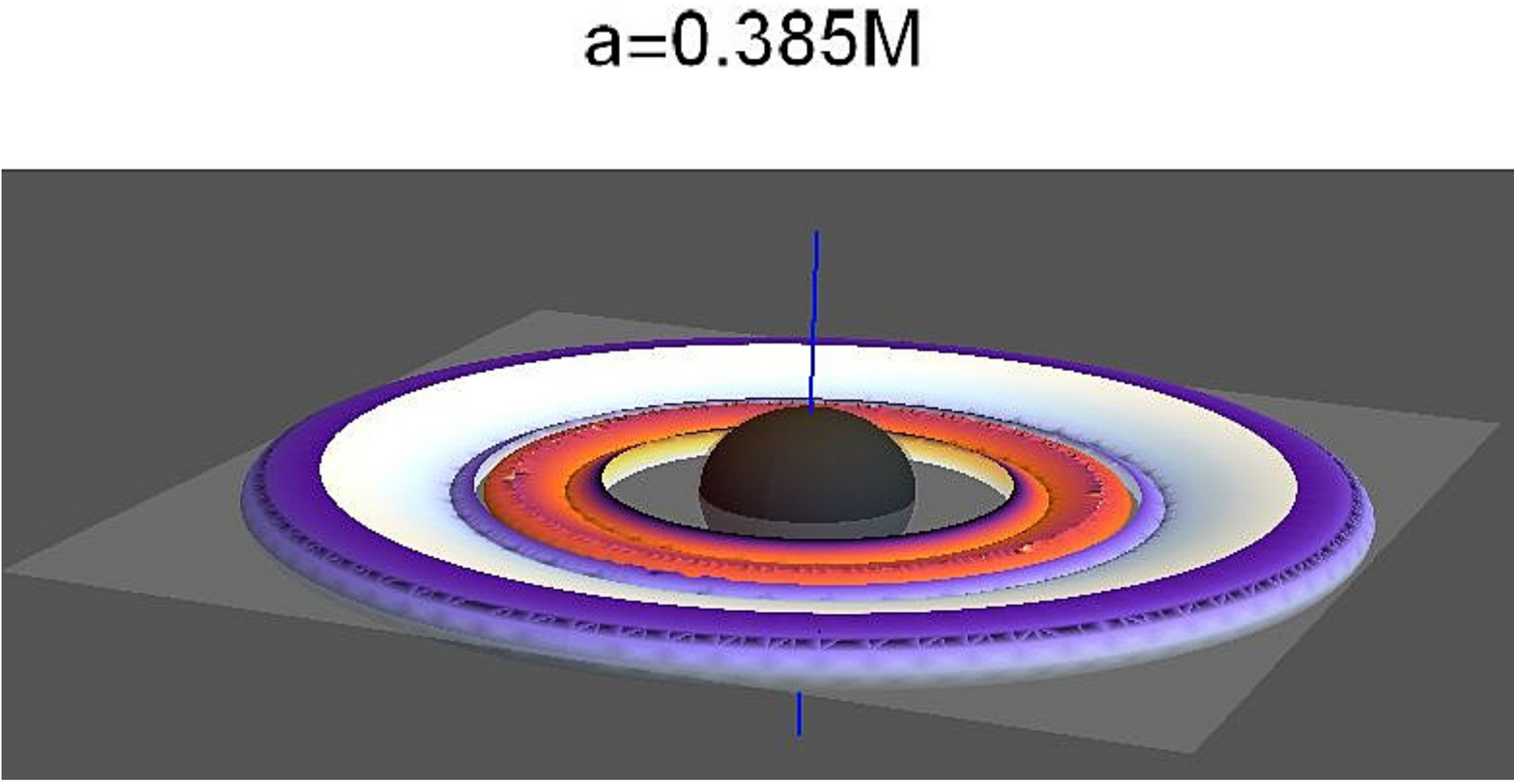}\\
\includegraphics[width=.571\columnwidth,angle=90]{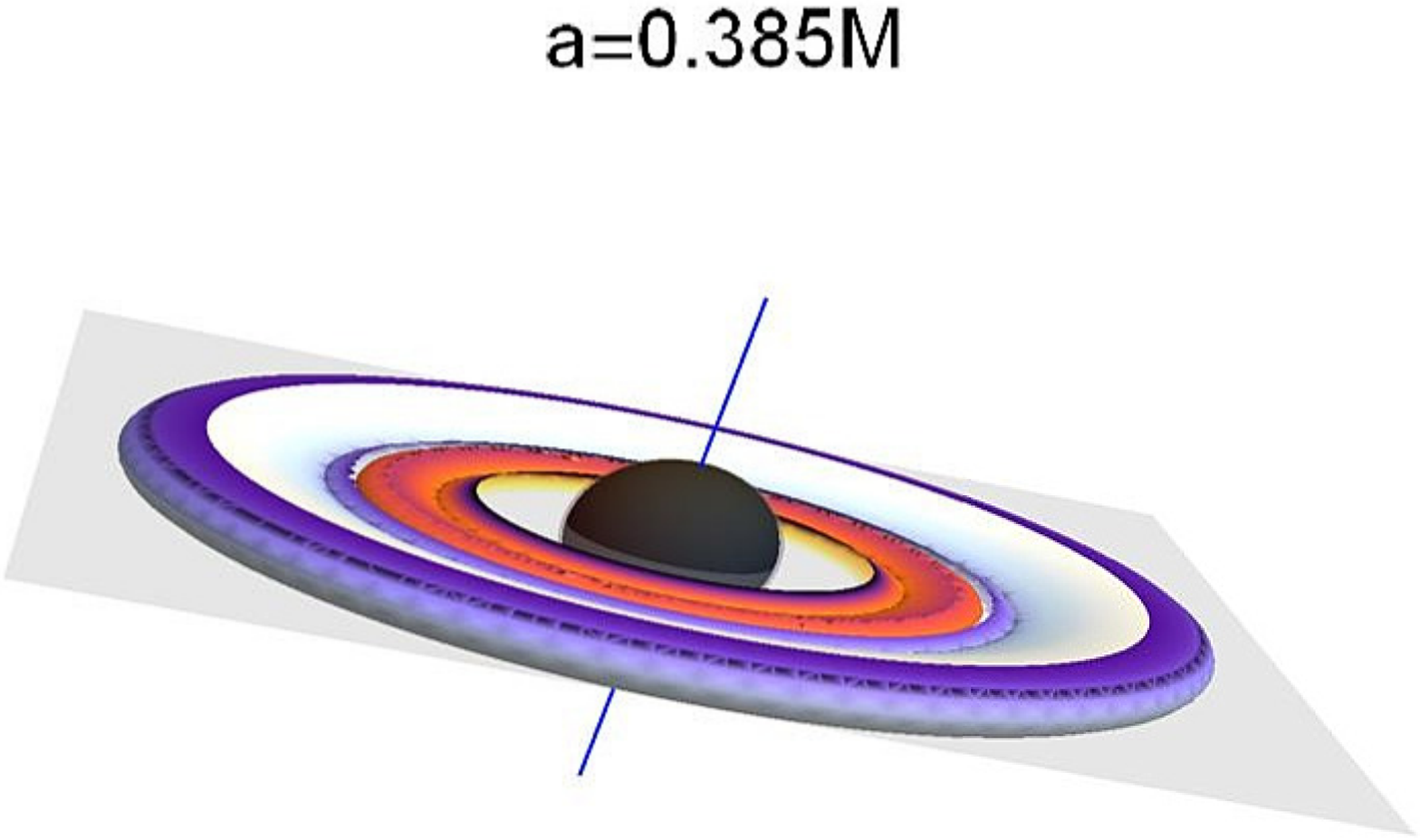}
\caption{\textbf{{GRHD}}-Numerical 3D integration of  tori density surfaces {(solutions of Eq.\il(\ref{Eq:scond-d})}). Colors are chosen according to improved visual effects. Integration is stopped at the emerging of tori collision. Black region is $r < r_+$, $r_+$ is the outer horizon
of the Black hole of spin $a = 0.385M$, gray region is the outer
ergosurface.  \emph{Top panel } pictures two $\ell$counterrotating tori $\cc^-_\times<\cc_\times^+$,  the inner corotating torus and the outer counterrotating torus  is accreting onto the central \textbf{BH}, accretion flux is stopped during \textbf{RAD}  integration.  \emph{Center panel}  focuses on a different \textbf{RAD} view,  to focus on the region close to the \textbf{BH} and to  colliding region. \emph{Bottom panel}  shows a second collision phase  when the matter flow from the outer counterrotating torus impacts on the inner corotating accreting torus. Tori, centered on the \textbf{BH}, are coplanar and  orbiting  on the equatorial plane of the  central  Kerr   \textbf{BH}. {The \emph{entire} \textbf{SMBH-RAD} system is tilted relative to the observer}.}\label{Fig:Fig3D}
\end{figure}
{In fact, the forces balance  condition for the accretion torus  can be encoded in     two functions defining each  \textbf{RAD} component:
the  torus  fluid (critical) specific angular momentum: $\ell^{\pm}_{crit}(a;r):\;\partial_r V_{eff}=0$,  defining the  critical points  of the hydrostatic pressure in the torus, see Eq.\il(\ref{Eq:continuus}) and the function
$K^{\pm}_{crit}(a;r,\ell): \; V_{eff}(a;r,\ell^{\pm}_{crit})$,    for  counterrotating and corotating fluids respectively,  $\ell_{crit}$ is present as a fundamental feature of the theory of accretion disks \cite{abrafra}. In the following, we use $\ell$ and $K$ generally  as model parameters and  parameter values, while  $\ell_{crit}$ and $K_{crit}$ are the functions of $a/M$ and $r$.   We consider these functions extensively in the following. Functions $\{ \ell^{\pm}_{crit}(a;r), K^{\pm}_{crit}(a;r)\}$  have been of great methodological importance in the development of the \textbf{RAD} model \cite{ringed}, through the definition of the set  of the functions $\{\ell_{crit}^i\}_{i=1}^n$, $\{K^i_{crit}\}_{i=1}^n$ (the \textbf{RAD} order $n$ is the number of its components) which is also the basis for the definition of effective potential  of the agglomerate \cite{ringed}. In the  model  adopted  for each component of the \textbf{RAD}  considered here, the specific angular momentum of the torus lies on a set of  level surfaces  $\ell^i=\ell_{crit}^i(a;r)=$constant, as  the parameter $K^i=K_{crit}^i(a;r)=$constant. Functions $\{\ell_{crit}^{\pm},K_{crit}^{\pm}\}$ are pictured in Figs\il\ref{Fig:Stopping}--while
further considerations are  reflected  in
Figs\il\ref{Fig:JohOCanonPro} and Figs\il\ref{Fig:Hologra}.
\begin{figure}
\includegraphics[width=1\hsize]{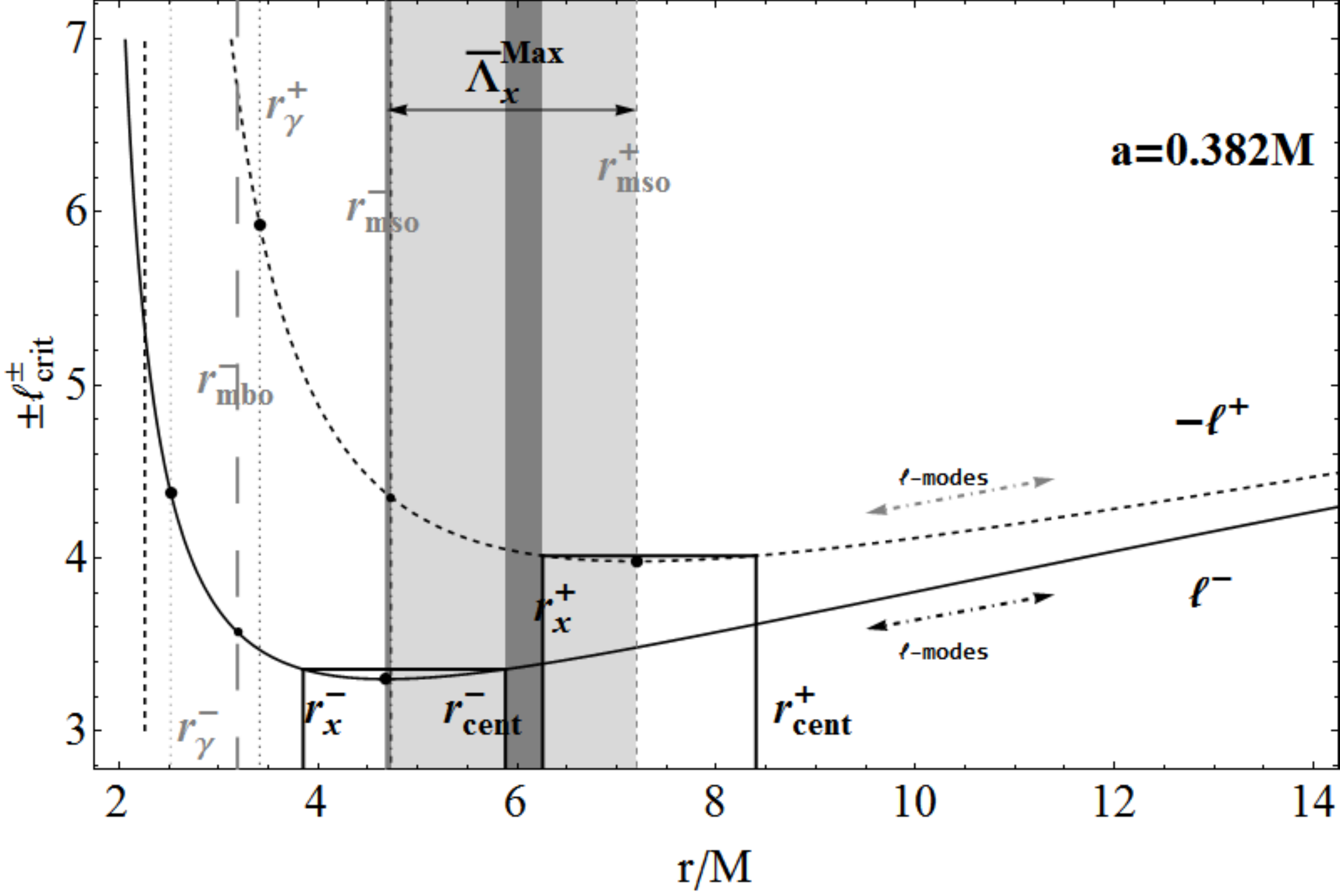}
\includegraphics[width=1\hsize]{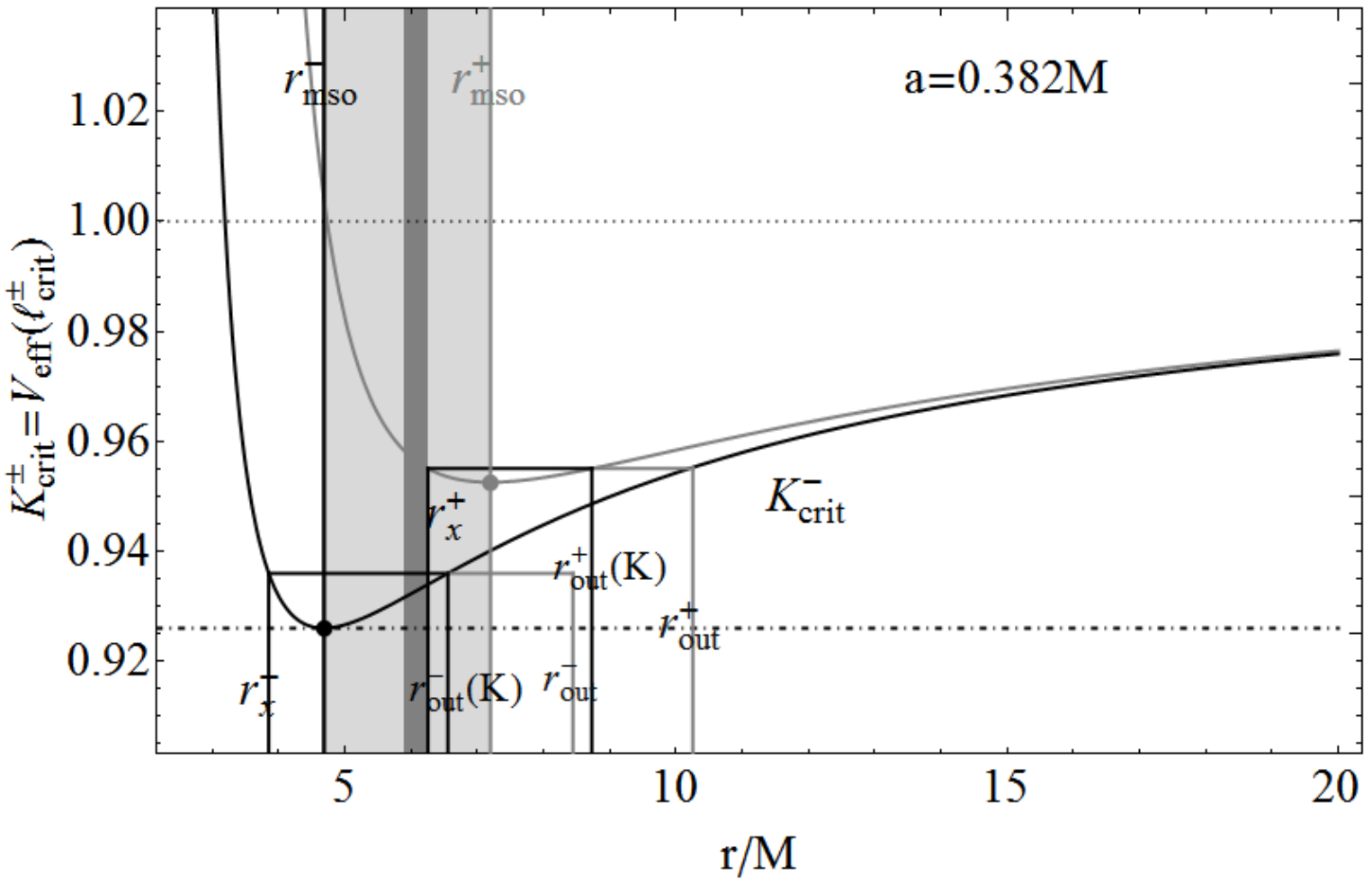}
\caption{\label{Fig:Stopping} $\ell$-modes representation. \emph{Upper panel}: fluid specific angular momenta $\mp\ell_{crit}^{\pm}$ for counterotating  and corotating tori respectively, indicating the pressure extreme points, in the spacetime of a \textbf{BH} with spin $a=0.382 M$ as function of $r/M$.
The \textbf{RAD} perturbation  $\ell$--modes reduces to shifts on the  each curve $\ell^{\pm}$.
Tori centers,  $r_{cent}^{\pm}$, and accreting points $r_{\times}^{\pm}$ of a   tori couple $\cc_{\times}^-<\cc_{\times}^+$ are shown  (lines $\ell=$constant),  $r_{\gamma}^{\pm}$ are the photon circular orbits--see Figs\il(\ref{Fig:Fig3D},\ref{Fig:teachploeff}) and Fig.\il(\ref{Fig:Hologra}) for related density surfaces. Light-gray region is the limiting  maximum spacing  of the two accreting tori in this spacetime, black continuum lines are the marginally stable orbits $r_{mso}^{\pm}$,  the gray  region is the limiting, maximum spacing for the specific tori  in the couple,   for growing $K_\ell$-modes of the inner corotating tori--see also Figs\il(\ref{Fig:Truem}) and (\ref{Fig:Truema}).
\emph{Bottom panel}: Plot of $K_{crit}^{\pm}$ as function of $r/M$,
inner edge $r_{\times}^{\pm}$ and outer edges of accreting tori  for a colliding couple are also shown, collision region is enlarged.
Light-gray region is the maximum limiting \emph{spacing}  for the two accreting tori.
each torus elongation on the equatorial plane is set by the    $K_{crit}=$constant lines. Gray  shades the same region as in upper-panel.
Note that for an accreting torus  plots $K_{crit}(r)$  do not directly locate the  torus center,  neither can be easily related to the effects of $\ell$ or $K$-modes.  For non accreting  (quiescent) tori, lines  $K_{crit}(r)=$constant locate the  inner edge of an accreting  torus $r_{\times}$  and the \emph{center}  $r_{cent}$ {(in figure $r_{out}(K)$)} an $\ell$corotating   couple of tori  with different  angular momentum,  see for a throughout discussion  and main applications \cite{ringed}.
Figs\il(\ref{Fig:Hologra}) and Figs\il(\ref{Fig:JohOCanonPro}) for further details.
Limiting  values $K_{mso}^-<K_{\lim}\equiv1$ are represented by    dot-dashed  and dotted lines  respectively.
}
\end{figure}
Curves $K^{\pm}_{crit}(a;r,\ell)$ locate the tori centers, provide information on torus elongation (see Fig.\il\ref{Fig:Quanumd}) and  density and,  for a  torus accreting onto the central \textbf{BH}, determine the inner and outer torus edges.
\begin{figure}
\includegraphics[width=9.510cm]{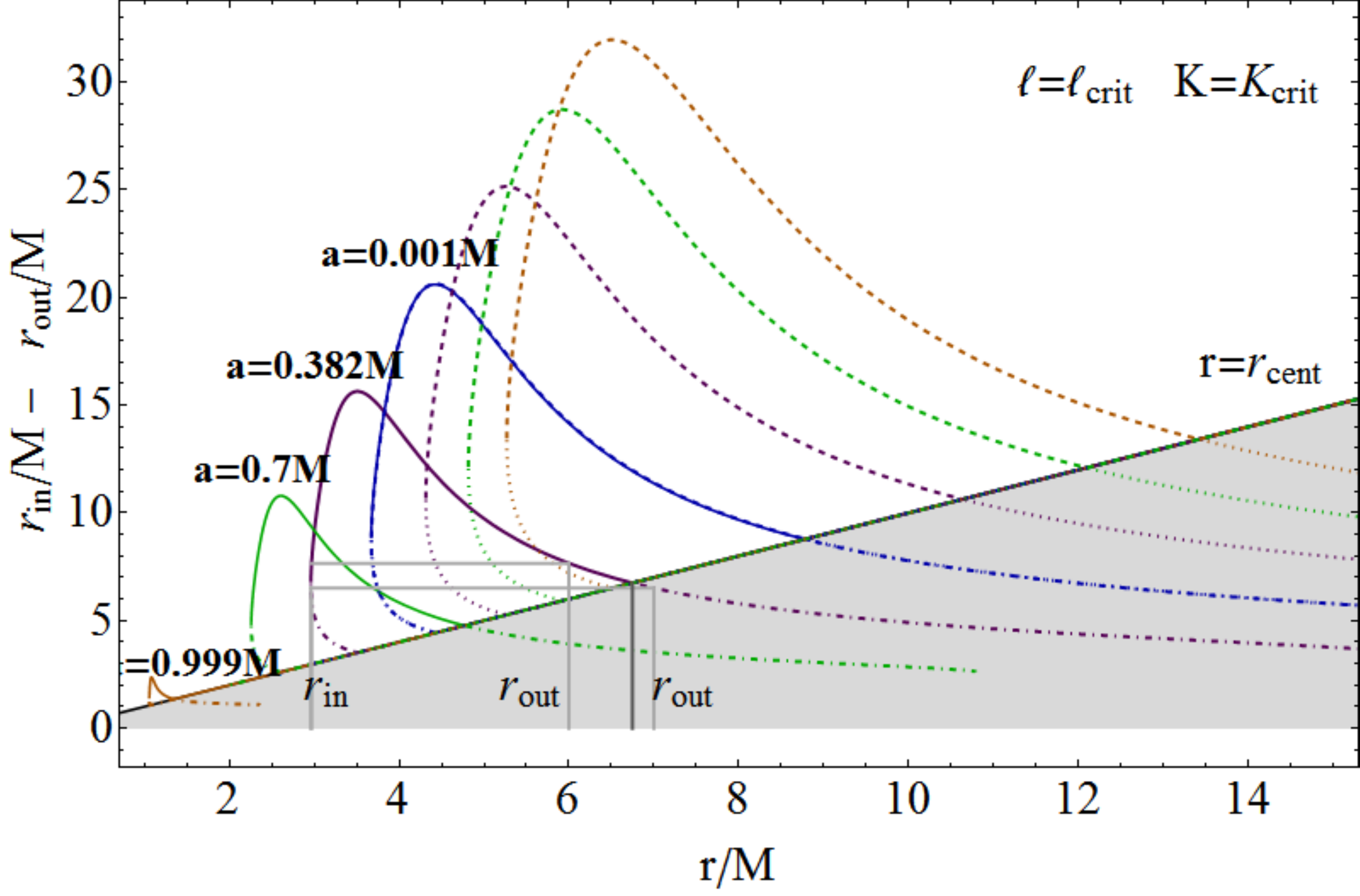}
\includegraphics[width=9.410cm]{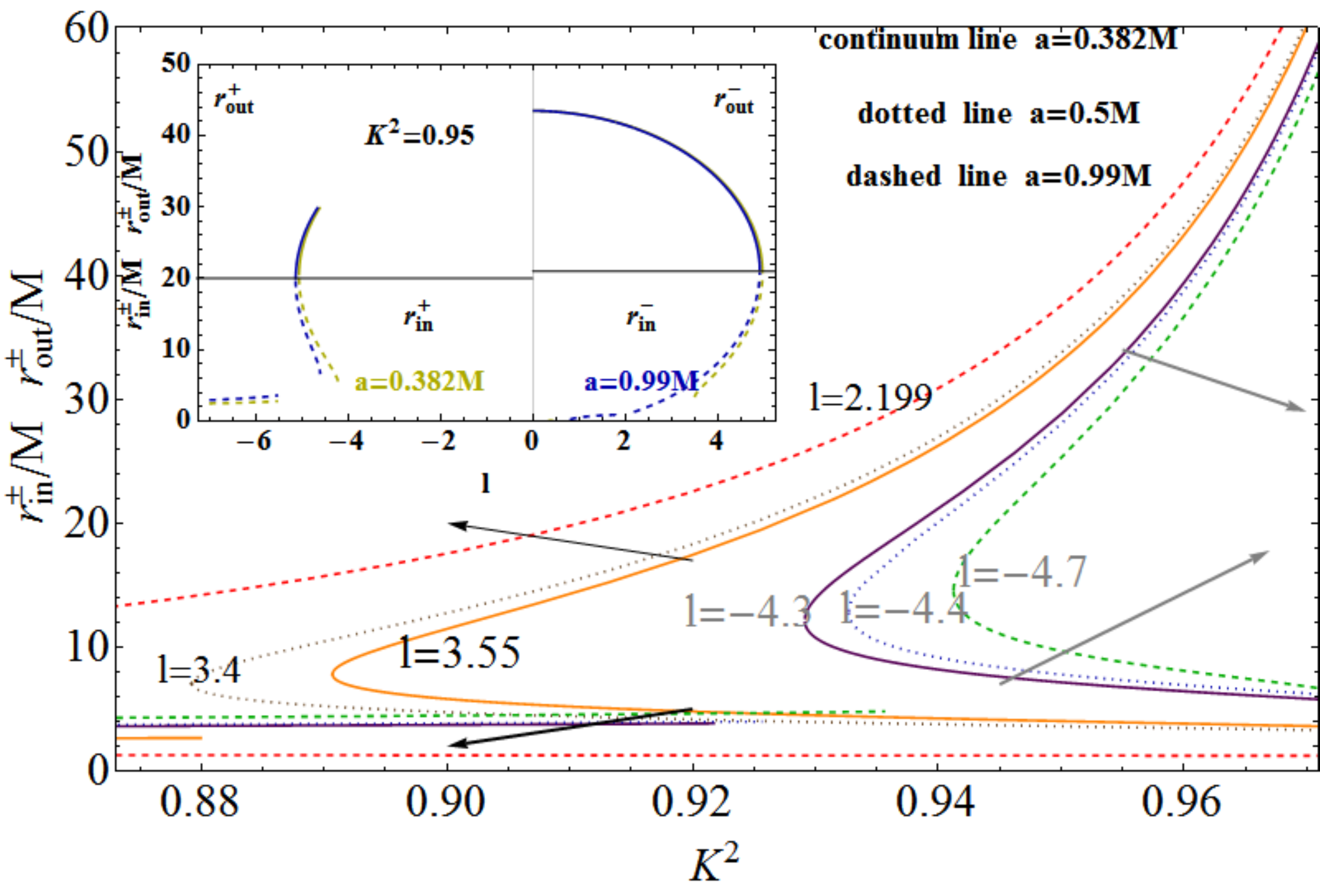}
\caption{\emph{Upper panel}: plots of  the inner $r_{in}$ and outer $r_{out}$ edges of \textbf{RADs} accretion  tori orbiting different \textbf{SMBHs}  (spins $a$ on curves), as functions of $r/M$.
$r_{in}$ and  $r_{out}$  are introduced in  Eqs\il(\ref{Eq:r-in-out-papers-out}). Shaded region is $r<r_{cent}$,
$r_{cent}$ is  the line of torus centers where   at fixed $r/M$ there is  $r_{cent}=r$.
Corotating tori are represented  in continuum  and dotdashed curves, thick dashed and thick dotted  curves are for counterrotating tori.
Horizontal lines on each curve
fix a torus (constant angular momentum), torus  inner and outer edges are  signed on the figure--see discussion on Eq.\il(\ref{Eq:r-in-out-papers-out}).
  \emph{Bottom panel}: Inner $r_{in}^{\pm}$ and outer edges $r_{out}^{\pm}$ of accretion torus, at different  spins and
 fluid specific angular momentum for corotating  (\textbf{-}) and counterrotating (\textbf{+}) fluids.
 as functions of $K^2$ where $K\in[K_{mso},1[$.
 Arrows  follow increasing values of the \textbf{SMBHs} spins.
 Inside panel: radii $r_{in}^{\pm}$ and  edges $r_{out}^{\pm}$  as functions of specific angular momentum $\ell$, for corotating and counterrotating  fluids, for
 \textbf{SMBH} spins $a=0.99M$ and $a=0.382M$ at $K^2=0.95$.
}\label{Fig:JohOCanonPro}
\end{figure}
\begin{figure}[h!]
\centering
\begin{tabular}{c}
\includegraphics[width=1\columnwidth]{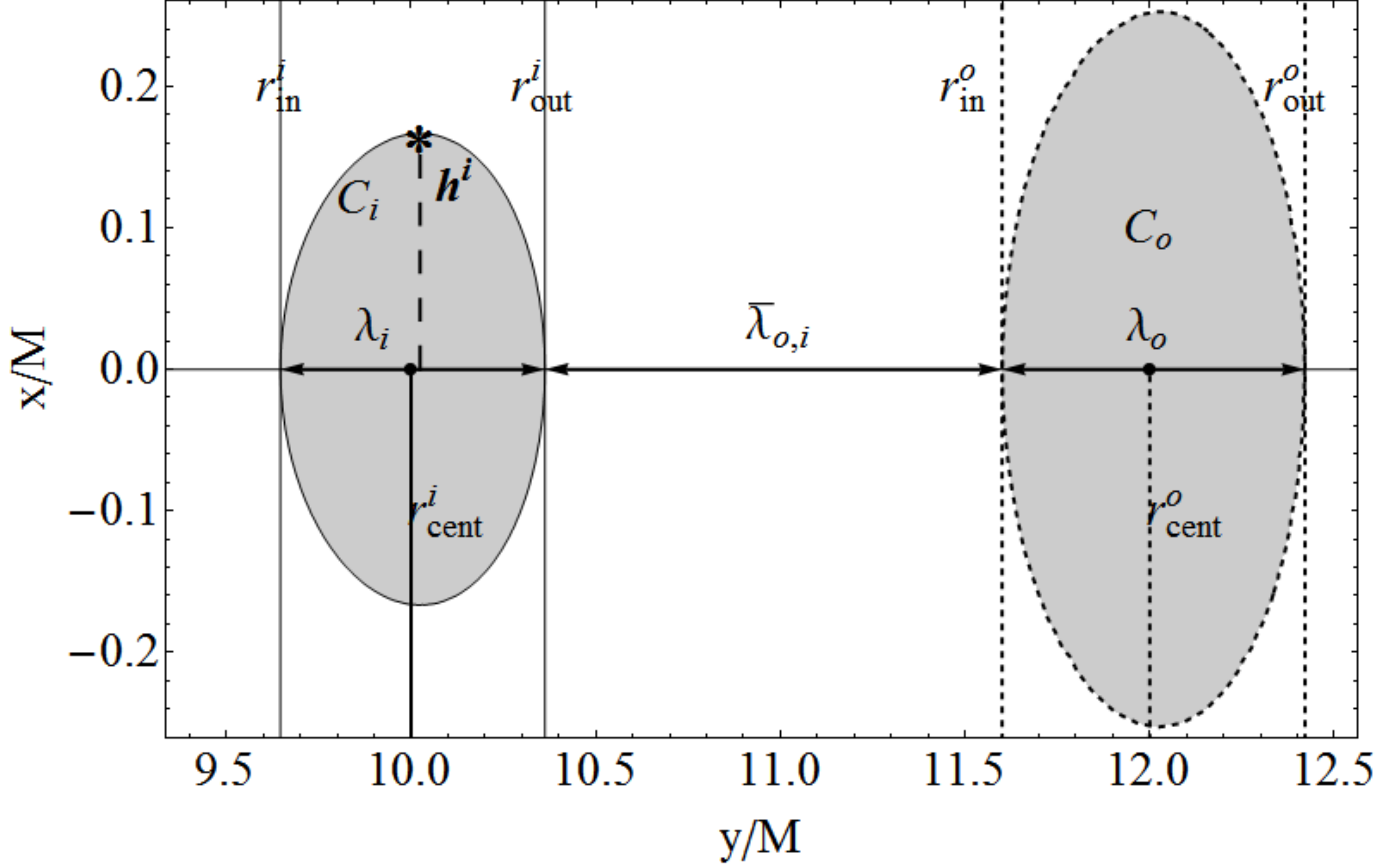}
\end{tabular}
\caption[font={footnotesize,it}]{\footnotesize{Cross sections on the equatorial plane of the consecutive   Boyer surfaces of two separated  rings $C_i<C_{o}$, with boundaries $\partial C_i$ and $\partial C_{o}$, the tori centers  $r_{min}=r_{cent}$ are signed with points and the lines $r_{cent}=$constant. With $r/M=\sqrt{x^2+y^2}$ and  $(x,y)$ are Cartesian coordinates. The inner $r_{in}$ and outer $r_{out}$ edge are also signed, $\lambda_a$ for $a\in\{i,o\}$ are the tori elongation and $\bar{\lambda}_{o,i}$ is the spacing among the tori. { $h^i$  is the  height associated to  the  $C_i$ torus, or   the maximum point  of the surface $\partial C_i$ } The $\ell$corotating  rings, rotate around a black hole  attractor with spin
$a=0.75M$ and specific angular momentum
$\ell_i=-4.2897$ and
$\ell_{o}=-4.41883$.}}\label{Fig:Quanumd}
\end{figure}
Angular momentum $\ell^{\pm}_{crit}(a;r)$ does \emph{not}  coincide with the Keplerian momentum  $L$, describing  the free particle angular momentum, but it is  related to this function through Eq.\il(\ref{Eq:continuus})--an extensive discussion on the role of $L$ and $\ell$ in the accretion tori and accretion processes can be found in  \cite{pugtot,PuMonBe12}, while
 $\ell^{\pm}_{crit}(a;r)$ and $K^{\pm}_{crit}(a;r)$  have been widely investigated in the developing of the \textbf{RAD} model in  \cite{ringed,dsystem}.
Our investigation is focused on the task to find  the  ranges of variation of the main fluid features, as the  specific angular momentum, or fluid density related to the $K$ parameter,  during different evolutionary phases of the \textbf{RAD}: from formation, growth, to  accretion of its components and tori collisions.  This aspect allows a  general applicability of the results found here eventually to other accreting disk models for the aggregate components.
 We  will briefly  resume this argument concerning tori collisions in Sec.\il(\ref{Sec:more-l}).
In general  model of accretion disks the angular momentum of matter in accretion disks  is considered to be sufficiently high   for  the centrifugal force to be  a predominant component of the four forces regulating the disks balance (centrifugal, gravitational, pressure and  magnetic forces), the  Bondi quasi-spherical  accretion constitutes an example of situation  when the condition ($|\ell|>|L|$) is not fulfilled.
In general    accretion disks, there must be an extended region where there is  $\mp\ell^{\pm}>\mp L^{\pm}$ (explicitly considering also the counterrotating fluids) in the same  orbital region. This is assumed to hold for a general  accretion torus with a general   angular momentum distribution. In this work we  provide limit to the inequality $|L|<|\ell|$ according to different  conditions  set on the \textbf{RAD}.}
Then, a particularly attractive feature of tori with constant specific
angular momentum $\ell$ is that   the  $\ell$corotating tori and  particularly  $\ell$counterrotating tori  are   constrained   by  the Kerr geometry \emph{geodesic structure}\footnote{It is worth specifying that this strong dependence of the model on the geometric properties  of spacetime induced by the central attractor enables us to  apply to a certain extent the   results found here to  different models of accretion disks \cite{abrafra}.}: this comprise   the \emph{{marginally stable circular orbit}}, $r_{\mso}^{\pm}$, the \emph{{marginally bounded circular orbit}}, $r_{\mbo}^{\pm}$ and  the \emph{{marginal circular orbit}} (photon orbit) $r_{\gamma}^{\pm}$-Fig.\il(\ref{PlotdisolMsMb}).
It is also necessary to  introduce also the   radii $r_{{(\mbo)}}^{\pm}$ and $r_{{(\gamma)}}^{\pm}$, defined as the solutions of  the  following equations:
\bea&&\label{Eq:system}
\text{\textbf{(--)}}\; r_{{(\mbo)}}^{\pm}:\;\ell_{\pm}(r_{{(\mbo)}}^{\pm})=\ell_{\pm}(r_{\mbo}^{\pm})\equiv \ell_{\mbo}^{\pm},
 \\\nonumber
 &&\text{\textbf{(--)}}\;r_{({{\gamma}})}^{\pm}: \ell_{\pm}(r_{({{\gamma}})}^{\pm})=\ell_{\pm}(r_{{{\gamma}}}^{\pm})\equiv \ell_{{{\gamma}}}^{\pm},
  \\&&
\text{  where  }\quad r_{{{\gamma}}}^{\pm}<r_{\mbo}^{\pm}<r_{\mso}^{\pm}<r_{{(\mbo)}}^{\pm}<
  r_{({{\gamma}})}^{\pm}
  \eea
--see Fig.\il(\ref{PlotdisolMsMb}). The radii $r_{(\gamma)}$ and $r_{(mbo)}$ are related to the radii $r_{\gamma}$ and  $r_{mbo}$      through the angular momenta   $\ell_{\gamma}$ and $\ell_{mbo}$ respectively.
Accordingly,  there are the following critical values of  the spin:
\bea\nonumber
&&a_{1}\equiv0.4740M:\quad r_{(\mbo)}^+=r_{({\mathrm{\gamma}})}^-,
\\\nonumber
&&
 a_{2}=0.461854M:\;r_{(\mbo)}^-=r_{\mso}^+,\\
 && \label{Eq:spins}
  a_{3}\equiv0.73688 M:\;r_{({\mathrm{\gamma}})}^-=r_{\mso}^+,
\eea
see Figs\il(\ref{PlotdisolMsMb}).
The role of  radii  $r_{(\gamma)}$ and  $r_{(mbo)}$ can be easily seen by considering the lines
 $\ell=$constant on $\ell(r)$ in Fig.\il(\ref{PlotdisolMsMb}). It is simple to see that,
consistently   with  most of the axi-symmetric  accretion tori  models,
the \emph{(stress) inner edge} of the  accreting  torus is at  $r_{\times}\in]r_{\mbo},r_{\mso}]$, while the torus    is centered at   $r_{cent}>r_{\mso}$
\cite{Krolik:2002ae,BMP98,Agol:1999dn,Afshordi:2002gi}.
This consistency with the relevant ranges for most of the  accretion tori is a good indication of the reliability of this approach to \textbf{RADs} at different systems.

In each couple, the two tori  are generally severely  constrained in  mutual relations of the ranges of  variation  of fluids  specific angular momentum, tori edges location,  and  the attractor spin: it can be shown that  the center of each torus  turns to be located in    $[r_{mso},r_{(mbo)}]$,  $[r_{(mbo)}, r_{(\gamma)}]$, $r>r_{(\gamma)}$, whereas  the inner edge  of each torus   lies  in these ranges or in  $[r_{mbo},r_{mso}]$, in dependence of the range of  the parameter $\ell$,  and if  the  tori are accreting or non-accreting.
{Any accreting torus is in fact   a toroidal solution of the Euler equation (\ref{E:1a0}) with  proper boundary conditions admitting a  cusp, a point of hydromechanical  destabilization. The vanishing  pressure implies  a non equilibrium   regime at radii $r\leq r_{\times}$ with the consequent dynamical mass loss from the disk  induced by the Roche lobe overflow (a  self-regulating process for   the accretion rate of  many kinds of accreting disks--see\cite{abrafra}). The range of fluid specific angular momentum values where closed cusped solutions can be found is  restricted to  $]\mp\ell_{mso}^{\pm},\mp\ell_{mbo}^{\pm}[$ respectively for counter-rotating and corotating tori.
 } It is straightforward to show that the location of the \emph{center} ({point of fluid maximum hydrostatic pressure and density}) is at $r_{cent}\in]r_{\mso},r_{(\mbo)}[$.
If $\pm\ell^{\mp}>\pm\ell_{\gamma}^{\mp}$, then the torus  center is located in $r_{cent}^{\mp}>r_{(\gamma)}^{\mp}$, this torus has no critical configuration (according to P-W mechanism)-\cite{ringed,open}.
We have therefore a way to place the critical pressure points  according to the range of variation of the fluid specific angular momentum. We refer to Table\il(\ref{Table:pol-cy-multi}) which  explicitly  introduces the   ranges $\mathbf{L1}$, $\mathbf{L2}$ and $\mathbf{L3}$ of the fluid momentum $\ell$.

\medskip

In the following, it will be convenient  to use the notation
$\cc^{\pm}$ ($\cc_{\times}^{\pm}$) for {non-accreting } (accreting) tori,    $\pp^{\pm}$  for a torus which may be  {non-accreting } or  accreting (according to hydro-gravitational destabilization   due to  P-W mechanism \cite{abrafra}).  Boyer surfaces and density plots with indication on the relative notation are   in  Figs\il(\ref{Fig:teachploeff}) and Figs\il(\ref{Fig:teachplo}).

 Also, symbols $\lessgtr$ $\left(\lessgtr_{\mathbf{\times}}\right)$
 for two tori refer to the relative position of the {tori centers}   $r_{cent}$ (accretion points or unstable points $r_{\times})$:
thus, for example, we use short notation $(\pp^-<\pp^+,\pp^-<_{\mathbf{\times}}\pp^+)\equiv \pp^{-}\ll_{\mathbf{\times}}\pp^+$  that means $r_{cent}^-<r_{cent}^+$  and  $r_{\times}^-<r_{\times}^+$--see Figs\il(\ref{Fig:teachploeff}) and Table\il(\ref{Table:pol-cy-multi}).

{We can  proceed  now by solving  the hydrodynamic equations (\ref{Eq:scond-d}) with a barotropic equation of state associated to  each torus of the four  couples    $\pp_i^{\pm}<\pp_o^{\pm}$ and  $\pp_i^{\pm}<\pp_o^{\mp}$, respectively  within the conditions  $r_{in}^o\geq r_{out}^i$  for the  inner $r_{in}^o$  (outer $(r_{out}^{i})$) edge of the outer (inner) torus.
Here the   fluid    specific angular momentum  varies in the  ranges    $\ell \in [\ell_{mso},\ell_{mbo}]$, $[\ell_{mbo},\ell_{\gamma}]$  or $\ell>\ell_{\gamma}$. }
We refer to Table\il(\ref{Table:pol-cy-multi}) for a brief summary of the introduced notation including brief description.
The analysis can be then  further simplified by considering appropriate boundary conditions on  a properly defined  effective potential for the couple of   tori.
In fact, we may introduce, as in \cite{ringed},
 an  effective potential $\left.V_{eff}^{\mathbf{C}^2}\right|_{K}$ and  also alternative $V_{eff}^{\mathbf{C}^2}$  potential for the system of the two tori
\bea\label{Eq:def-partialeK}
&&\nonumber\left.V_{eff}^{\mathbf{C}^2}\right|_{K}\equiv V_{eff}^{i}\Theta(-K_i)\bigcup V_{eff}^{o}\Theta(-K_o)
\\\nonumber
&&\mbox{and}
\\&&\nonumber
V_{eff}^{\mathbf{C}^2}\equiv V_{eff}^{i}(\ell_i)\Theta(r_{cent}^{o}-r)\Theta(r-r_+)\bigcup
\\\label{Eq:def-partialeK}
&&\nonumber \qquad \qquad V_{eff}^{o}(\ell_o)\Theta(r-r_{cent}^{i}),
\eea
where $\Theta$ is  the Heaviside (step) function  such that  for example $\Theta(-K_i)=1$  for $V_{eff}^{i}<K_i$ and $\Theta(-K_i)=0$ for $V_{eff}^{i}>K_i$.
The constraints in  $V_{eff}^{\mathbf{C}^2}$ are provided through the parameters $\{K_i,K_o\}$, while the potential  $V_{eff}^{\mathbf{C}^2}$ is determined by the fixed centers $(r_{cent}^{i},r_{cent}^{o})$ only.  The effective potential of the couple can  therefore  be written as a  coupling of the effective potentials for each torus.  In both cases, with $V_{eff}^{\mathbf{C}^2}$ and $\left.V_{eff}^{\mathbf{C}^2}\right|_{K}$, the two tori  are determined by the couple of the parameters   $K_i=$constant and $K_o=$constant (where $\ell_i$ and $\ell_o$ are fixed).

The results  show that
two   {non-accreting} $\ell$corotating and $\ell$counterrotating tori might orbit any Kerr  attractor,  if their specific angular momenta are properly related.
However, the emergence of  accretion  occurs in the following   four  couples only:
 \bea&&\label{Eq:mack-asc}
 {\textbf{(i)}}:\;  \cc_{\times}^{\pm}< \cc^{\pm}, \quad{\textbf{(ii)}}:\; \cc_{\times}^{+}< \cc^{\pm},
 \\
 &&\nonumber{\textbf{(iii)}}:\; \cc_{\times}^{-}< \cc^{\pm},\quad {\textbf{(iv)}}: \;\cc_{\times}^{-}< \cc_{\times}^{+}
 \eea
see Figs\il(\ref{Fig:teachploeff}) and Figs\il(\ref{Fig:teachplo}).

We can   describe  results as it follows:
for a $\ell$corotating couple ({\textbf{(i)}}), or  a Schwarzschild (\emph{static} ($a=0$)) attractor,    only the inner torus of  a couple can be  accreting (this  can be also seen by considering   the curves $\ell$=constant for $\ell(r)$ Fig.\il(\ref{PlotdisolMsMb})-right).

In the  $\ell$counterrotating  case, an accreting   \emph{corotating} torus {must  be}  the inner one of the couple while the outer counterrotating torus can be non-accreting or in accretion-Fig.\il(\ref{Fig:teachploeff}).

If there is a  $\cc_{\times}^-$ torus, or   if the attractor is {static},  then  no  inner (corotating or counterrotating) torus {can} exist, and  then  $\cc_{\times}^-$ is part of $\cc_{\times}^-<\cc^-$  couple as in Figs\il(\ref{Fig:teachplo}) or of a $\cc_{\times}^-<\pp^+$  one as in Figs\il(\ref{Fig:teachploeff}), {that is the inner accreting torus of a couple where the outer is quiescent (in this case it can be corotating or counterrotating) or also in accretion and in this case it has to be counterrotating.}

A  corotating torus can be the outer  of a couple of tori with an   inner counterrotating accreting torus. Then the outer torus may be corotating (non  accreting), or counterrotating in accretion or  {non-accreting}--Figs\il(\ref{Fig:teachplo}).    Both  the inner corotating  and the outer counterrotating  torus of the couple  can accrete onto the attractor.
A counterrotating  torus can therefore reach the instability being the inner one of a $\ell$corotating or $\ell$counterrotating couple as in  Figs\il(\ref{Fig:teachploeff}), or the outer torus of a  $\ell$counterrotating   couple as shown in Figs\il(\ref{Fig:teachplo}).

\begin{table*}
\caption{{Lookup table with the main symbols and relevant notation  used throughout the article.}}
\label{Table:pol-cy-multi}
\centering
\resizebox{.9\textwidth}{!}{%
\begin{tabular}{ll}
 \hline \hline
$ \cc$&   cross sections of the closed Boyer surfaces (quiescent non accreting  torus)\\
$ \cc_{\times}$&   cross sections of the closed cusped  Boyer surfaces (accreting torus)\\
$ \oo_{\times}$&   cross sections of the open cusped  Boyer surfaces (proto-jet)
 \\
 $\pp$& any of the  topologies $(\cc, \cc_{\times},\oo_{\times})$
 \\
 $(r_{in}, r_{out})$& inner and outer edge of $\cc_i$ ring
 \\
 $\pm$&counterrotating/ corotating
  \\
 $\ell$counterrotating /$\ell$corotating tori& $\pp^{\pm}-\pp^{\mp}$ /$\pp^{\pm}-\pp^{\pm}$  tori
 \\
$r_{cent}$& center of outer Roche lobe (torus center)
\\
$r_{\times}$& accretion point (stress inner edge of accreting  torus)
\\
$r_{\jj}$& unstable point in open configurations
\\
$r_{coll}$& contact point in collisions among two  quiescent tori
\\
$r_{{(\mbo)}}^{\pm},r_{{(\gamma)}}^{\pm}$& radii of the complement geodesic structure in Eq.\il(\ref{Eq:system})
\\
$\mathbf{L1}$ (or $\mathbf{L1}^{\pm}$)& specific angular momentum range: $\mp\ell^{\pm}\in[\mp\ell_{mso}^{\pm},\mp\ell_{mbo}^{\pm}[$
\\
$\mathbf{L2}$ (or $\mathbf{L2}^{\pm}$)& specific angular momentum range: $\mp\ell^{\pm}\in[\mp\ell_{mbo}^{\pm},\mp\ell_{\gamma}^{\pm}[$
\\
$\mathbf{L3}$ (or $\mathbf{L3}^{\pm}$)& specific angular momentum range: $\mp\ell^{\pm}\geq\ell_{\gamma}^{\pm}$
\\
$\lessgtr$& rings sequentiality according to the centers $r_{cent}$ (inner/outer tori) i.e. $r_{cent}^i\lessgtr r_{cent}^i$
\\
$\pp^i,\pp^o$& generally inner and outer configurations of a couple i.e. $\pp^i<\pp^o$
\\
$\lessgtr_\times$ &rings sequentiality according to the critical points $(r_{\times},r_{\jj})$ i.e. $r_{\times}^i\lessgtr r_{\times}^i$
\\
$\ell_{i/i+1}\equiv\ell_i/\ell_{i+1}$&
 ratio in  specific  angular momentum of $\cc_i$ and $\cc_{i+1}$
 \\
 ${\mathbf{C}}_{coll}/\hat{r}$ &colliding couple  with $r_{coll}$ located on a radius $\hat{r}$ outer: torus is quiescent
 \\
$\mathbf{{{C}}}_{\times}$& colliding couple: the outer torus is accreting
 \\
  $\mathbf{C}_{coll}^{\times}$&  colliding couple:    combination of the two processes  ($\mathbf{C}_{coll}$, $ \mathbf{C}_{\times}$), where  $r_{{out}}^{i}=r_{\times}^o=r_{coll}$
 \\
 $\mathbf{{{C}}}_{\times}/r_{\mso}^+$&  colliding couple:  the outer torus is accreting where $r_{\times}^+\equiv r_{\mso}^+$
 \\
 $\mathbf{{{C}}}_{\times}/r_{\mbo}^+$& colliding couple:  the outer torus is accreting where $r_{\times}^+\equiv r_{\mbo}^+$
 \\
\hline\hline
\end{tabular}}
\end{table*}
Then it is worth noting that if  the \emph{accreting} torus is \emph{counterrotating} with respect to the Kerr attractor, i.e.  a $\cc_{\times}^+$, then  there is \emph{no} inner  counterrotating torus, but a couple may be formed as a   $\cc_{\times}^+<\cc^{\pm}$ or as a $\pp^-<\cc_{\times}^+$ one.

Couples
$\cc_{\times}^{\pm}<\cc^{\pm}$  and  $\cc_{\times}^-<\cc^{+}$  having only the inner torus in accretion, while the outer torus is \emph{quiescent},  may form around any attractors with $a\in[0, M]$. 
A   $\cc_{\times}^{-}< \cc_{\times}^{+}$  couple, featuring the occurrence of double accretion, as shown in  Figs\il(\ref{Fig:teachploeff}),  is possible  in all Kerr spacetimes where   $a\neq0$. However, the  couple is subjected to  several constraints on the fluid specific  angular momentum:  if the  \textbf{BH}  dimensionless spin $a\lessapprox a_{1}$, and the
  lower must be the specific angular momentum $\ell^-$ of  the inner corotating torus.
{
Thus, if a counterrotating   torus is   accreting onto the central black hole,   there could be an inner corotating  torus, which may also accrete onto the spinning attractor,  acting as a screening torus for the matter flow of the accreting counterrotating outer torus.  However, the lower  is the Kerr black hole  dimensionless spin, say  $a\lessapprox a_1$,
 the lower must be the corotating torus specific angular momentum--see
\cite{dsystem}.
Nevertheless, a counterrotating  torus can therefore reach the instability being the inner or the outer ring of an $\ell$counterrotating couple.}
In fact,  couples $\pp^+<\cc^-$, with an outer corotating  and quiescent torus,  have to be observed  in  any spacetime $a\in[0,M]$, but only for slower spinning \textbf{BHs} with
   $a\in[0,a_{2}[$,  the corotating non-accreting (quiescent) torus  $\cc^-$   approaches  the instability  (i.e. the inner edge  is $r_{\times}\gtrapprox r_{\mso}^-$).
Note that the  {faster}  is the Kerr attractor ($a\gtrapprox a_{3}$),   the {farther away}  should be the outer torus to prevent collision (i.e. the torus center $r_{cent}>r_{({\mathrm{\gamma}})}^-)$.
{The system consisting  of an  inner accreting  counterrotating  torus and an outer equilibrium corotating torus may be formed
in any spacetime,  but the  {faster}  is the attractor,  the {farther away}   should be the outer torus and this also implies the outer torus has large specific angular momentum \cite{ringed,open,long,Pugliese:2018zlx,multy}.}
Finally, we  emphasize that  for the tori  couples with inner corotating tori,  in particular for the cases   $\mathbf{(ii)}$ and $\mathbf{(iii)}$ considered  in  Eq.\il(\ref{Eq:mack-asc}),  the possibility of penetration of matter into ergoregion  $\Sigma_{\epsilon}^+$,  or the formation of  extended toroidal matter configurations contained in this region, can occur;  we refer to \cite{pugtot,ergon,ringed,open,long,multy,Proto.Jets} for a detailed discussion on this possibility, while  in Sec.\il(\ref{Sec:coroterf}) we briefly  add  further considerations on this aspect.

\medskip
\textbf{General comments on methods}
{We close  this section with some methodological notes, introducing next section\il(\ref{SeC:coll}) dealing with tori collisions. In  the investigation of \textbf{RAD} agglomerate and its internal dynamics, from its formation and evolution   to  the   \textbf{RAD} tori collisions,  we have  taken into account \emph{all} the evolutive possibilities of  the model through the parameter analysis, constraining ultimately the \textbf{RADs} order and the distribution of momentum depending on the central \textbf{SMBH} spin  \cite{ringed,multy}.
Here, having fixed the \textbf{RAD} order as $n=2$, we still  have different degrees of freedom provided by the    $K$ and  $\ell$ parameters; we will better look  at this aspect in Sec.\il(\ref{Sec:const})  introducing   the \textbf{RAD}  $\ell$-modes and $K$-modes--\cite{ringed}}.
More specifically,  as also mentioned  above, tori collisions and \textbf{RAD} formation are strongly constrained by the geodesic structure of the Kerr spacetime, this fact has been highly emphasized in \cite{ringed,open,dsystem}.
This feature  constitutes indeed a methodological advantage of the model, as these  limits, because of their geometric nature due to the spacetime   properties of  the central \textbf{BH} attractor, have in fact a major role in determining the force  balance in the \textbf{RAD}  in  even more complex situations where, for example, other effects are taken into account, as the magnetic field distribution,  which makes these models highly predicting as emerges  in   agreements with of many GRMHD  analysis.  Considerations developed here   considerably constrain some important features of the an accretion tori formation \cite{long,open}, and the possibility of  the formation of a tori agglomerate.
A further significant  consequence is that such kind of investigation  provides  \emph{ranges}  of variation  of certain parameter set. For the tori collisions to occur, we have to fix    parameter values for one specific situation represented by (arbitrarily chosen) initial data, $(\ell_i,\ell_o,K_i,K_o)$, for the tori couple orbiting a specific \textbf{SMBHs} with fixed dimensionless spin parameter $a/M$.
 The spacetime geodesic structure and its complement structure as defined in Eq.\il(\ref{Eq:system}),  as founded in \cite{ringed,open} in the developing of the \textbf{RAD} model,  actually influence the aggregate composition through the function $\ell_{crit}$. We note that in the \textbf{RAD} context the  two functions  $(\ell_{crit}^i, \ell_{crit}^o)$  must be simultaneously considered in the study of the  couple. Figures \il(\ref{Fig:Stopping}), (\ref{Fig:teachploeff}), (\ref{Fig:Truem}) and (\ref{Fig:Truema})  well show the different situations for  corotating and counterrotating fluids, while further details  on this aspect are explored in  Sec.\il(\ref{Sec:theory-weell}).  The analysis of fluid specific angular momentum ranges solves only a part of the problem to constraint the \textbf{RAD}, because together with $\ell_{crit}$, it is necessary to consider   $K^{\pm}_{crit}$, and more generally the values of $K$-parameters which fix in fact many important accretion tori features, from the elongation in the equatorial plane  being related to tori density and, for example  the mass-accretion-rate--see Sec.\il(\ref{Sec:theory-weell}).
 The conditions on  the collisions  have much freedom, depending, at fixed $a/M$, on  the couple $(\ell,K)$ for the two tori.
From a methodological viewpoint, we have reduced the freedom represented by the choice of $\ell$ and $K$ parameter for any \textbf{RAD} component, by
 using  constraints  on  the fluids  critical pressure points,  locating the torus inner edge, $r_{in}$,   the torus center, $r_{cent}$, and the outer edge, $r_{out}$. The explicit use of these constraints has been also detailed in \cite{open,long}.
However, fixing only one or two of these radii, does not respond to the problem of the construction and constraining of the \textbf{RAD} structure and   evolution. In order to respond to this issue, the whole  set  of points generated by each toroidal  component of the \textbf{RAD},  must be \emph{simultaneously} considered  during the analysis of \textbf{RAD} constraints (see also the \textbf{RAD} function Eq.\il(\ref{Eq:def-partialeK})). In Sec.\il(\ref{Sec:const}),   constraints on the collision emergence will be provided.
The   points $\{r^j_{in},r^j_{cent},r^j_{out}\}$, for any torus $j$ of the \textbf{RAD}, are then studied in a  model defined by  $2n+1$  parameters, where $n$ is the \textbf{RAD} order (number of the tori composing the  \textbf{RAD} aggregate).
In this investigation  there is  $n=2$ and the  model parameters are then given by the pairs $\mathbf{p_i}\equiv (\ell_i,K_i)$ and $\mathbf{p_o}\equiv(\ell_o,K_o)$ and by the dimensionless spin $a/M$ of the central  \textbf{SMBH}--this is a $4+1$ parameters model\footnote{Then, we  are actually  interested in identifying the \emph{ranges}  of variation  of the $\{\mathbf{p_i},\mathbf{p_o},a/M\}$ parameters, while the exact values of there parameters for one special torus model could be easily found, by  fixing $ a / M $ and appropriate initial conditions \cite{dsystem}.}.
To simplify the analysis, it was convenient to introduce the notion  of the  \textbf{RAD} \emph{state}, consisting  in the  precise arrangement of the following characteristics of the couple: parameters   $(\ell_i, \ell_o)$ and $(K_i, K_o)$,  and relative location of the tori edges, topology  (if accreting or non accreting torus),  if in collision or not.
Then, a   ringed disk of the  order  $n=2$, with fixed  critical topology   can be generally in $n=8$ different states according to the relative position of the centers and rotation: $n=4$ different states if  the rings are $\ell$corotanting,  and $n=4$ for  $\ell$countorrotating rings. Considering also the relative location of points of minimum pressure, then the couple $(\mathbf{\pp_a-\pp_b})$  with different, but fixed topology,  could  be in $n=16$ different states. To this number of possibilities we have to add the different cases generated by the  $K$-modes, resulting in the variation of  elongation  and thickness of each torus. In this work, we reduce these possibilities--\cite{dsystem}.
This study applies also to  the case of  $n>2$:
 the $4$ couple  in  Eqs\il(\ref{Eq:mack-asc}) constitute the \textbf{RADs seeds}, and
the study of the case of larger  \textbf{RAD},  with $n>2$,  can be carried out by composing  the  four  couples considered here-\cite{ringed,open}.  In Sec.\il(\ref{SeC:poly})  the constraints are reformulated  in terms of fluid density.
In Sec.\il(\ref{SeC:coll}),  we investigate in detail conditions for the emergence of the collisions of the tori in  the cases presented  in Eq.\il(\ref{Eq:mack-asc}), considering the  relations between the fluid specific angular momentum
\section{Tori edges  and elongation}\label{Sec:theory-weell}
Tori elongation on the equatorial plane     for the  \textbf{RAD} ringed structure are illustrated in Figs\il(\ref{Fig:Quanumd}) with   other relevant quantities of the tori agglomerates, while in
 Figs\il(\ref{Fig:callneoc}) and  Figs(\ref{Fig:RTMe100dol})  we show  the elongation, and the tori inner and outer edges as functions  on  torus parameters $\ell$ and $K$  for corotating and counterrotating fluids.
As a sideline result to  the \textbf{RAD} investigation   a  comparative analysis of the corotating   and counterrotating  tori is obtained  in dependence  of the  central Kerr black hole spin-to-mass ratio.
The hypothesis that counterrotating tori can  be formed around supermassive  Kerr \textbf{BHs} has already been addressed in literature, for example in \cite{Carmona-Loaiza:2015fqa,Dyda:2014pia,Volonteri:2002vz,Stu:conto}.
However,   we have been carried out here a  focused investigation, providing a detailed   comparative analysis of the two classes of accreting tori.
In  the \textbf{RAD} framework,  some of more intriguing consequences of the presence of  ringed structure orbiting the central Kerr \textbf{BH} have place for  $\ell$counterrotating tori, thus  in this section we provide a perspective of some important aspects of the counterrotating tori with respect to corotating tori.
We address a morphological analysis  focusing on the  elongation and location of inner edge of corotating and counterrotating tori, these results are important also to fix the \textbf{RAD} $K$ and $\ell$-modes and  the tori collision.
In Sec.\il(\ref{Sec:more-l}) we will consider   the  mass  accretion rates and torus luminosity along with other significant features of the accretion tori for corotating and counterrotating fluids in dependence of the \textbf{BH} spin.

Torus elongation $\lambda$ on the equatorial plane can be written as function of  the fluid  specific angular momentum and the $K$-parameter as follows:
\bea&&\nonumber\lambda(a;\ell,\Qa)=\frac{2}{3}   \left[\sin \left(\frac{\arcsin\upsilon}{3}\right)+\cos \left(\frac{\arccos \upsilon }{3} \right)\right]A,
\\\nonumber
&&\mbox{where}
\\
&&\nonumber
\upsilon\equiv-\frac{9 (\Qa-1) \left[3 \Qa^2 (a-\ell)^2-2 \Qa \left(\ell^2-3 a \ell+2 a^2\right)+a^2\right]+8}{(\Qa-1)^3 A^3},
\\\label{Eq:lambda-qub}
&&
A\equiv\sqrt{\frac{3 \ell^2 (\Qa-1) \Qa-3 a^2 (\Qa-1)^2+4}{(\Qa-1)^2}} \quad \mbox{and}\quad \Qa\equiv K^2
\eea
\begin{figure}
\includegraphics[width=7.71cm]{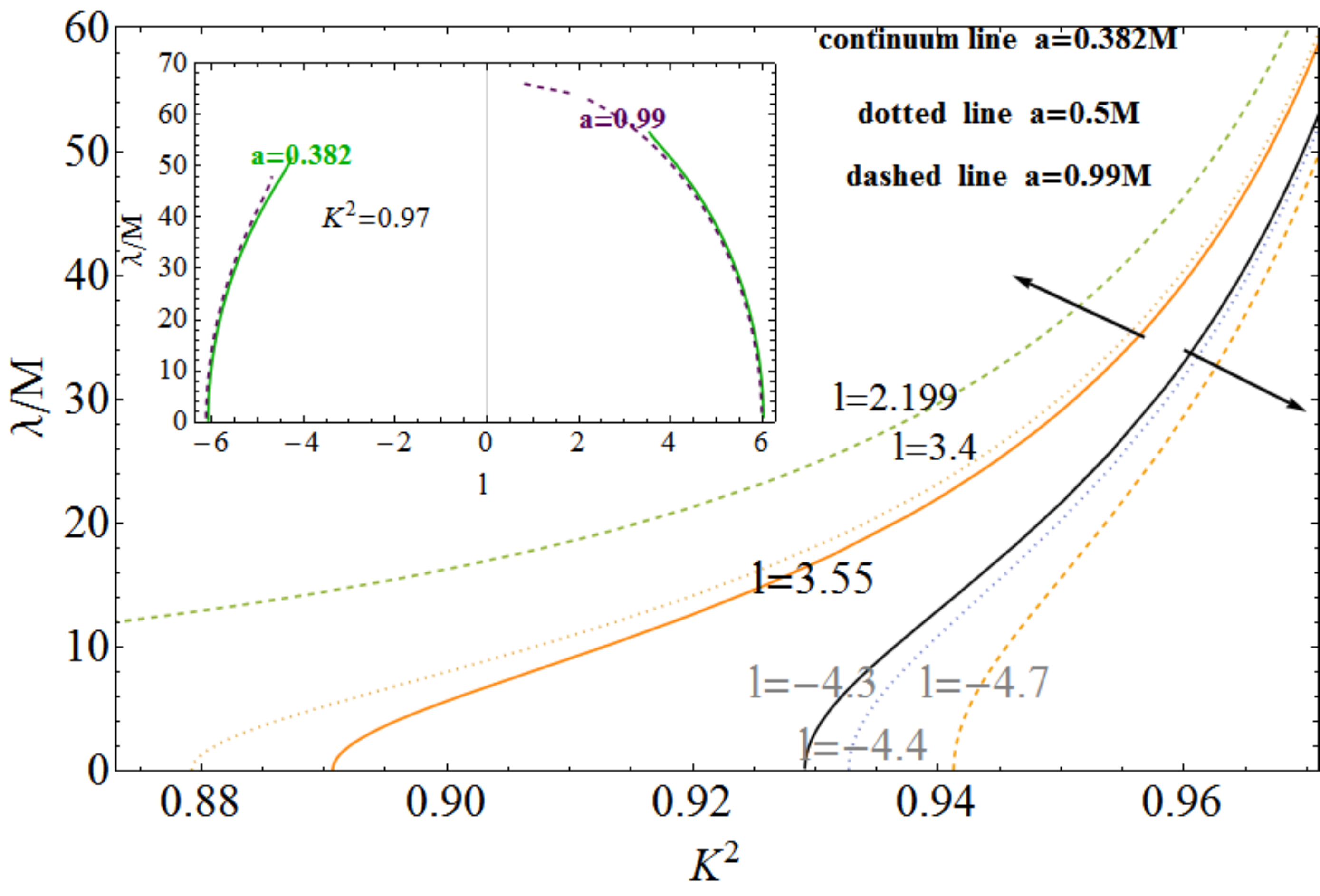}
\includegraphics[width=7.71cm]{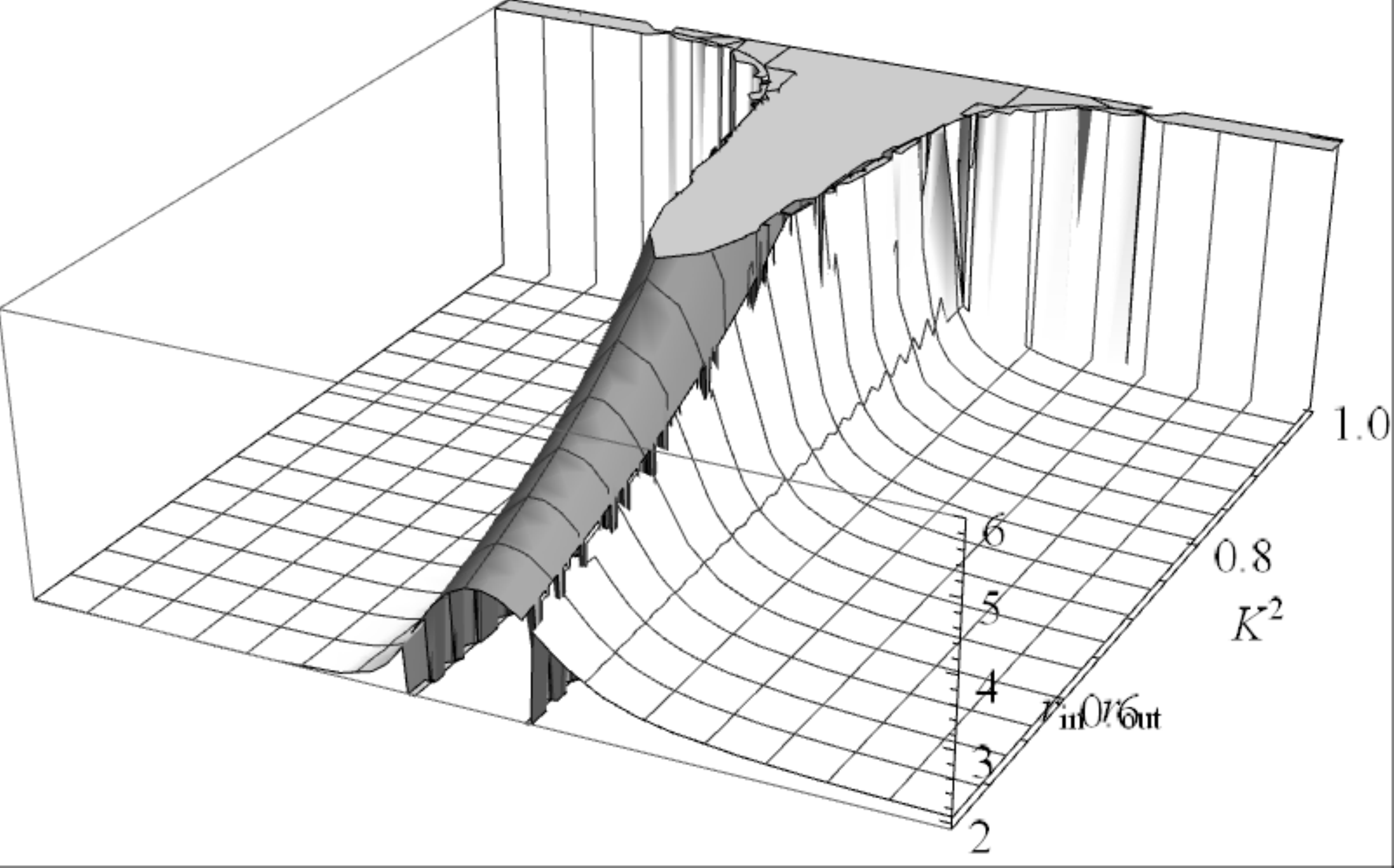}
\caption{\emph{Upper panel:} Plots of the tori elongation  $\lambda$, defined in Eq.\il(\ref{Eq:lambda-qub}),  for different values of the fluid specific angular momentum and \textbf{SMBH} spins    as function of the $K^2$ where $K\in]K_{mso}^{\pm},1[$ parameter  for counterrotating  and corotating  tori respectively.
Arrows indicate  the increasing \textbf{BH} spin $a/M$ for the corotating  and the counterrotating curves. \emph{Inside panel}, tori elongation $\lambda$  at fixed $K^2$, for different \textbf{SMBH} spins, as function of fluid specific angular momentum $\ell$. It is clear the symmetry of between the $\ell$corotating and $\ell$counterrotating  configurations. \emph{Bottom panel},  torus inner edge $r_{in}$ (white) and outer edge $r_{out}$ (gray) in Eq.\il(\ref{Eq:r-in-out-papers-out}) as functions of $K^2$ and fluid specific  angular momentum, for \textbf{SMBH} spin $a/M=0.382$.}\label{Fig:callneoc}
\end{figure}
\begin{figure}
\includegraphics[width=8.71cm]{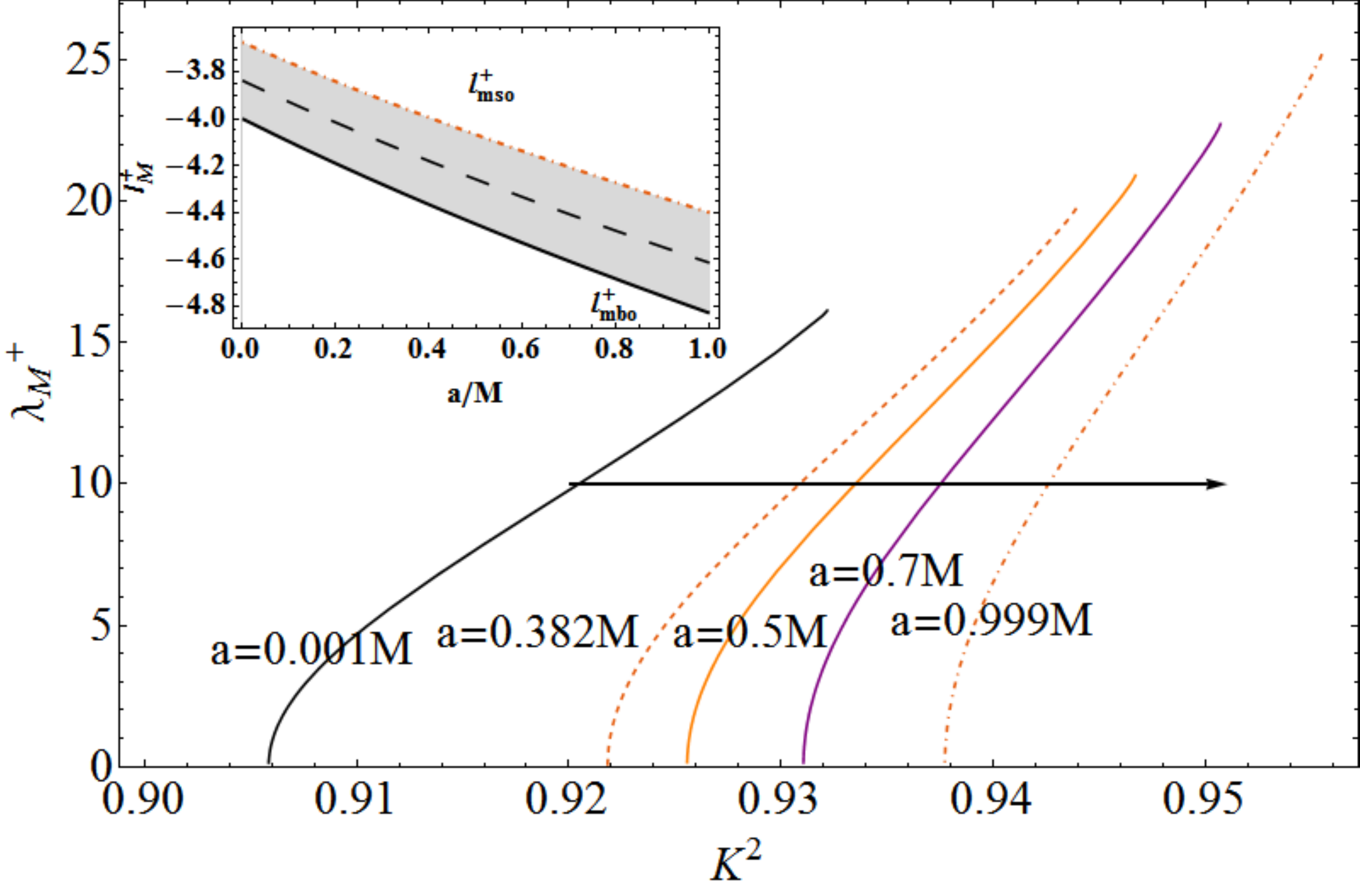}
\includegraphics[width=8.71cm]{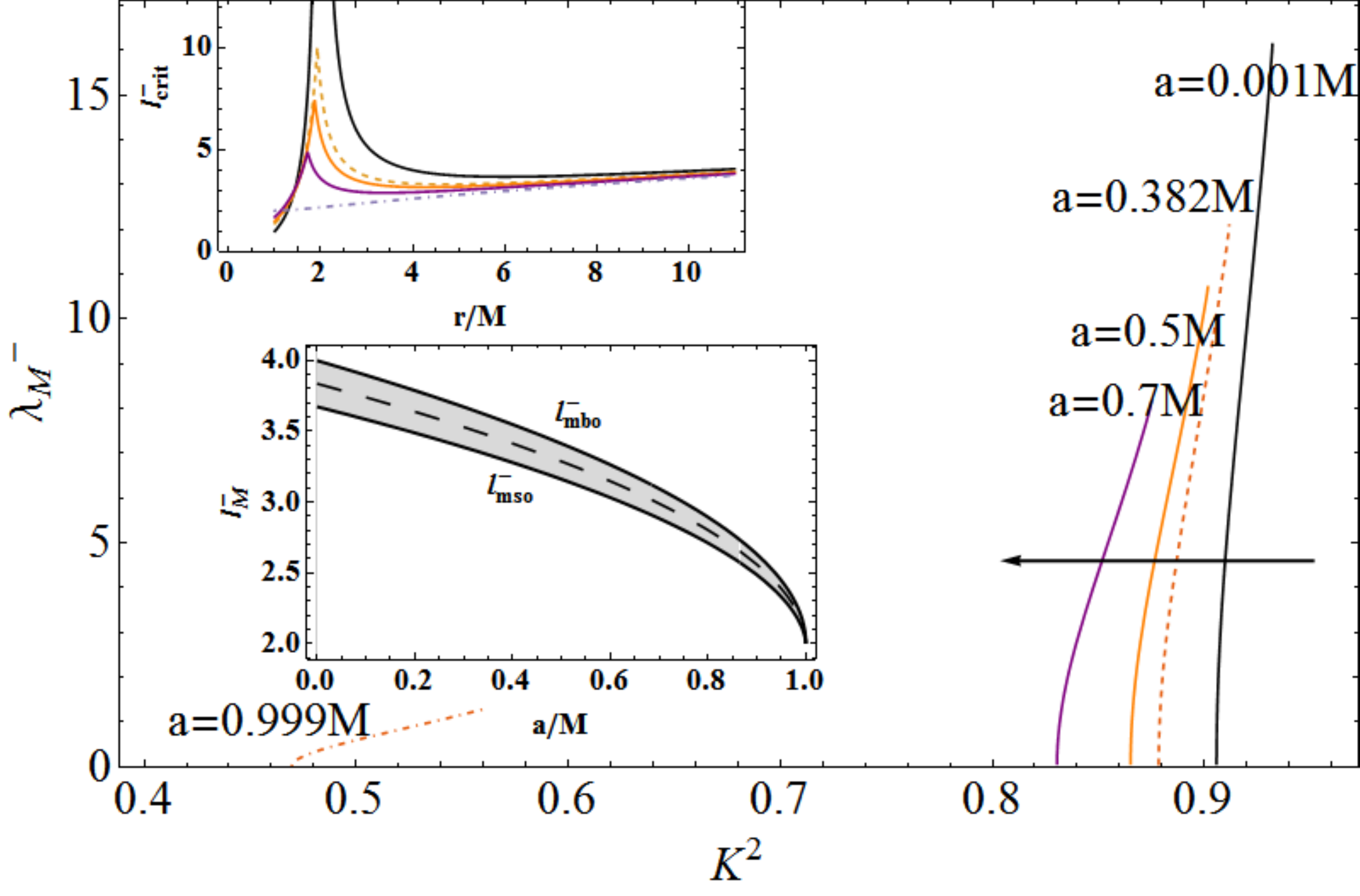}
\caption{Torus elongation $\lambda_{M}(a;\ell_M,K^2)$  in Eq.\il(\ref{Eq:lambda-qub}) on the equatorial plane  for corotating $\lambda_{M}^- $  (\emph{left panel})  and counterrotating fluids $\lambda_{M}^+$ (\emph{right panel}) for fixed values of \textbf{SMBH} spin $a/M$ (as singed on each curves) as function of $K^2$, where $K\in[K_{mso},1[$. Arrows  follow increasing values of \textbf{SMBHs} $a/M$.
There is $\ell_M^{\mp}\equiv\pm (\ell_{mbo}^{\mp}-\ell_{mso}^{\mp})/2$, dashed black curve in inside panels, gray region sets the the entire range of variation of  tori specific fluid angular momentum $\ell$ depending on the central \textbf{SMBH} dimensionless spin $a/M$, and the range for the function $\ell_{crit}$. Right inside panel shows the curves $\ell_{crit}^-$ as function of $a/M$. }\label{Fig:RTMe100dol}
\end{figure}
dimensionless units, $\lambda\rightarrow\lambda/M$ and $a\rightarrow a/M$, are considered. Note the dependence from the quantities  $(K^2-1)$ and $(\ell-a)$--see also \cite{pugtot}.
Figs\il(\ref{Fig:callneoc})  enlighten   the symmetries  between the tori of the $\ell$counterrotating couples.
 As expected, the elongation $\lambda^{\pm}$ increases with $K^2$. However,
for fixed  $K$, increasing the dimensionless \textbf{BH} spin  and decreasing the angular momentum magnitude, there is an increase of the torus elongation for corotating fluids, while  $\lambda^+$ decreases when matter is counterrotating  with respect to the central attractor, corresponding also to an increase of angular momentum magnitude. Generally, at a fixed $a/M$, larger values of $ K_+^2$ are needed  for the counterrotating fluids with respect to the corotating ones, implying  that the counterrotating tori are expected to be characterized  generally by larger ``density-parameter'' $K$ and larger elongation $\lambda$ then the corotating tori. In Sec.\il(\ref{SeC:poly}) we shall see  how  the torus density $\varrho$ and $K$-parameter can be related; in this way,  the results of this analysis set in terms of the  $K$-parameter  could be  directly read as  constraints on  the tori fluid density.
Figs\il(\ref{Fig:callneoc}) however represent the two  accretion  tori,  $\cc^{+}$ and $\cc^-$, by considering  the separation of the two set of curves.
   The inner panel  better shows such symmetry by  considering the elongation $\lambda^{\pm}$ versus  $\ell^{\pm}$, at fixed $K^2$, holding for  the set of corotating and counterrotating  curves included in the frame panel.
This analysis also proves
the different role played by the \textbf{BH}  spin for the evolution of  the corotating and counterrotating  tori:
for  greater values of the \textbf{BH} spin $a/M$, larger  values of  $K$ and  $|\ell|$ are generally expected.
Figs\il(\ref{Fig:RTMe100dol})  then show better this situation, by
  focusing on a more accurate analysis of the two different situations represented by the  $\ell$counterrotating couples and evidencing
differences and symmetries  of the   two toroids.
To obtain a  comparative analysis between the corotating and counterrotating fluids in dependence of the \textbf{BH} spin, we set the specific angular momentum as $\ell_M^{\mp}(a)\equiv\pm (\ell_{mbo}^{\mp}-\ell_{mso}^{\mp})/2$, this is a function of the \textbf{BH} spin.  Each curve is bounded  by the case of a particle string ring (with null elongation) to the  limiting open configuration of proto-jet. As expected, the $K^+$-range lingers in  values greater then  the $K^-$ parameter, and $\lambda^+$ increases with  $K^+$.
However,   tori with equal elongation may have  different angular momentum and they can even orbit \textbf{BHs} with different spin-to-mass ratio; as such the tori elongation (with some exceptions for the  corotating cases) cannot be used as tracers for the \textbf{BH} dimensionless spin or  to reveal the tori  fluid rotation with respect to the central Kerr attractor.
For the counterrotating fluids, increasing the \textbf{BH} spin, and following the    $\lambda^+=$constant lines, there is an increase of $-\ell^+$ (inner panel) and $K^+$. At  $K$=constant, on the other hand,  some regions of $K$-values do not allow the tori formation,  but with increasing $a/M$, $-\ell^+$ increases, while the elongation  $\lambda^+$ decreases. Considering the vertical line at $K$=constant, the curves  for the counterrotating fluids are more spaced with  respect to the corotating case.
However,   in general  the elongation can be relatively very large up to $\lambda^+\gtrsim 15 M$.
Focusing on the analysis for the corotating case, we noted that
 the curves are more close to each others at small spins, viceversa, there is a pronounced difference between   $a=0.7M$ and  $a=0.999M$ attractors.  This is because the corotating case is, at high \textbf{BH}  spin, very much dependent on the  variation of the  \textbf{SMBH}  spacetime  structure, particularly for very high spin the frame-dragging becomes  relevant and the tori may approach the ergoregion\footnote{We should note also that we use the Boyer-Lindquist frame where $r_{+}=r_{mbo}^-=r_{mso}^-=r_{\gamma}^-=M$ at $a=M$ (extreme Kerr \textbf{BH}).} --see also discussion in Sec.\il(\ref{Sec:coroterf}).
The curves representing the elongation of corotating   tori are  closer  in the $\lambda-K^2$ plane  then for  the counterrotating ones, this implies, for sufficiently small \textbf{BH} spin, a  significant similarity of certain characteristics of such tori with respect to parameters $K^2$, which  renders more difficult to distinguish different \textbf{SMBH}-\textbf{RAD} system.  Noticeably, it is  possible, however,  to distinguish tori orbiting  around slow rotating attractor and those around  faster rotating Kerr \textbf{SMBHs}.
 The elongation $\lambda^-$ is  generally rather smaller, then the counterrotating case being     $\lambda^-\lesssim14M$.
 Moreover $K^-$, $\ell^-$ and $\lambda^-$ decrease with $a/M$,  contrary to the counterrotating case
 The variation of the curves of  $K^2$, with the respect to the counterrotating case,  suggests that at fixed $a/M$, for a small variation of $K$ ($K_{\ell}-$modes)  $\lambda$ increases rapidly, and this is more clear for the slow rotating \textbf{BHs}. These results are finally confirmed also from the  the inner panel plots. This is an  interesting difference in the $\ell$counterrotating tori having potentially  significant implications on  tori collisions.
 In the characterization of the  $\ell$counterrotating systems, these doubled   analysis has to be considered simultaneously.

Similarly,  we can evaluate the inner and outer edge of an accretion torus as follows:
\bea\label{Eq:r-in-out-papers-out}
&&
r_{in}(a;\ell,\overline{Q})=-\frac{2}{3}\left[\mathbf{\alpha} \cos\left[\frac{1}{3} (\pi +\arccos\beta)\right]+\frac{1}{\overline{Q}} \right],\\
&&\nonumber
r_{out}(a;\ell,\overline{Q})=\frac{2}{3}\left[\alpha\cos\left(\frac{1}{3} \arccos\beta \right)-\frac{1}{ \overline{Q}}\right],
\\
&&\nonumber
\text{ where }
\\
&&
\alpha=\sqrt{\frac{4+3 \overline{Q} \left[(\overline{Q}+1)(\ell^2-a^2) +a^2\right]}{\overline{Q}^2}},
\\
&&
\beta =-\frac{9 \overline{Q} \left[(\overline{Q}+1)(\ell^2-a^2) +a^2\right]+8+27(\overline{Q}+1) \overline{Q}^2 (a-\ell)^2}{\alpha^3\overline{Q}^3}\\
&&\nonumber
and \quad \overline{Q}\equiv(K^2-1)<0
  \eea
(dimensionless units have been used). Figs\il(\ref{Fig:JohOCanonPro})  and Figs\il(\ref{Fig:Twomodestate})  explicitly show the variation of these quantities with respect to the fluid rotation  $\ell$ and the  $K^2$ parameter,  a further analysis, showed in  Figs\il(\ref{Fig:Hologra}), pictures  a more complex situation  featuring the special case of couples  $\cc_{\times}^-<\cc_{\times}^+$, where the dependence of these quantities is shown in the case of accreting tori.
\begin{figure}
\includegraphics[width=\columnwidth]{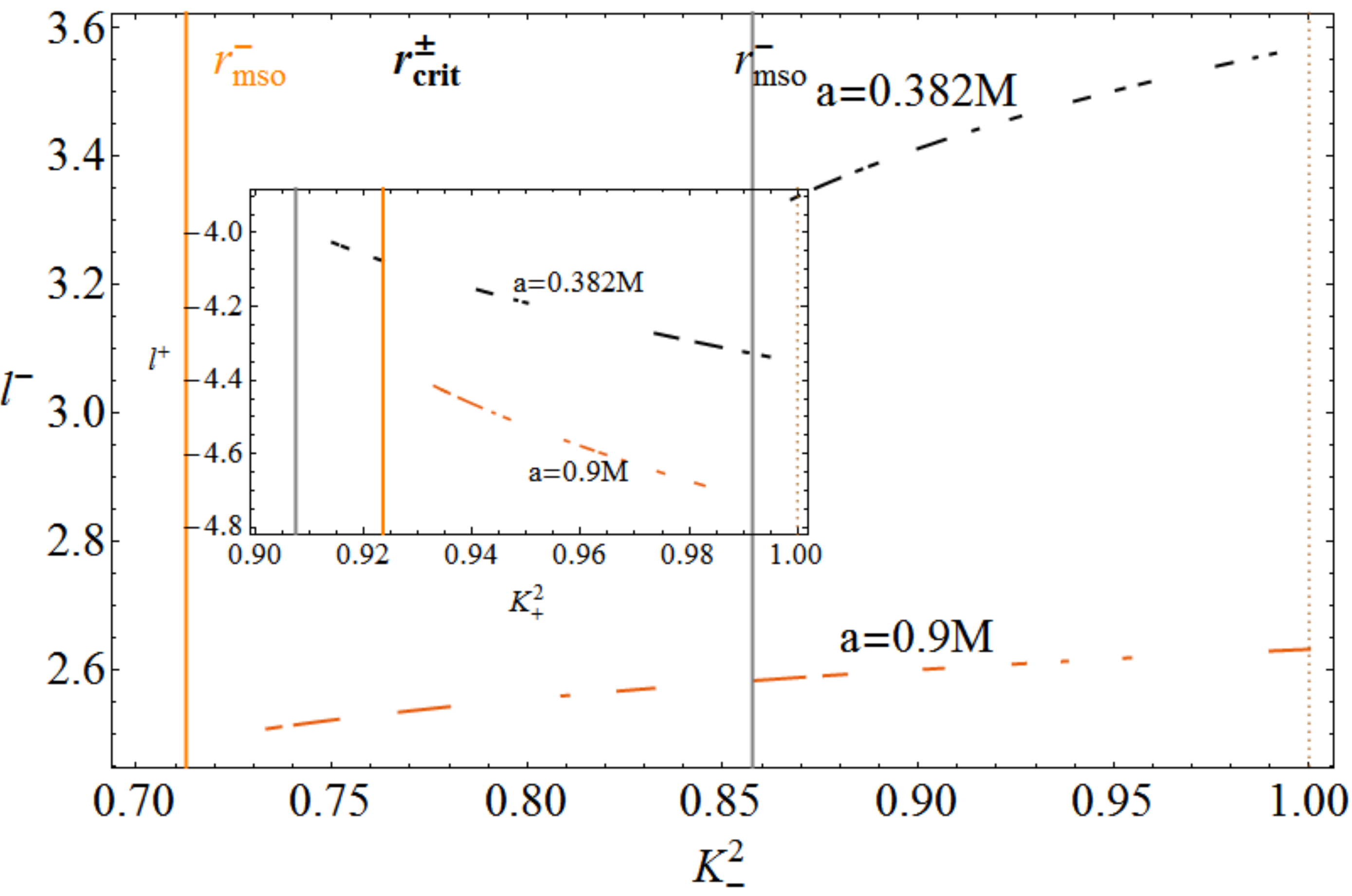}
\caption{Fluid angular momentum $\ell$ versus parameter $K^2$ for corotating $\ell^-$ and counterotating fluids $\ell^+$ (inset plot), orbiting a \textbf{SMBH} with spin $a=0.382M$ (black line) and $a=0.9M$ (orange line).   Curves  of the planes $\ell^{\pm}-K^2$ sets the inner edges of  the accreting  \textbf{RAD} tori.  Note that this information is provided by combining $\ell_{crit}$ and $K_{crit}$ at fixed spin in $r\in]r_{mbo},r_{mso}]$ with $\ell_{crit}\in]\ell_{mbo},\ell_{mso}]$ and $K_{crit}\in[K_{mso},1[$.  Radii $r_{mso}^{\pm}(a)$ for $K_{mso}^{\pm}$, and $K=1$, correspondent to $r_{mbo}^{\pm}(a)$ are also  plotted. Regions bounded by these radii  define the tori stability properties and their topology--see discussion in Sec.\il(\ref{Sec:Kerr-2-Disk}).}\label{Fig:Twomodestate}
\end{figure}
\begin{figure*}
\includegraphics[width=\columnwidth]{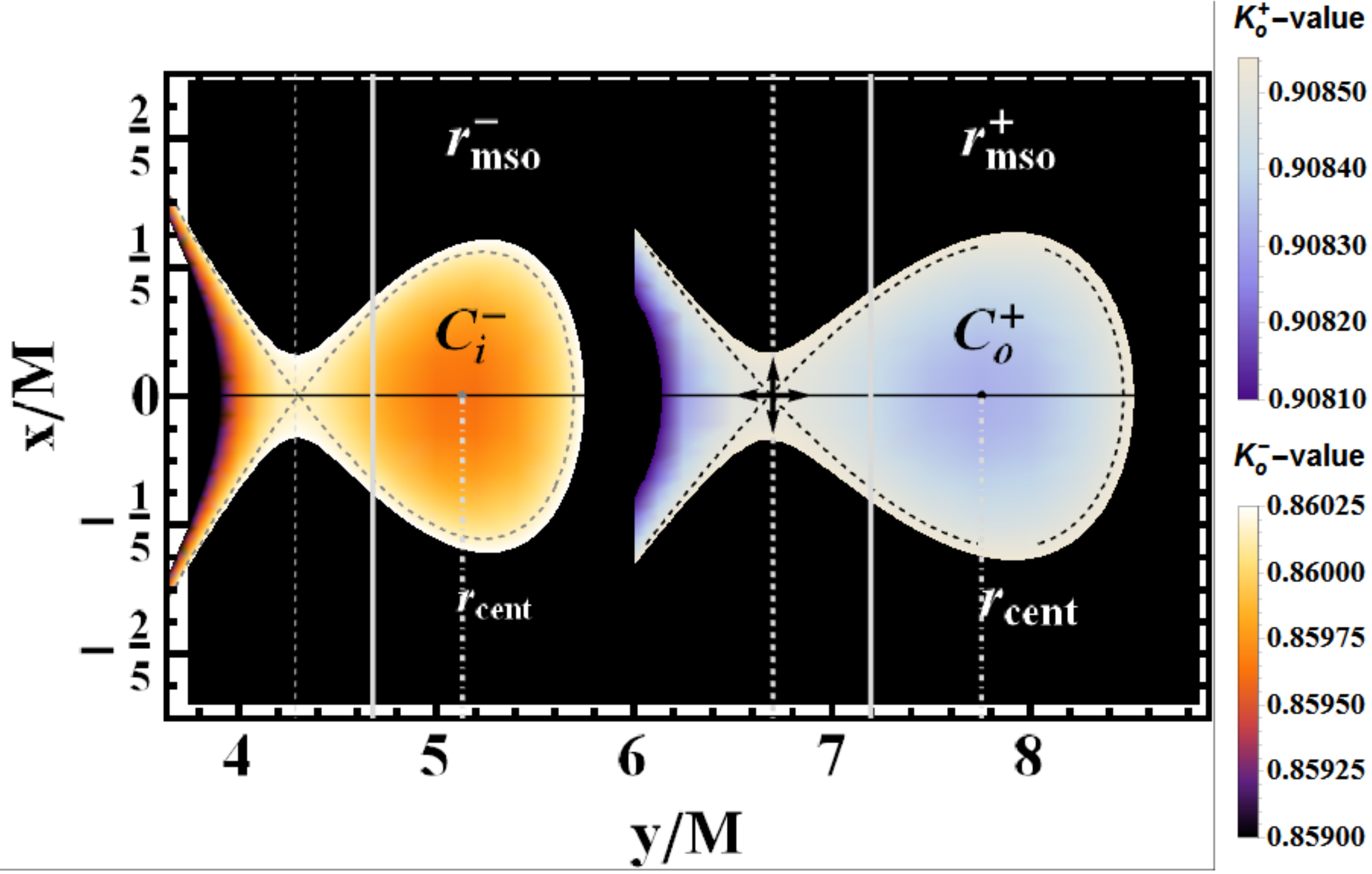}
\includegraphics[width=\columnwidth]{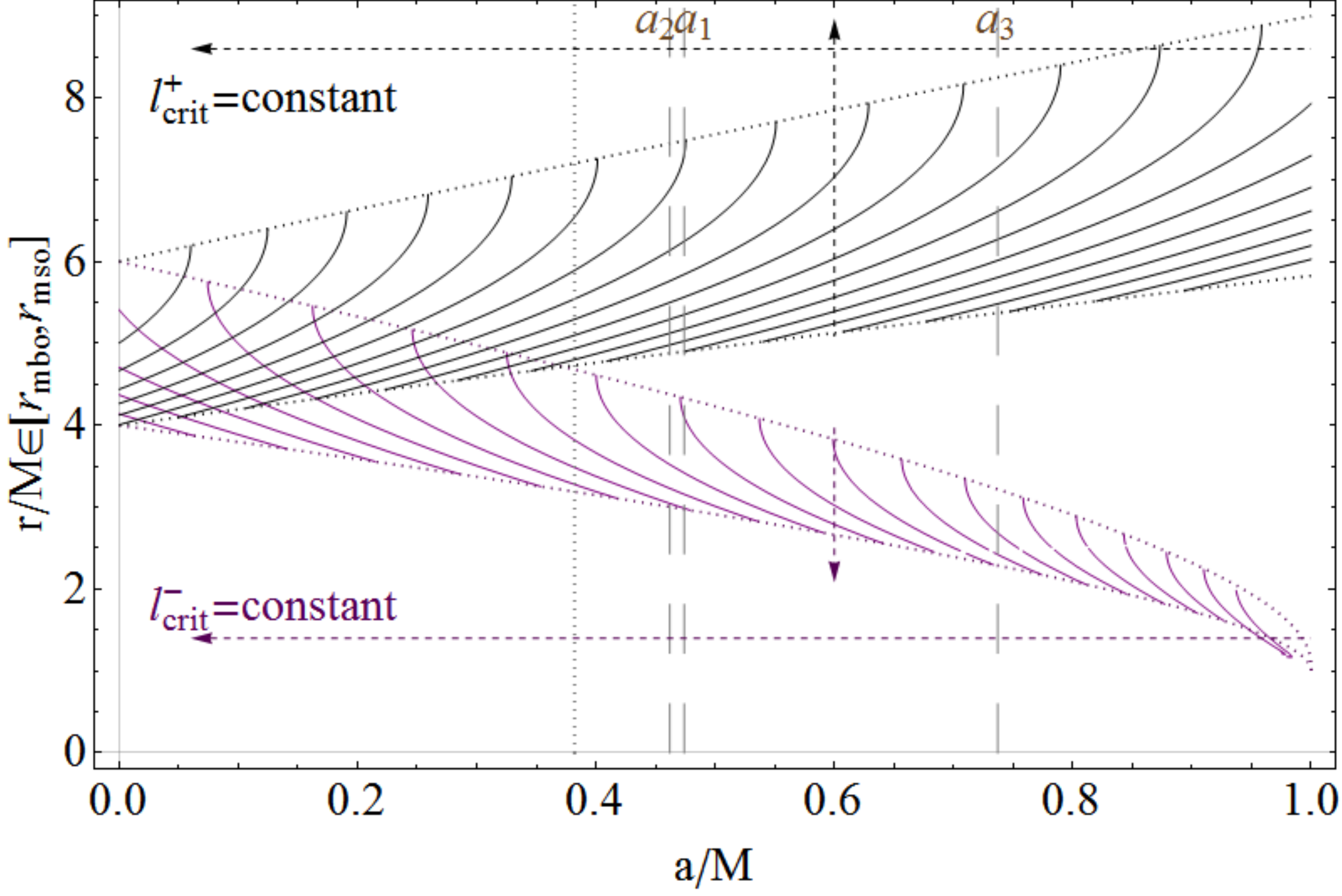}
\includegraphics[width=\columnwidth]{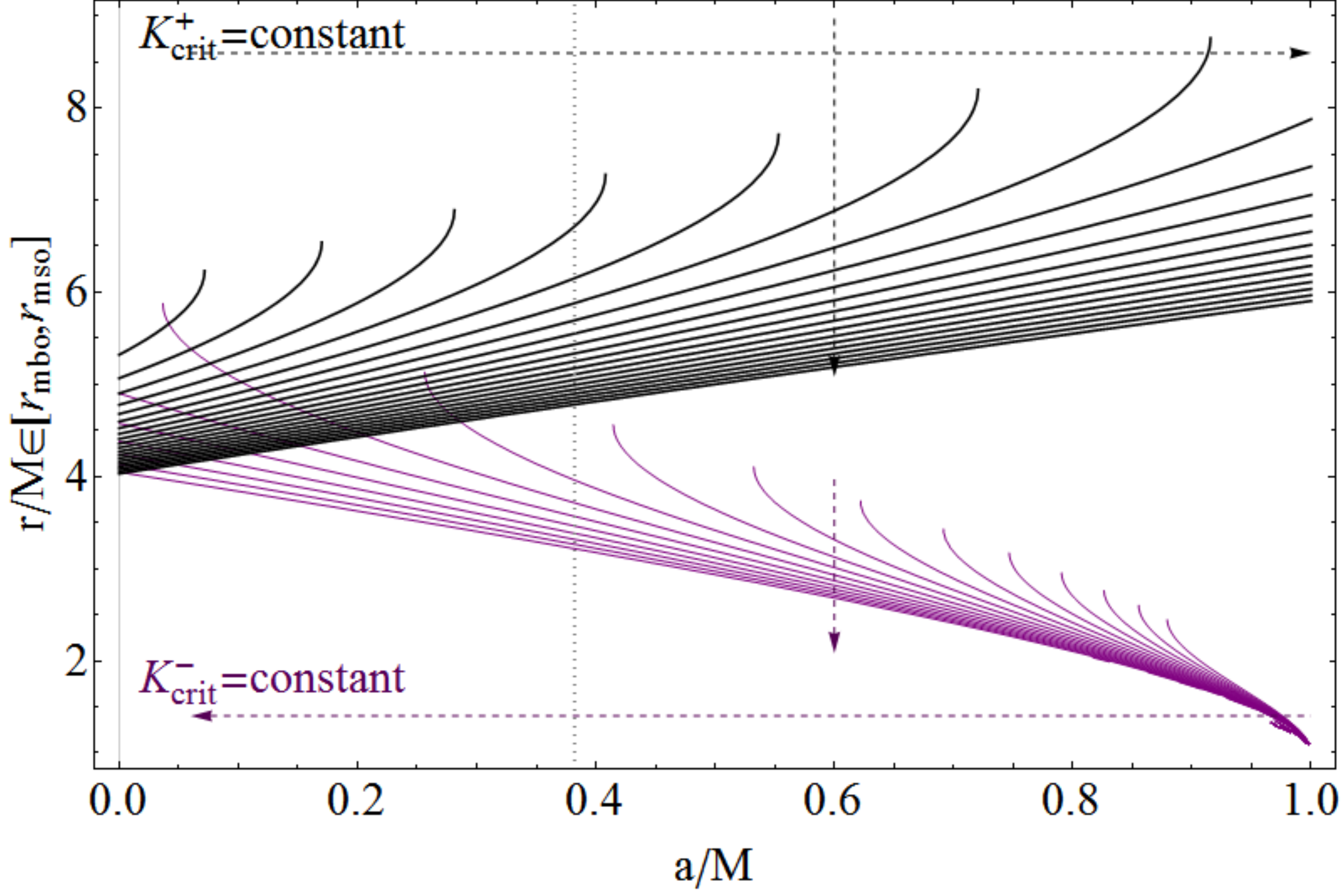}
\includegraphics[width=\columnwidth]{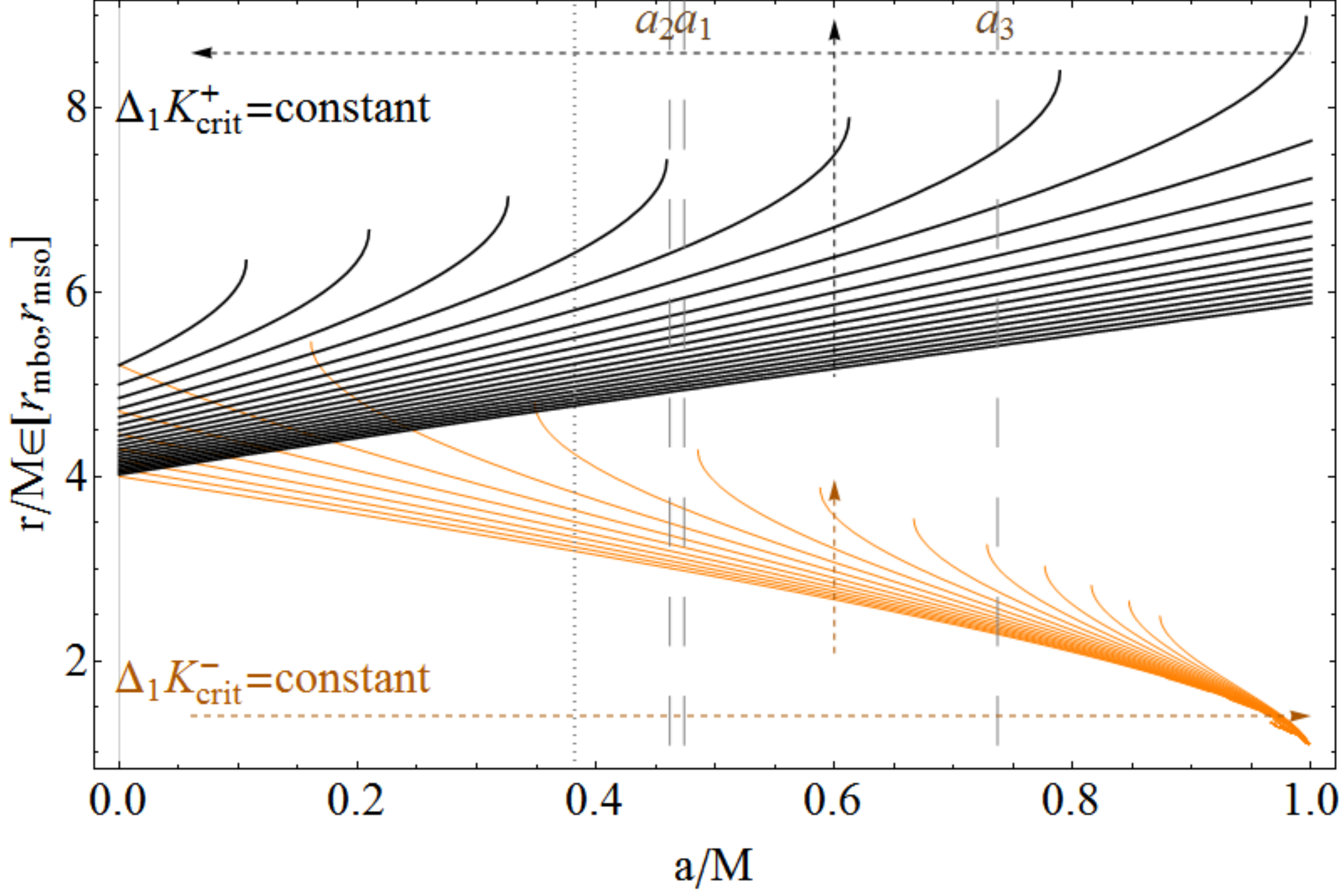}
\includegraphics[width=\columnwidth]{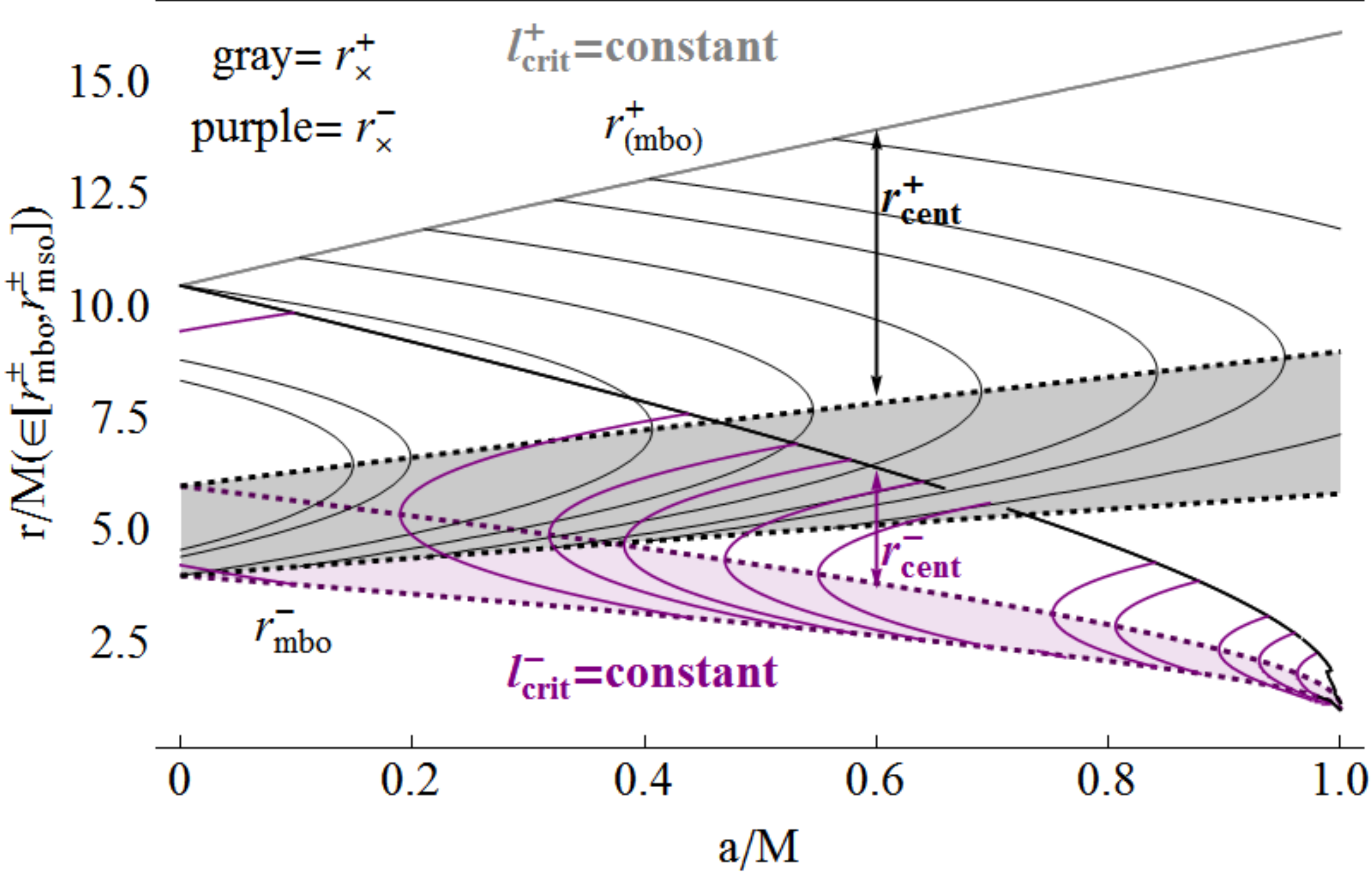}
\includegraphics[width=\columnwidth]{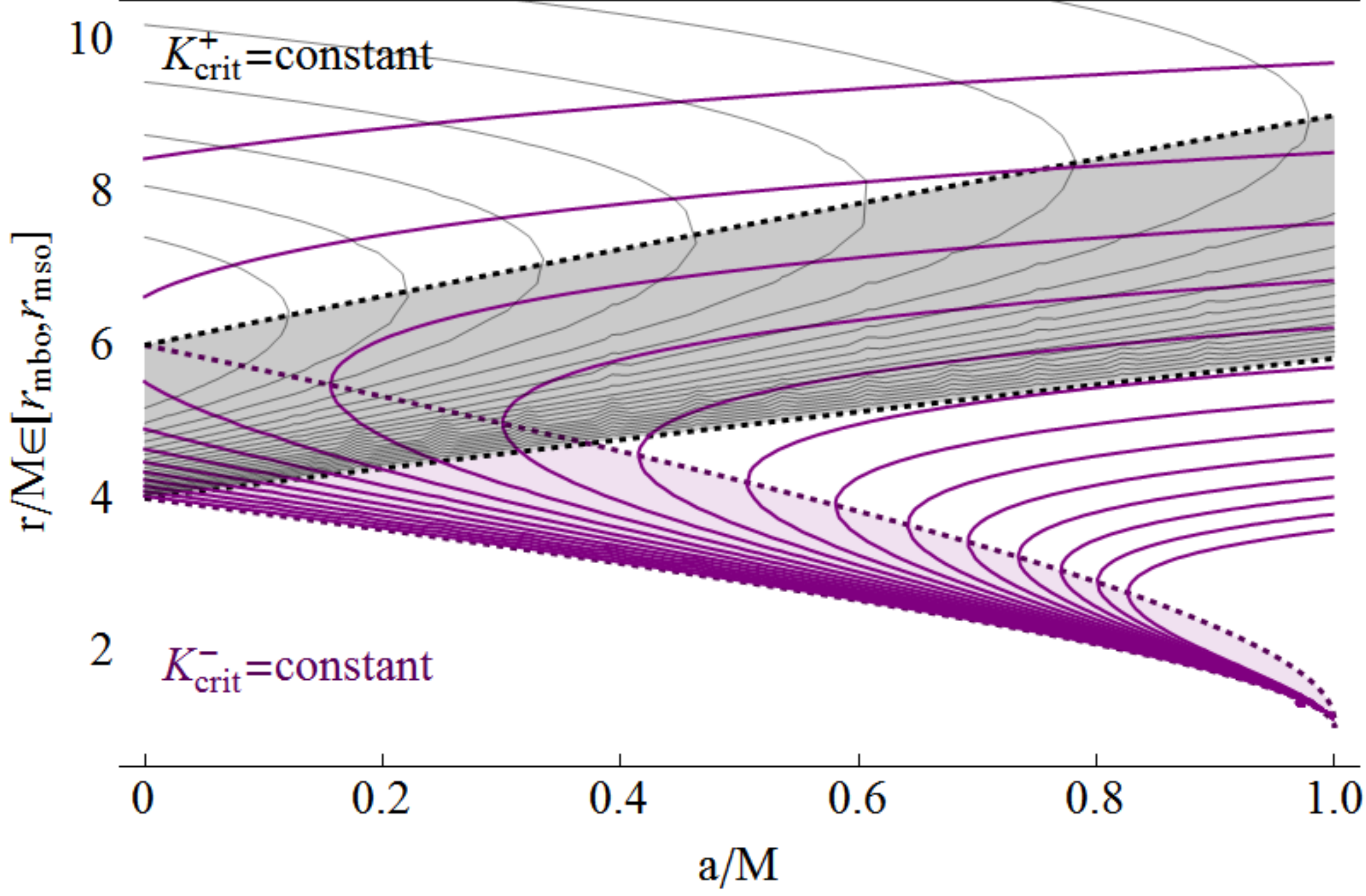}
\caption{ \emph{Upper left panel}:   \textbf{GRHD} numerical 2D integration {of Eq.\il(\ref{Eq:scond-d}) for a } double toroidal configurations (\textbf{RADs} of the order $n = 2$) There is \textbf{BH} spin $a=0.382M$,
$\ell^o_+ =- 3.99$ and
$\ell^i_- = 3.31$, vertical arrows set the thickness
${\widehat{\mathbf{h}}^+_s}$ of the accreting matter  flow--see also analysis in Figs\il(\ref{Fig:Manyother}). \emph{Upper right panel}:
Fluid specific angular momentum $\ell^{\pm}=$constant in the plane
$r\in[r^{\pm}_{mbo},r^{\pm}_{mso}]$ and $a\in[0,M]$,
for corotating (purple curves) and counterrotating  fluids (black curves).
\emph{Center left panel}:Curves $K_{crit}^{\pm}=$costant  for $a\in[0, 0.9998M]$  and $r\in[r^{\pm}_{mbo},r^{\pm}_{mso}]$ for corotating (purple curves) and counterrotating (black curves) fluids.
\emph{Center right panel}:
difference $\Delta_1 K_{crit}^{\pm}\equiv (K_{crit}^{\pm}(r_{mbo}^{\pm})-K_{crit}^{\pm})$=constant (maximum location of inner edge is  $r_{\times}\lessapprox r_{mso}$), as function of $a/M\in [0,0.9998]$   and $r\in[r^{\pm}_{mbo},r^{\pm}_{mso}]$,  for corotating (\textbf{-})
(orange), and  counterrotating (\textbf{+}) fluids  (black).
 Arrows  in the plots follow  increasing values of the plotted functions. Spins $(a_1
a_2,a_3)$ are also plotted.  Dotted vertical line  is $a=0.382 M$.\emph{ Bottom panels} show  extended plots with  indication of the torus centers location (region where $r_{cent}^{\pm}$ exist are pointed with doubled arrows).
}\label{Fig:Hologra}
\end{figure*}
Figs\il(\ref{Fig:JohOCanonPro})-\emph{left}
combines  the information provided by the two functions  $K_{crit}$ and $\ell_{crit}$ together.
Using {$K=K^{\pm}_{crit}(a;r,\ell)$ and  $\ell=\ell^{\pm}_{crit}(a;r)$} in   $r_{in}$ and $r_{out}$ of  Eq.\il(\ref{Eq:r-in-out-papers-out}), we obtain functions $\mathbf{{\hat{r}_{in}}}^{\pm}(a;r)$ and $\mathbf{\hat{r}_{out}}^{\pm}(a;r)$ for counterrotating and corotating fluids respectively, providing $r_{in}, r_{out}$ and $r_{cent}$ for each torus, therefore including also  the information provided by $\ell_{crit}$ and $K_{crit}$. These quantities are essential to determine the \textbf{RAD} stability properties, and  particularly for the  tori collision, and in the study of the  $ K $-modes and $\ell$-modes.
Increasing the \textbf{BH} dimensionless spin, the elongation $\lambda^-$, the    inner edge $r_{in}^-$ and torus center $r_{cent}^-$  decreases (due to the increasing Lense--Thirring effect). Viceversa, in the counterrotating case, increasing the spin $a/M$, the $(r_{in}^+, r_{out}^+)$ curves   move outwards and the elongation increases, being larger than for the  corotating ones (counterrotating tori are generally  bigger, with larger equatorial elongation, than the corotating ones). The situation is in fact opposite with the respect to the $\cc^-$ tori. Furthermore, the larger is the \textbf{BH} spin, the greater is the possibility to form a \textbf{RAD} with  an inner  corotating  torus (this can be seen for example by comparing  the case for \textbf{SMBH} with spin  $a=0.7M$ or $a=0.999M$).
The analysis  in Figs\il(\ref{Fig:JohOCanonPro}) shows also the limit of small $r/M$, corresponding  to the limiting case of a string of test  particle with null elongation.
Second panel enlightens   further aspect of the  symmetry  of the $\ell$counterrotating couples, by considering the dependence on  the central Kerr \textbf{BH}, and the functions  $r_{in}$ and $r_{out}$ in dependence on $K^2$ and $\ell$.
In the study of tori collision emergence, we  need to  consider this doubled analysis simultaneously.
At fixed $K^2$, increasing  $a/M$ and $\ell^{+}$ in magnitude, the tori inner edge $r_{in}^+$ increases, while the outer edge $r_{out}^+$ decreases. Viceversa, for the  corotating tori, at fixed $K^2$, increasing $a/M$ and  decreasing $\ell^{-}$, the tori inner edge $r_{in}^-$ decreases, while the outer edge $r_{out}^-$ increases. Curves are  bounded  by the  limiting  string particle ring of matter.
Increasing $K^+$, there is an decreases of $\lambda^\pm$ (within this special choice of the parameters, $a/M$ and $\ell$), while  increasing $(|\ell^\pm|, a/M)$, the  $K^\pm $ parameter  increases.
 For a $\ell$counterrotating couple, there is in general  $r_{in}^-<r_{in}^+<r_{out}^+<r_{out}^-$.
The separations of the set of curves increases  at large spin, and $K^-$ decreases, giving possible $\cc^-$ tori all contained in $\Sigma_{\epsilon}^+$.
From the location of $r_{in }$  it is therefore possible to provide an estimation of the location of  $r_{out}$  for  accreting tori, and the $\textbf{BH}$ spin  $a/M$,  with the tori rotation with respect to the attractor.
Inner plot confirms the symmetric situation occurring for corotating and counterrotating fluids at fixed $K$,
 with different values of the \textbf{SMBH} spin.
Fig.\il(\ref{Fig:Twomodestate}) on the other hand shows the  inner edge of accreting tori varying
  $K$ and  $\ell$ at different \textbf{BH} spin $a/M$.
In the corotating case, the difference between  the tori inner edge locations, $r_{in}$,  for tori   orbiting around \textbf{SMBHs} with different  spins, is greater then for the  counterrotating ones. The corotating fluids are better traces of  the \textbf{BH} spins. On the other hand, the different behavior with respect  to the  fluid rotation makes determination of   the inner edge of accreting tori a promising  method also to  eventually  distinguish   the corotating and counterrotating tori.
Figs\il(\ref{Fig:Hologra})
 show surfaces $\ell^{\pm}_{crit}=$constant  in the  $r/M$--$a/M$ plane, each curve sets a special torus. We see  different situation for a \textbf{BH} spin shift,  moving along each curve contained  in the possible range of parameter values where
 accretion onto the central \textbf{BH} can occur, considering then  locations of the inner edges $r_{\times}$. Below panels show also the location of  the tori centers $r_{cent}$.  The $\ell$-modes are confined on a segment  bounded by  $r=r_{mbo}^{\pm}$ and $r=r_{mso}^{\pm}$ in each strip in the plot, each of them featuring the corotating or counterrotating fluid,  on a vertical curve $a/M=$constant.
 $K-$-modes are not described by this analysis.
A \textbf{BH} spin shift corresponds  to the translation from a vertical line to another.
At fixed spin,  the $\ell$corotating couples are fixed moving along  the curves  crossing points,  in the same strip, with a fixed vertical line.
For the $\ell$counterrotating  couples, the   tori associated with   the crossing points of  the vertical lines with the curves contained in the  two strips have to be considered together.
We limited our analysis  to the  $\textbf{L1}$ range of specific angular momentum; we note that for greater magnitude of specific momentum,  $\ell\in \textbf{L2}$,  the centrifugal component of the force is very high compared with the gravitational one in the  force balance, giving rise to the unstable phases of   proto-jets configurations. In order  to  let  torus accrete onto the central \textbf{BH}, the angular momentum magnitude  has to decrease to  reach the values of the   $\textbf{L1}$ range. As  already mentioned, we find indications  that $\pp^-<\pp^+$ couple, i.e. a couple formed by an   external counterrotating  and inner corotating torus, are expected to be the favorable $\ell$counterrotating couples to be  observed.
The $\ell$-modes are realized as a shift,  on a vertical line, form one curve to another on the same strip. 
A further analysis considers $K_{crit}$ values with a similar approach,  this analysis  will be  also resumed in Sec.\il(\ref{Sec:more-l}) in the evaluation of the \textbf{RAD} tori accretion rates.
%
\section{Emergence of collision}\label{SeC:coll}
We identify  the following three     situations where  a collision arises: \textbf{1.}  The  $\mathbf{C}_{coll}$  couple  with  contact  in $r_{{in}}^o=r_{{out}}^{i}\equiv r_{coll}$,  for    ({non-accreting or accreting) $\ell$corotating or $\ell$counterrotating tori. In this colliding couple  the outer  torus is non-accreting Fig.\il(\ref{Fig:teachplo})-right.   $\mathbf{C}_{coll}$     eventually leads   to tori merging. \textbf{{2.}} The $\mathbf{C}_{\times}$   couple featured in Fig.\il(\ref{Fig:teachploeff}). This process   implies   emergence of the outer torus  instability. This situation is   possible only for  the    ${ \pp}^+\ll_{\mathbf{\times}} \cc_{\times}^-$ tori  in spacetimes with  $a\neq0$,  within the necessary conditions  discussed below in Sec.\il(\ref{Sec:const}). The fluid accreting onto  the central \textbf{BH} necessarily impacts on the {(non-accreting or accreting)} inner torus.
\textbf{3.}  Finally, the  $\mathbf{C}_{coll}^{\times}$ colliding couple    would emerge as a combination of the two processes considered before,  this is therefore  a  ($\mathbf{C}_{coll}$, $ \mathbf{C}_{\times}$) combination, where  $r_{{out}}^{i}=r_{\times}^o=r_{coll}$.  Collision is  combined with  an  hydro-gravitational destabilization   (Paczy{\'n}ski-Wiita mechanism)   or   any other   local instability   of the outer torus.

Tori collisions  arise   due to different causes; in general,  $\mathbf{C}_{coll}$ collision   occurs when    the outer torus   grows  or loses   its angular momentum,  approaching the accretion phase.
Collisions due to \emph{inner} torus growing   is a particularly constrained case.
In general, the inner torus evolution affects the   $\mathbf{C}_{coll}$  emergence     mainly in the early stages  towards the accretion; the inner torus  instability   in several cases   would prevent  collision.
In Sec.\il(\ref{Sec:const}) we discuss the constraints on different couples of tori enabling emergence of  collision, and    in Sec.\il(\ref{Eq:CM-ene}) we provide an   evaluation of the      center-of-mass  energy $(\mathrm{E}_{\mathrm{CM}})$      for two  colliding particles  from the  pair of interacting tori.
\subsection{Constraints}\label{Sec:const}
Emergence of tori collision is  featured  by several constraints. Here we consider the constraints on the tori angular momentum.

First, the   $\ell$corotating tori are characterized by the relation $\ell_{o}/\ell_{i}>1$ (this may be verified   by drawing the line $\ell=$constant for each curve $\ell(r)$ in Fig.\il(\ref{Fig:teachploeff})). For the  $\ell$counterrotating  tori, the situation is much more complicated;  in this case we have   to analyze simultaneously the  curves $\mp\ell^{\pm}(r)$.
The fluid specific angular momentum range of   colliding configuration  $(\mathbf{C}_{coll},\mathbf{C}_{\times},\mathbf{C}_{coll}^{\times})$   is constrained by the following relations:
\bea\nonumber
&&\mbox{for}\quad { \pp}^-<\pp^+\quad\mbox{there is}\quad\;|\ell^-/\ell^+|<1
\\
&&\label{Eq:conf-decohe}\mbox{and}\quad
\ell^-\in]\ell_{\mso}^-, \ell^-(r_{cent}^+)[.
\eea
Also this result can be verified by drawing the line $\ell^-=$constant and $-\ell^+=$constant for each curve $\ell(r)$ in Fig.\il(\ref{Fig:teachploeff}). Relation (\ref{Eq:conf-decohe}) implies   that  the angular momentum of   matter in the outer   counterrotating torus is \emph{always} greater in magnitude than the angular momentum on the inner corotating one.

 Furthermore, relation \il(\ref{Eq:conf-decohe}) restricts the location of the center $r_{cent}^-$ of the corotating tori of the couple in the following way
\bea&&\nonumber
r_{cent}^-<\bar{r}\quad\mbox{where}\quad
\bar{r}>r_{\mso}^-:\quad \ell^-=-\ell^+>-\ell_{\mso}^+\\
&&\nonumber\mbox{ and}
\\\nonumber
&& \mbox{if there is  }\quad\pp^-=\cc_{\times}^-\quad\mbox{ then }\quad r_{cent}^-\in]r_{\mso}^-,r_{(\mbo)}^-[.
\eea
However, not  all the $\pp^-<\pp^+$ couples  satisfy the  conditions for  the outer torus  accretion:   in many cases an  outer torus instability  would be preceded by collision and merging with the inner torus.

In fact, the \emph{necessary} (but not sufficient) conditions  for  the  outer torus accretion  are provided by the relation  $\pp^{-}\ll_{\mathbf{\times}}\pp^+$,  with      $-\ell_o^+\in]-\ell_{\mso}^+,-\ell^+(r_{\mso}^-)[$   and  $\ell_i^-\in]\ell_{\mso}^-,\ell^-(r_{\mso}^+)[$.
It is easy to show that this also implies the remarkable conclusion  that there are no more than $n_{\max}=2$ tori satisfying these properties, while any further outer torus (or even a torus internal  to the couple)  must be  {non-accreting}.

Conversely, particularly  for  $a\gtrsim0$, some of these solutions do not allow an outer torus accretion: in fact
 the accretion takes place {only if} there is   $ \ell^-\in]\ell^-(r_{\times}^+),\ell^-(r_{cent}^+)[$,  where  $\pp^-\ll_{\mathbf{\times}}\cc^+$.
On the other hand, then for $\pp^+=\cc^+$ (i.e. a non-accreting  counterrotating tori), there is $\ell^-(r_{\times}^+)\in]\ell_{\mso}^-,\ell^-(r_{cent}^+)[$. Particularly,  for  $\ell^-\in]\ell_{\mso}^-, \ell^-(r_{\times}^+)[$, with $\pp^->_{\mathbf{\times}} \pp^+$,  more then $n_{\max}=2$ tori   may satisfy these conditions ($\ell_{o/i}=-1$, 
$\ell^-/-\ell^+\gtrless1$).
\begin{figure}
\includegraphics[width=8.4cm]{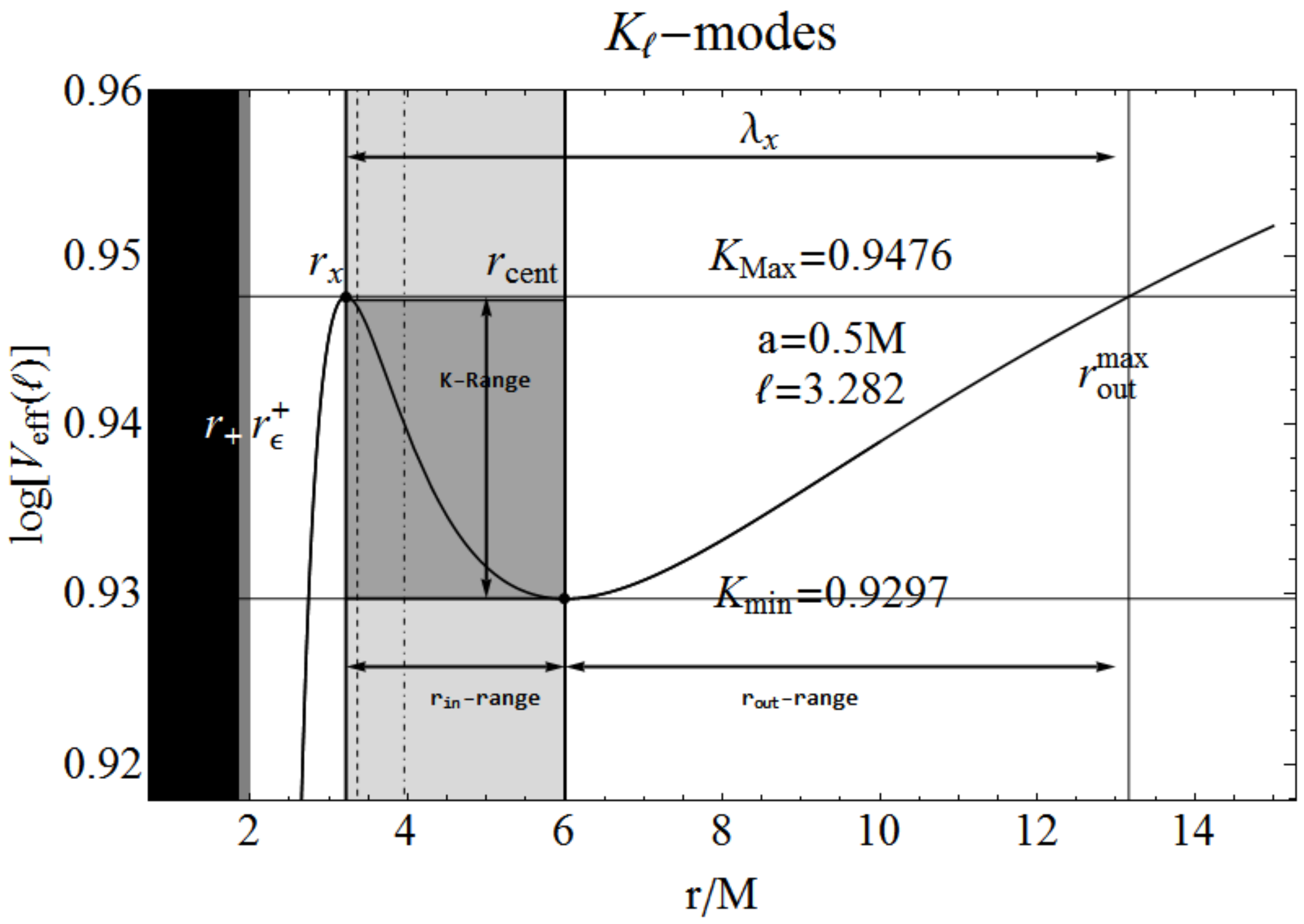}
\includegraphics[width=8.4cm]{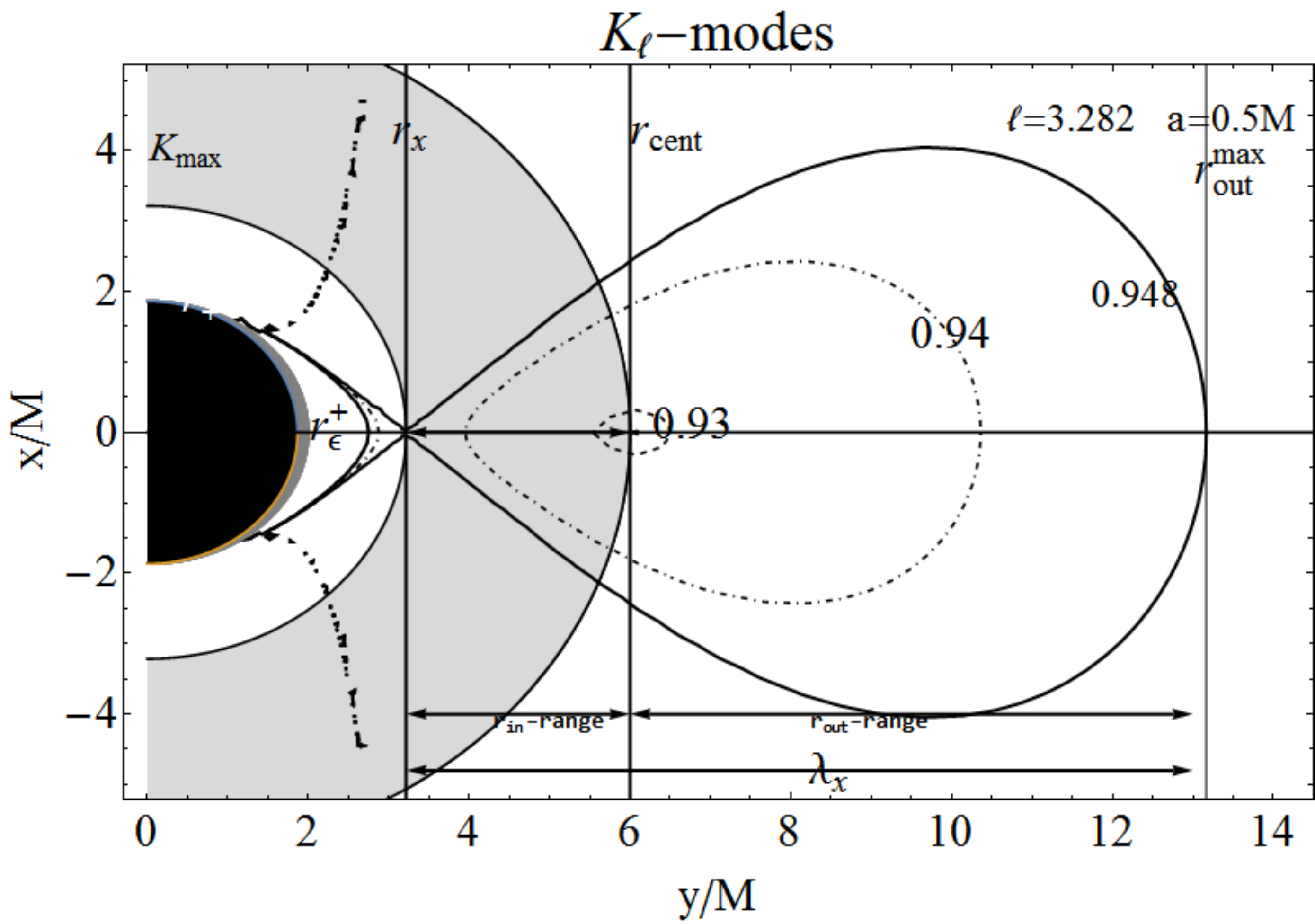}
\caption{Illustration of $K_{\ell}$-modes. \emph{ Upper panel} shows  the  function $W_{\bar{\ell}}=\left.\ln[V_{eff}(\ell)]\right|_{\bar{\ell}}$,  in Eq.\il(\ref{Eq:scond-d}), at constant  angular momentum $\bar{\ell}$.
The effects of modes, $K_{\ell}\in[K_{min},K_{\max}]$ are illustrated by the range of parameters and variables   variations (shaded-labeled regions): $K$-range (dark gray), inner torus edge $r_{in}$-range (light-gray), outer torus margin  $r_{out}$-range (white). Arrows indicate the ranges. The torus maximum elongation  $\lambda_{\times}$,  following  the $K_{\ell}$-modes is also shown. $r_{cent}$ is the  maximum pressure point of the torus, $r_{\epsilon}^+$ is the outer ergosurface, black region marks the range $r\leq r_+$ where $r_+$ the the outer \textbf{BH} horizon. In the \textbf{RAD} context, the study of $K_{\ell}$-modes for one torus,   must be coupled with the $K_{\ell}$-modes for any other component of the  aggregates and with the $\ell$-modes--Fig.\il(\ref{Fig:Truema}).
\emph{Bottom panel}: Some torus profiles corresponding to different values of $K$.
Light-gray  circular region corresponds to the $r_{in}$-range.--See also Figs\il(\ref{Fig:Stopping}).
\label{Fig:Truem} }
\end{figure}
\begin{figure}
\includegraphics[width=8.4cm]{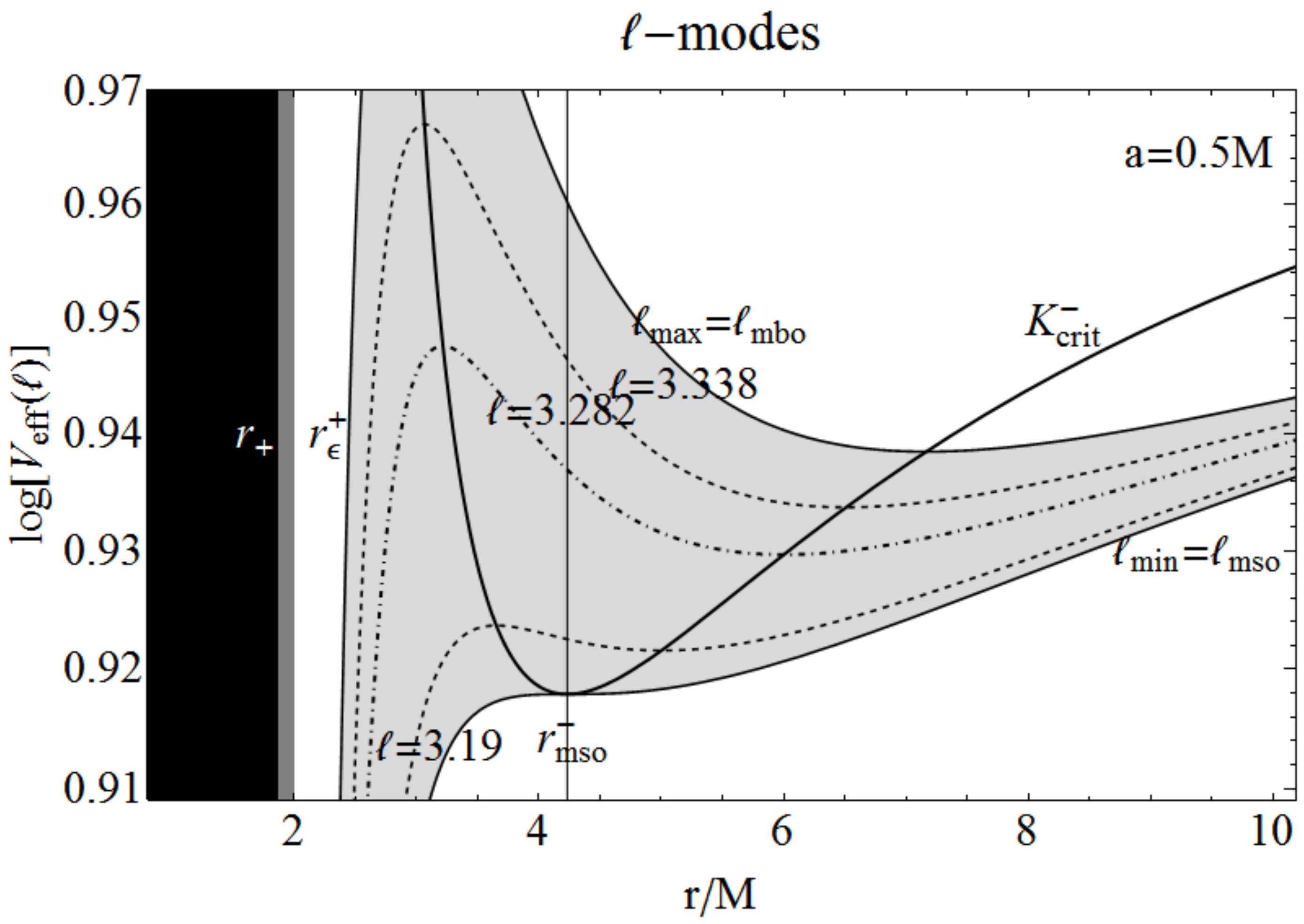}
\includegraphics[width=8.4cm]{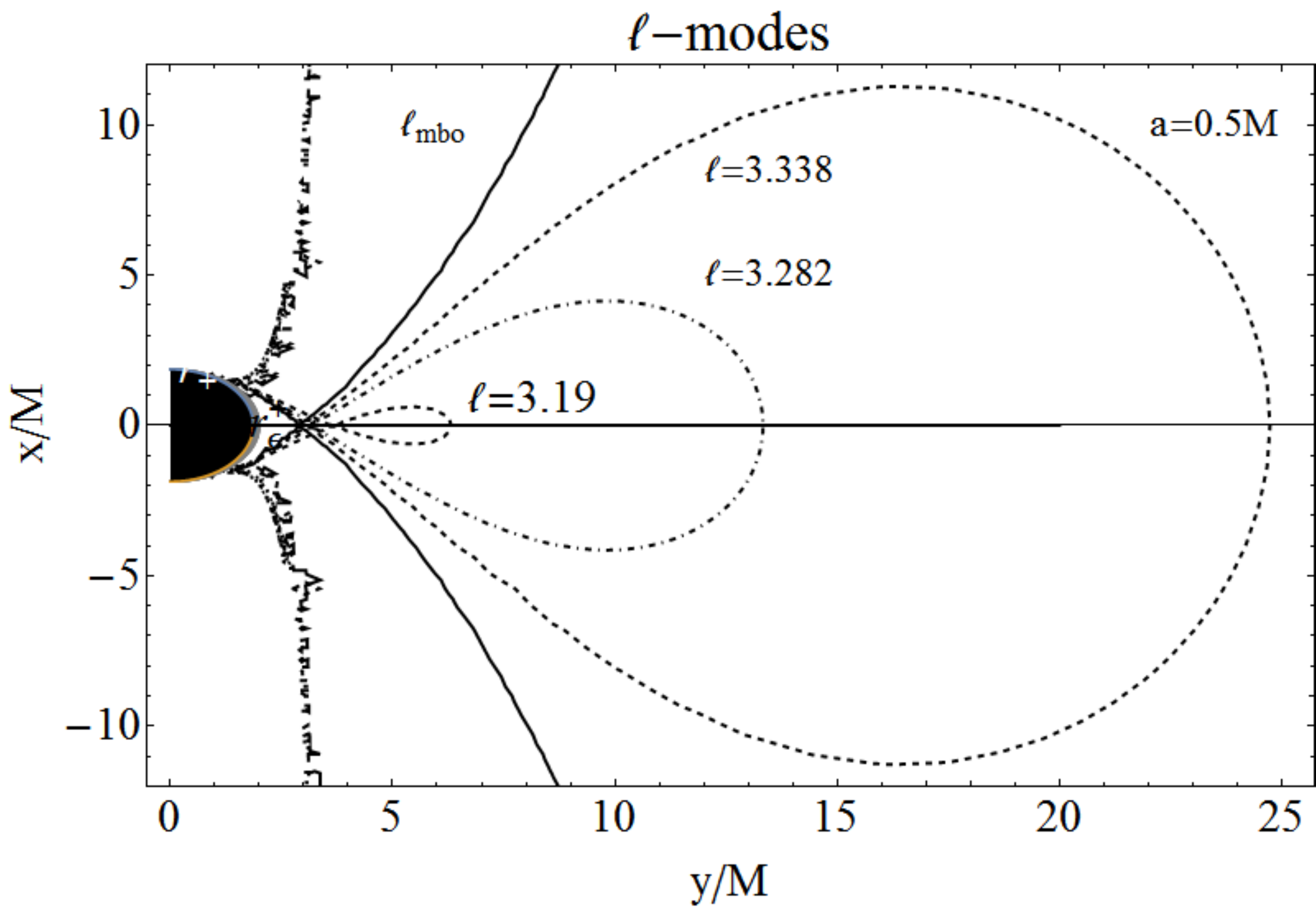}
\caption{Illustration of ${\ell}$-modes. \emph{Upper panel}- The  function $W=\ln[V_{eff}(\ell)]$,  in Eq.\il(\ref{Eq:scond-d}), for different values of the fluid specific  angular momentum $\ell$.
$\ell$-range, maximum range of variation  of the $\ell$ parameter, $\ell\in[\ell_{mso},\ell_{mbo}[\equiv[\ell_{\min},\ell_{\sup}[$, is  light-gray  colored.
Black-thick curve marks $K_{crit}$ for the fixed value  \textbf{BH} dimensionless spin $a/M$--see also Fig.\il(\ref{Fig:Stopping}).
\emph{{Bottom panel}}-torus profiles, corresponding to different values of $\ell\in\ell$-range. $r_{\epsilon}^+$ is the outer ergosurface, black region marks the range $r\leq r_+$ where $r_+$ is the outer \textbf{BH} horizon. The analysis  of ${\ell}$-modes of each torus of the \textbf{RAD} must be combined with $K$-modes ${\ell}$-modes for any other \textbf{RAD} component; ${\ell}$-modes are associated to shifts related points on the  $K_{crit}$ curve (at fixed $a/M$) (each point sets a different minimum and maximum pressure point, each point of the such a couple on $K_{crit}$ sets different angular momenta. On the other hand only for accreting tori, curves $K_{crit}=$constant fix $r_{\times}$ and $r_{out}$ of the  accreting torus correspondent to one fixed value of specific angular momentum-see also Figs\il(\ref{Fig:Stopping}) and Figs\il(\ref{Fig:callneoc},\ref{Fig:RTMe100dol},\ref{Fig:JohOCanonPro},\ref{Fig:Twomodestate},\ref{Fig:Hologra}).
\label{Fig:Truema} }
\end{figure}
Finally the conditions  for  the  $\pp^+<\pp^-$  couples in $\mathbf{C}_{coll}$  collision are more  articulated. We proceed from the analysis of radial $\ell^{\pm}$ profiles,
 and the following two cases are possible
\bea&&\nonumber\mathrm{\mathbf{(1)}}\; \pp^+\ll_{\mathbf{\times}}\cc^-,\; (a\gtrapprox0)\quad\mbox{where there is}\quad n_{\max}=2,\\
&&\nonumber \quad\mbox{and}
\\
&&\nonumber
\mathrm{\mathbf{(2)}}\;\pp^+ <\cc^-, \quad\mbox{with}\quad   \pp^+>_{\mathbf{\times}}\cc^-.
\eea
The first case-$\mathrm{\mathbf{(1)}}$ then implies the maximum number of tori $n_{\max}=2$.

{Then, if we define the radius  $ \bar{r}^-:\;\ell^-(\bar{r}^-)=-\ell^+$,  then there are the two possibilities for $\mathrm{\mathbf{(1)}}$ and $\mathrm{\mathbf{(2)}}$: }
\bea\label{Eq:wi-rddr}&&
\mbox{\textbf{--}there is}\quad |\ell^-/\ell^+|>1\quad{if}\quad r_{cent}^->\bar{r}^- \;
\\
&&\nonumber\mbox{which implies case}\quad \mathrm{\mathbf{(b)}}\; n_{\max}\geq2.\quad \mbox{Or}\\
 &&\mbox{\textbf{--} there is }\quad|\ell^-/\ell^+|<1 \quad{if}\quad r_{cent}^-\in]r_{cent}^+,\bar{r}^-[.
\eea
{In various  aspects of this work it was stressed that  we adopted a  constraints analysis, through the use of quantities defined in Eqs\il(\ref{Eq:def-partialeK}), choosing a
family of parameters $\{\mathbf{p^i},\mathbf{p^o},a/M\}$ for the  description of the \textbf{RAD}     orbiting around a \textbf{SMBH} with dimensionless  spin $a/M$.
 For a wide discussion on the application  of this method  in the collision emergence we refer to the \cite{dsystem}, while here we  point out  that tori collision is essentially determined, at \emph{fixed} value of the specific angular momentum, $(\ell_i,\ell_o)$ and of the \textbf{SMBH} parameter $a/M$, by   the \textbf{RAD} $K_{\ell}$-modes featuring, for increasing values of $K$,  the torus growth and  the increase of torus elongation $\lambda$ on the equatorial plane--this quantity pictured in Fig.\il(\ref{Fig:Quanumd}) and considered also in Fig.\il(\ref{Fig:callneoc}) in given explicitly in Sec.\il(\ref{Sec:theory-weell}).
 The $K$- and $\ell$ and $K_{\ell}$-modes are some of the  modes for \textbf{RAD} associated with the various perturbative approaches explored in \cite{ringed}, here we do not go into the details of this issue, being   beyond the purpose of  the present    work, but   we limit ourselves  to the description of some modes   adapting the discussion to the purposes faced here.
First, the  perturbations arising in the ringed disk structure are generally meant to be  perturbations of  its toroidal components,  subjected to different  constraints and  generated by perturbing     the \textbf{RAD} effective potentials in
Eq.\il(\ref{Eq:def-partialeK}) or, equivalently,  perturbations of the sequences of parameters $\mathbf{p}=\{\mathbf{p}_i\}_{i=1}^n$. There are several sub--modes associated with two main mode classes  $\ell$- and $K$ modes.
Keeping the other parameters  fixed, we could  speak  of \emph{$K$-mode} or  \emph{$\ell$-mode} of the perturbation  respectively. The \textbf{RAD}  can be perturbed in one of these modes or in a combination of  them, and each mode can have some  sub-modes defined   by the restrictions  on the perturbation, bounded by specific relationships on the spacings and elongations.
Keeping the sequence of parameters $\{\ell_i\}_{i=1}^n$   fixed,   the  perturbations will be only on the Heaviside functions in the effective potential (\ref{Eq:def-partialeK}). This is a   one dimensional problem (actually this is entirely fixed by the radial direction only, thought the \textbf{RAD} is generally a ``knobby'' accretion disk--see discussion in Sec.\il(\ref{Sec:theory-weell})). If each $\ell_i$ is fixed, then the range of variation for $K$ is comprised in a limited ranged defined by the value of $\ell_i$.  It should be noted that this kind of perturbation is actually a   (rigid) perturbation of the inner and  outer edges of each torus (or also the elongations $\lambda_i$ at fixed  center $r_{cent}^i$): i.e.,  the perturbation of an edge is transmitted rigidly  to the other range, with fixed center $r_{cent}$. The elongation $\lambda_i$ is however not conserved (when this is imposed one has  the  $K_{\lambda}-$modes).
The $K-$modes  may lead  effectively from a quiescent torus with angular momentum in  the range $\textbf{L1}$ to an accreting phase or, viceversa, to block  the accretion (for example this may occur in the so called ``drying--feeding''-processes considered in \cite{dsystem}). Nevertheless both these modes   preserve the   symmetry for reflection on the equatorial plane of the \textbf{RAD}, but  the perturbation generally leads  also  to a change in the vertical direction with a change of the height $h_i$ in each sub-configuration--see also Fig.\il(\ref{Fig:Quanumd}).
Varying  $\ell_i$ at fixed $K_i$, the Boyer surfaces   are not rigidly translated (a shift in the torus center $r_{cent}^i$) on the radial direction, but the torus  morphology  and, in particular,  its thickness changes.
In the  $\ell$-modes  an increase in  the specific angular momentum magnitude means a  perturbation
outwards of the torus center (which viceversa  is kept fixed by $K$-modes), leading to a radial movement
outwards.
The $\ell$-modes can also lead  (even at  fixed $K$) to a change of the tori equilibrium from a quiescent phase to the accretion onto the central \textbf{SMBH}  with the decrease of specific angular momentum magnitude.
 As seen in Figs\il(\ref{Fig:Stopping},\ref{Fig:Truem}) and (\ref{Fig:Truema}),  $K$-parameter varies in the range    $[K_{\min},K_{\lim}[$, i.e.,  $K$ is bounded from below by the limiting case of  test  particle ring  with zero pressure contribution  and from above by the limit  $K_{\lim}$, i.e.  $K_{\max}$ correspondent to the point of  minimum pressure,  or  $K_{\lim}=1$, depending on the angular momentum  in  $\mathbf{L1}$ or $(\mathbf{L2},\mathbf{L3})$ respectively. Actually, for  the $\mathbf{j}$-torus of the aggregate, $K^\mathbf{j}_{\lim}$   is bounded by the presence of the  adjacent tori \cite{ringed}--see also Eqs\il(\ref{Eq:def-partialeK}).  $\ell$-modes are considered in Figs\il(\ref{Fig:Stopping}) and Figs\il(\ref{Fig:Truema}), while
$K-$modes are  in Fig.\il(\ref{Fig:Truem}).
A significant part of this investigation   accurately   evaluates  the totality of these possibilities for  $\ell$corotanting  and the $\ell$counterrotating cases complectly describing the model --Fig.\il(\ref{Fig:callneoc},\ref{Fig:RTMe100dol},\ref{Fig:JohOCanonPro},\ref{Fig:Twomodestate}). }
\subsection{Center-of-mass (CM) energy }\label{Eq:CM-ene}
We  can now evaluate     CM energy $(\mathrm{E}_{\mathrm{CM}})$      for two colliding particles  from the tori in $\mathbf{C}_{coll}$, $\mathbf{C}_{\times}$ or $\mathbf{C}_{coll}^{\times}$ couples, in the test  particle approximation, with  four-momenta $p_{j}^a$     and  rest masses $\mu_{j}$. There is:
\bea\nonumber
&&\mathrm{E}^2_{\mathrm{CM}}=-p^a_{tot}p_{tot\phantom\ a}= \mu_{i}^2+\mu_{o}^2
-2g_{ab}p_{i}^ap_{o}^b,
\\
&&\label{Eq:pd-cong}\mbox{where}\quad
p_{tot}\equiv p_{i}+p_{o},
\eea
 as
 observed by a local observer which is at rest in the CM
frame of the two particles.

We  consider   the cases with  at least one circularly orbiting particle (equatorial motion), see \cite{Harada:2010yv,Tursunov:2013zha,1980BAICz..31..129S,Blaschke:2016uyo,Stuchlik:2013yca,Stu-Sche:2012:CLAQG}. Then  Eq.\il(\ref{Eq:pd-cong}) reads
\bea
&&\nonumber
\mathrm{E}^2_{\mathrm{CM}}= \frac{2p_t^{\pi} [r^3+a^2 [2-2 \overline{\ell}_{T}+(2-r)\overline{\ell}_{\pi} +r]]}{r(r^2-2 r +a^2)}+2\mu^2,
\\\label{Eq:CMr}
&& \mbox{where}\quad\overline{\ell}_{T}\equiv \overline{\ell}_{i}+ \overline{\ell}_o,\quad \overline{\ell}_{\pi} \equiv \overline{\ell}_{i}\overline{\ell}_o,\quad p_t^{\pi}\equiv p_t^{i} p_t^{o}
\eea
($M=1$, $\mu_{i}=\mu_{o}\equiv\mu$). We note that  $\mathrm{E}^2_{\mathrm{CM}}$ in Eq.\il(\ref{Eq:CMr})  is function of the  ratio $\overline{\ell}_{i}\equiv\ell_i/a$ (for $a>0$)      \cite{pugtot}  within Eqs\il(\ref{Eq:conf-decohe},\ref{Eq:wi-rddr}).
Whereas the dependence  on  the torus masses and elongations on the equatorial plane derives   from    $p_t^{\pi}$  related  to the contact point $r$ through, at least, $p_t^{i}$.
By considering  the    test free particle    limit we use the definition of $\ell$ with $p_{\phi}$ and
$p_{t}$ constant--see also discussion in Sec.\il(\ref{Sec:more-l}). 

The contact (collision) point $r$ in Eq.\il(\ref{Eq:CMr}) is    $r_{coll}$ for the case  $\mathbf{C}_{coll}$,   where  there is $u_r^i(r_{coll})=u_r^o(r_{coll})=0$ ($u_r$ is the radial velocity),
it is $r_{out}^{i}$ for the case   $\mathbf{C}_{\times}$   (where there is $u_r^i(r_{out}^{i})=0$), and  it is $r_{coll}^{\times}=r_{out}^{i}=r_{\times}^{o}$  (with $ u_r^i(r_{coll}^{\times})=u_r^o(r_{coll}^{\times})=0$) for  $\mathbf{C}_{coll}^{\times}$.

Note that in these three cases, there is always  $u_r^{i}=0$ (circular motion assumption).
For  $(\mathbf{C}_{\times},\mathbf{C}_{coll}^{\times})$ (with $\pp^-\ll_{\mathbf{\times}}  \cc_{\times}^+$), there is    $r=r_{\times}$   and fluid specific angular momentum $\ell_{o}=\ell^+(r^{o}_{\times})$,  $\ell_{i}=\ell^-(r_{cent}^{i})$.

We consider the energy $\mathrm{E}_{\mathrm{CM}}$  for a free falling particle approximation from  $r_{\times}^+=r_{\mso}^+$  on a thin ring of matter close to $ r_{\mso}^{-}$ (i.e. $\mp\ell_{\pm}\gtrapprox \mp\ell_{\mso}^{\pm}$, and  $r_{cent}^+= r_{\mso}^+$), reducing  formally to the problem  of ``on-ISCO'' collision from an ``in-ISCO'' particle considered in \cite{Harada:2010yv,Stuchlik:2013yca,Stuchlik:2017rir}.

However, even within the test particle approximation, in the  cases  $\mathbf{C}_{\times}$, $\mathbf{C}_{coll}$ and $\mathbf{C}_{coll}^{\times}$,  we retain the constraints on  $(\overline{\ell}_{T}, \overline{\ell}_{\pi})$  in  $\mathrm{E}^2_{\mathrm{CM}}$ for the torus formation and emergence of   collisions--see also Table\il(\ref{Table:pol-cy-multi}) for a summary of the notation introduced in this section.

In general, {for a $\mathbf{C}_{\times}$ collision  in a couple  $\pp^-<\cc_{\times}^+$ (with outer counterrotating torus in accretion and inner corotating quiescent or in accretion onto the central \textbf{BH})},  we  use the   assumption of a free particle at $r_{\times}$ and    $p_{\phi}^{o}=$constant. {In this tori couple inner torus acts as a ``screening'' torus for the outer accreting torus.}
For the inner torus, the situation is much  more complex and we could not use the free particle approximation except  in the case of an  inner thin ring  of matter.
Leaving the treatment of the general situation for  future analysis, in Fig.\il(\ref{Fig:AliceB})  we restrict our consideration to some simple assumptions: in the $\mathbf{C}_{coll}$ case, we evaluate  $\mathrm{E}_{\mathrm{CM}}$ for two circularly orbiting particles with constants  $(-p_t,p_{\phi})$ at $r_{coll}$. This situation may be applied for $\ell$corotating and $\ell$counterrrotating  cases, eventually we could consider the inner torus in accretion.

Then, a particularly interesting  situation occurs when  the inner torus of $\mathbf{C}_{coll}$ is  accreting,   or the outer torus  of $\mathbf{C}_{coll}^{\times}$  is in accretion  or finally, the two tori are accreting in a $\pp^-< <_{\mathbf{\times}}\cc_{\times}^+$ couple.
 Within  the assumption of circular motion in $r_{coll}$
for the outer torus,  there is  $\ell^{o}
>
p_{\phi}^{o}({r_{coll}})$  in magnitude, while for the inner torus the situation is more complicated.
 {By solving  the hydrodynamic equations (\ref{E:1a0}) for each torus,
  with
$r_{in}^o\geq r_{out}^i$,  it is possible to find that for
$\pp^-< <_{\mathbf{\times}} \cc^+_{\times}$, {see Table\il(\ref{Table:pol-cy-multi})},   the  points $(r_{\times}^+,r_{coll}^{\times})$ are in the range $]r_{\mbo}^+,r_{\mso}^+]$ only  for $a\in[a_{0},M]$, where $a_{0}\equiv0.372583 M$--Fig.\il(\ref{Fig:teachploeff}) and Figs\il(\ref{PlotdisolMsMb}) for $a_{0}$. {This means that collision with an outer counterrotating accreting torus can occur in these geometries \emph{only}.}
In fact:   this can be also seen, by noting that a $\pp_{\mathrm{I}}^+$ configuration (with $\mp\ell^{\pm}\in ]\mp\ell_{\mso}^{\pm},\mp\ell_{\mbo}^{\pm}[$) cannot cross the orbit $r_{\mso}^-$ ($r_{\mso}^- \not{\in} \pp_{\mathrm{I}}^+$) for
 $a>a_{0}$, while for $a <a_{0}$ there is
$r_{\mso}^-{\in}\cc_{\times}^+$, and  $r_{\mso}^-{\in}\cc^+$, if  $\ell^+\in[-\ell^+(r_{\mso}^-),-\ell_{\mbo}^+]$.

Crossing of  $(r_{\mso}^+,r_{\mbo}^+)$  in $\pp_{i}^-$ assures  that the  $\mathrm{E}_{\mathrm{CM}}$   evaluation    in Fig.\il(\ref{Fig:AliceB}) is well grounded.  Figure\il(\ref{Fig:AliceB})   shows   evaluations of  the energy $\mathrm{E}_{\mathrm{CM}}$  for  different coupled of tori in dependence on the \textbf{BH} spin $a/M$.

A rapid look at Figure\il(\ref{Fig:AliceB})  reveals  that for the  collisions $\mathbf{C}_{\times}$, the energy center-of-mass energy  $\mathrm{E}_{\mathrm{CM}}$ can  be arbitrarily large due to increasing  \textbf{BH} dimensionless   spin up  to $a=M$.
\begin{figure}
\includegraphics[width=7.71cm]{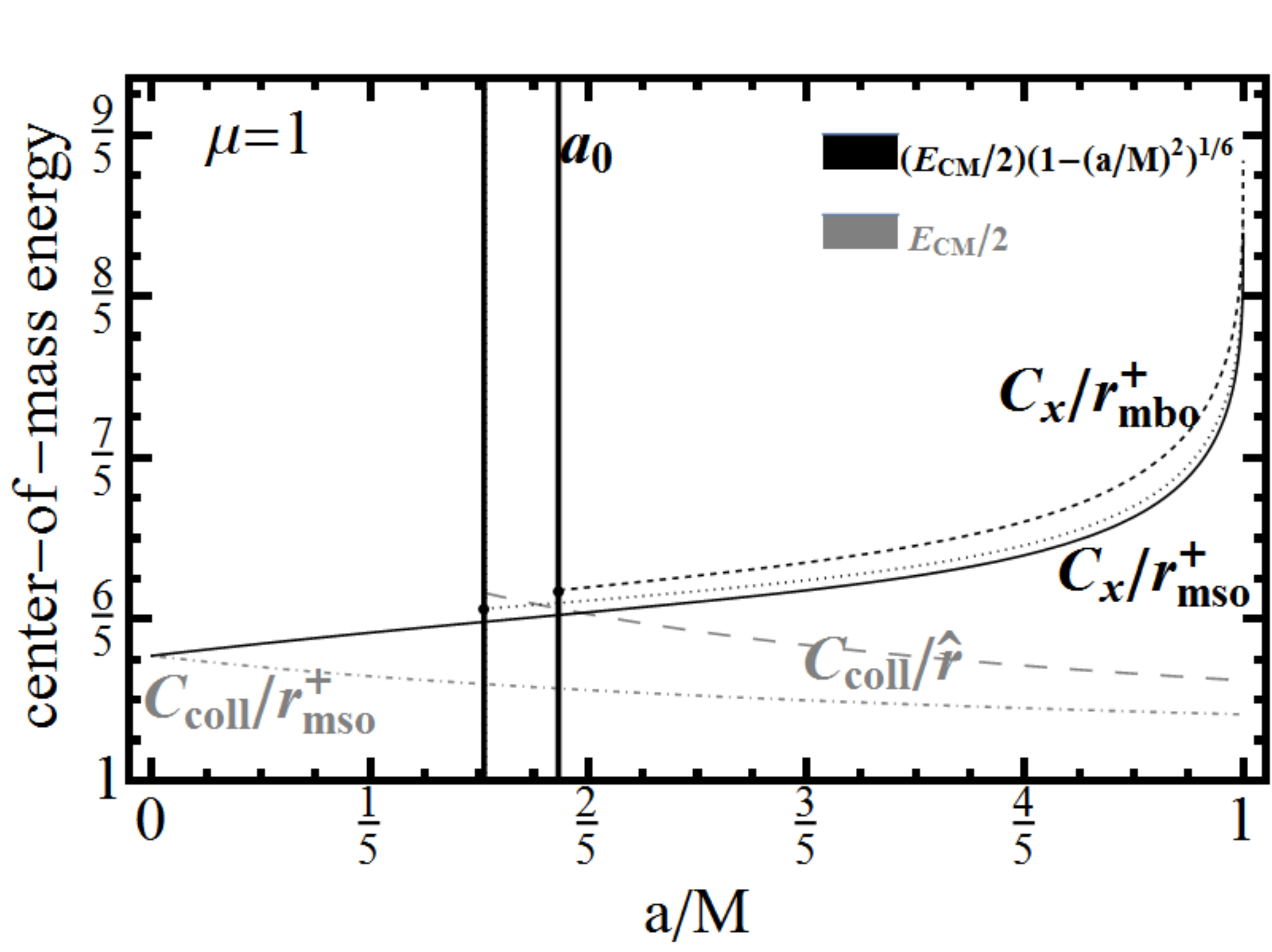}\\
\includegraphics[width=5.1cm,angle=90]{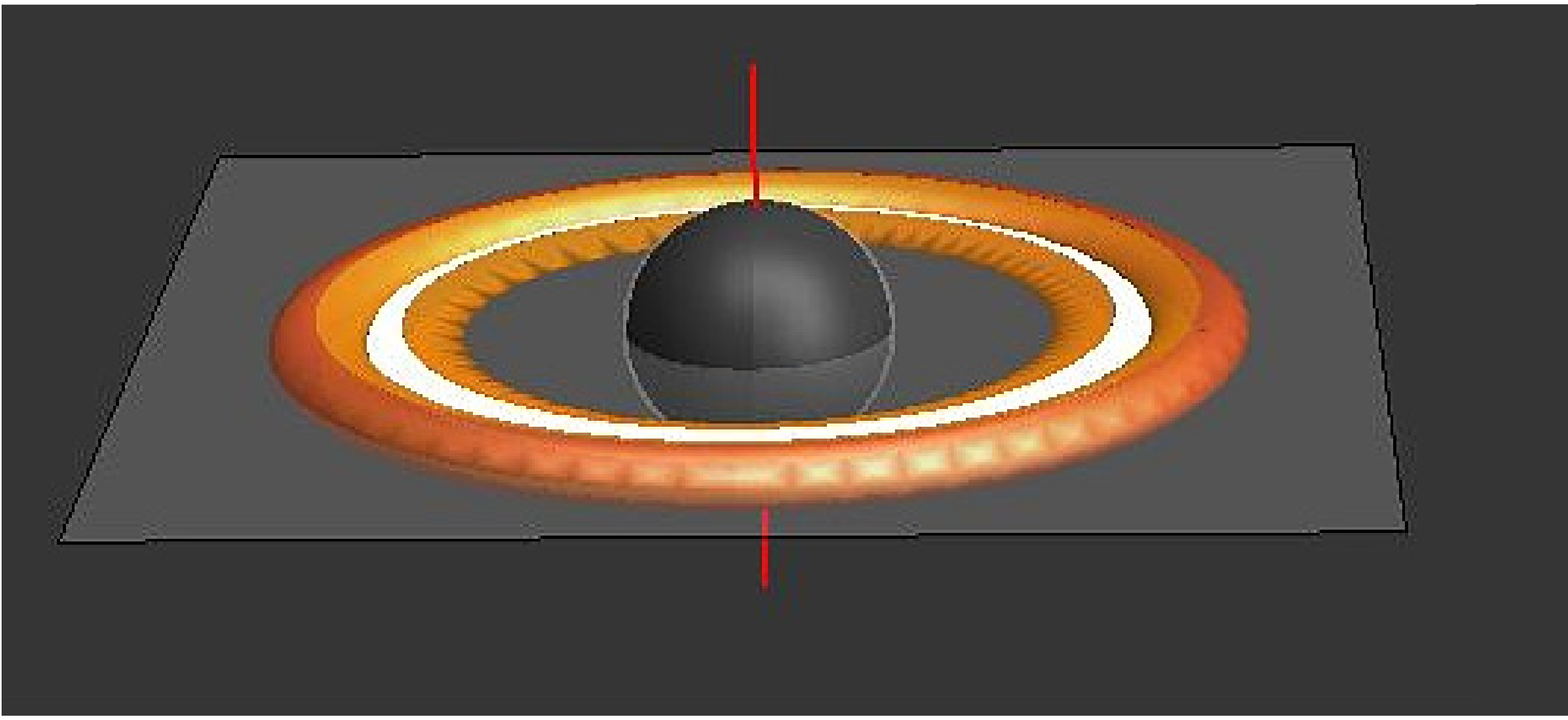}\\
\includegraphics[width=5.1cm,angle=90]{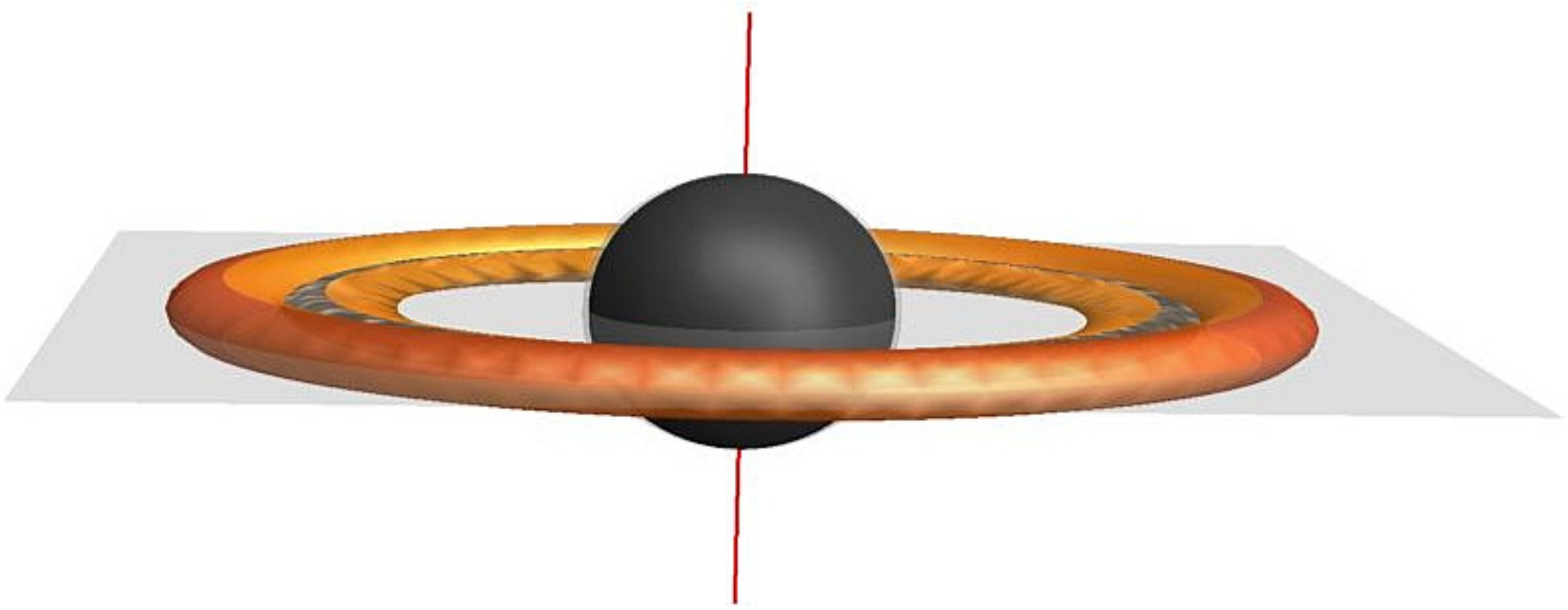}
\caption{\emph{Upper} panel: Center-of-mass energy  $\mathrm{E}_{\mathrm{CM}}/2$ in the  test particle approximation,  versus $a/M$, for  the  couples $\pp^-< <_{\mathbf{\times}} \cc^+_{\times}$--Figs\il(\ref{Fig:teachploeff}) and Figs\il(\ref{Fig:teachplo}). We  consider: infalling particle from $r_{\mso}^+$ black curve $(r_{\mbo}^+$-dashed-black curve) of $\mathbf{{{C}}}_{\times}$ colliding configuration.  They are labeled respectively as  $\mathbf{{{C}}}_{\times}/r_{\mso}^+$ (black curve)   and $\mathbf{{{C}}}_{\times}/r_{\mbo}^+$ (dashed-black curve)). Dotted black curve is for  a particle  from orbit $\hat{r}\equiv r_{\mso}^+-2M$  colliding    on a circular orbiting particle located  at  $r_{\mso}^-$-(for the asymptote-factor  $(1-(a/M)^2)^{1/6}$).
Collision  of counterrotating particles  on an inner corotating thin ring  at   $r_{\mso}^+$ ($\hat{r}$) are indicated with $\mathbf{C}_{coll}/r_{\mso}^+ $(dotted-dashed gray curve)  and ${\mathbf{C}}_{coll}/\hat{r}$ (dashed gray curve) respectively, -see also \cite{Harada:2010yv}.
Main notation is also listed in Table\il(\ref{Table:pol-cy-multi}). \emph{Center panel:}
 \textbf{{GRHD}}-numerical 3D integration   {of Eq.\il(\ref{Eq:scond-d}) proving the } density surfaces of colliding $\ell$corotating $\cc_i^-<\cc_o^-$ couple of corotating ($\mathbf{C}_{coll}$) tori,  with specific angular momentum $\ell_i^-=3.3$ and $\ell_o^-=3.4$ and $K$ parameter $(K_i^-=0.927, K_o^-=0.935)$ respectively. Black region is $r<r_+$, $r_+$ is the outer horizon of the black hole of spin  $a=0.385 M$, gray region is the outer ergosurface, color are chosen according to improved visual effect, integration is stopped at the emerging of tori collision--see also Fig.\il(\ref{Fig:teachplo}). \emph{Bottom panel}, different view of the colliding couple $\mathbf{C}_{coll}$ to enlighten the colliding region.  Tori are coplanar and  centered on the central \textbf{BH} orbiting on its  equatorial plane.}\label{Fig:AliceB}
\end{figure}
\subsection{Evidences of double accretions}\label{Sec:more-l}
From observational view-point a relevant aspect of the \textbf{BHs} astrophysics related to the interaction of central \textbf{BH} with  the  \textbf{RADs}   consists in the possibility that a \textbf{RAD} would have been indeed
observed so far and identified   as the case of  one  accretion torus  orbiting its central \textbf{BH} attractor:  this possibility, also discussed in \cite{multy}, challenges the description of \textbf{RAD} as an whole  orbiting toroidal structure  despite its internal complex dynamics. In order to do this in \cite{ringed}, the   definitions of  \textbf{RAD} elongation and thickness, and   the \textbf{RAD} distribution of angular momentum were provided,  while  in \cite{multy} these systems were carefully related to  the characteristics of the \textbf{SMBHs} attractors.
This new framework  can be particularly interesting  and can have important implications
 in the processes analyzed in   \cite{apite1,apite2,apite3,Li:2012ts,Oka2017,Kawa,Allen:2006mh}, where the hypothesis of different accretion stages   are claimed  as ground  to  justify the \textbf{SMBH} masses at high redshift,  and can have a relevant role  in the  studies of the X-ray screening emission where an accreting torus is supposed to be adjacent  to an obscuring torus     (i.e. a $\pp^-<\cc_{\times}^+$ couple, or a combination of these in the \textbf{RAD} model). However, our analysis strongly constrains  this possibility. Different accreting phases may be explained  in the \textbf{RAD} model for example by a simultaneous accretion phases in the couples  $\cc_{\times}^-<\cc_{\times}^+$, or a discontinuous accreting phase induced by the internal \textbf{\textbf{RAD}}   evolution (for example a ``drying-feeding'' effect described in \cite{dsystem} and considered in Sec.\il(\ref{Eq:CM-ene})).
Our analysis shows how the \textbf{RAD} formation is actually highly constrained and  strongly limited by several factors, such as  its  angular momentum distribution, particularly  by the dimensionless spin of the central attractor, rendering the  contexts where to potentially  observe these objects sharply focused and precise
  \cite{dsystem,multy,long}. Moreover, the possible  inter-disk activity in the agglomerate, such as the arising  of proto-jets  \cite{long,open}, collision or double accretion here analyzed in details, shows a quite large
 template of  phenomenology associated to the \textbf{RAD}.

It is then important to note that in general the  \textbf{RAD} is in fact a  ``knobby'' accretion disk; in the  $n=2$ case, the  $K$ and  $\ell$ modes and  the evolution of each \textbf{RAD} component is determined by a variation of the distribution of the matter in the \textbf{RAD}.
 Note that  an  $\ell$counterrotating \emph{accreting} couple has no special {constraints} on the relative height of the tori. This obviously implies a very wide set of  possibilities for  a knobby
 \textbf{RAD} disk.
First, as demonstrated in  \cite{ringed}, a \textbf{RAD} is always a geometrically thin accretion disk.
In order to fix some ideas
 we provide here some specific considerations for a   $\cc_{\times}^-<\cc_{\times}^+$ couples of Figs\il(\ref{Fig:Hologra}).
We can easy evaluate many  of the  \textbf{RAD}   characteristics, as the  \textbf{RAD} thickness $(h)$ and elongations $(\lambda)$, tori spacings $({\bar{\lambda}_{i,o}})$ showed in Fig.\il(\ref{Fig:Quanumd}), through an assessment of the  $ K $-parameter only--for a precise definition and discussion on this quantities we refer to \cite{ringed}.
 We focus in particular on the case of Fig.\il(\ref{Fig:teachploeff})  and Fig.\il(\ref{Fig:Hologra}) where there is
 $a = 0.382 M$,  $\ell^+_o = -3.99$ and $\ell^-_i = 3.31$. In this specific case,  the  \textbf{RAD}  aggregates a set of rather small tori. In fact:
  for the counterrotating  torus, $\cc_{\times}^+$, there is
$r_{cent}^+= 7.75994M$,
$r_{\times}^+=6.70523M$,
$r_{out}^+= 8.4422 M$, with elongation
$\lambda^+= 1.73696M$, while the torus height  is
$h^+/2= 0.223684M$;
for the corotating torus, $\cc_{\times}^-$, the torus center, the inner and outer edges are located at
$r_{cent}^-= 5.13459M$,
$r_{\times}^-=4.28942 M$,
$r_{out}^-= 5.73609M$, while the torus elongation is
 $\lambda^-= 1.44667 M$, and the high
$ h^-/2=0.230375 M$.
The spacing $\bar{\lambda}_{i,o}$  between two consecutive tori regulates the collision, the  \textbf{RAD}-modes,  its formation and in the analysis of the  stratified \textbf{RAD}  X-ray emission spectra, and can determine the effects of the  X-ray screening.
In fact, in the couple $\cc_{\times}^-<\cc_{\times}^+$, we also recall that  screening effects are only possible with the corotating tori as  follows $\cc_{\times}^-<\cc^-<...<\cc_{\times}^+<\cc^{\pm}$. The  inner accreting corotating torus may act as  a screening torus for the accreting outer counterrotating  torus together with any inner, quiescent corotating torus, formed in the spacing region.  It is therefore clear that this situation may occur only in the first phases of evolution for the outer torus, depending on the spacing in the tori couple.
The spacing for the special   $\cc_{\times}^-<\cc_{\times}^+$ couple  of  Fig.\il\ref{Fig:teachploeff} is $\bar{\lambda}\equiv r_{\times}^{o}-r_{out}^i=0.969137M$. Note that the space-scales are in units of \textbf{SMBH} masses (i.e. $10^6 M_{\odot}$--$10^9 M_{\odot}$-- $M_{\odot}$ being solar mass)-- \cite{ringed}.
Such a couple can be observable  orbiting any  Kerr \textbf{BH}, and the larger is the \textbf{BH} dimensionless spin, the bigger the tori can grow,  while a larger spacing is required. In fact,
an immediate evaluation  shows  that the maximum spacing  possible for  a double accreting couple is  $\bar{\lambda}_{\max}\lesssim 8M$, in the case of a near extreme Kerr \textbf{BH} where  $a\approx M$.
This also  shows how  \textbf{SMBHs}and high spin \textbf{BHs}  are the  most promising attractors for  \textbf{RAD} formation and particularly for the most articulated  structures as the $\ell$counterrotating tori, with a double accretion or  screening tori.

Below we provide an investigation of mass accreting rate, related to \textbf{RAD} tori, and  other significant functions of the accreting disks physics focusing on their dependence on the \textbf{SMBH} spin, and the differences between corotating and counterrotating fluids.

\subsection{Tori from polytropic fluids}
We consider   \textbf{RAD} tori   with  polytropic fluids:  $p=\kappa \varrho^{1+1/n}$ (here $n$ should not be confused with the  \textbf{RAD} order). In Sec.\il(\ref{SeC:poly})  we shall discuss  this hypothesis more closely with some additional considerations,  here we note that
this hypothesis allows us to provide estimates of different quantities of each torus, which are extremely important for the evaluation of the  energy release from \textbf{RADs}. In particular the mass-flux,  enthalpy-flux (evaluating also the temperature parameter),
and  the flux thickness--see \cite{abrafra} and \cite{Japan}. In details, these quantities  are listed in Table\il(\ref{Table:Q-POs}).
\begin{table*}[ht!]
\caption{Quantities $\mathcal{O}$ and $\mathcal{P}$.  $\mathcal{L}_{\times}/\mathcal{L}$ stands for  the  fraction of energy produced inside the flow and not radiated through the surface but swallowed by central \textbf{BH}. 
Efficiency
$\eta\equiv \mathcal{L}/\dot{M}c^2$,    $\mathcal{L}$ representing the total luminosity, $\dot{M}$ the total accretion rate where, for a stationary flow, $\dot{M}=\dot{M}_{\times}$,  $W=\ln V_{eff}$ is the potential  of Eq.\il(\ref{Eq:scond-d}), $\Omega_K$ is the Keplerian (relativistic)  angular frequency,
$W_s\geq W_{\times}$ is the value of the equipotential surface, which is taken with respect to the asymptotic value, $ W_{\times}=\ln K_{\max}$  is the  function at the cusp (inner edge of accreting torus), $\mathcal{D}(n,\kappa), \mathcal{C}(n,\kappa), \mathcal{A}, \mathcal{B}$ are functions of the polytropic index and the polytropic constant.
}
\label{Table:Q-POs}
\centering
\begin{tabular}{|l|l|}
 \mbox{\textbf{Quantities}}$\quad  \mathcal{O}(r_\times,r_s,n)\equiv q(n,\kappa)(W_s-W_{\times})^{d(n)}$ &   $\mbox{\textbf{Quantities}}\quad  \mathcal{P}\equiv \frac{\mathcal{O}(r_{\times},r_s,n) r_{\times}}{\Omega_K(r_{\times})}$\\\hline\hline
$\mathrm{\mathbf{Enthalpy-flux}}=\mathcal{D}(n,\kappa) (W_s-W)^{n+3/2},$&  $\mathbf{torus-accretion-rate}\quad  \dot{m}= \frac{\dot{M}}{\dot{M}_{Edd}}$  \\
 $\mathrm{\mathbf{Mass-Flux}}= \mathcal{C}(n,\kappa) (W_s-W)^{n+1/2}$& $\textbf{Mass-accretion-rates }\quad
\dot{M}_{\times}=\mathcal{A}(n,\kappa) r_{\times} \frac{(W_s-W_{\times})^{n+1}}{\Omega_K(r_{\times})}$
 \\
 $\frac{\mathcal{L}_{\times}}{\mathcal{L}}= \frac{\mathcal{B}}{\mathcal{A}} \frac{W_s-W_{\times}}{\eta c^2}$&     $\textbf{Cusp-luminosity}\quad  \mathcal{L}_{\times}=\mathcal{B}(n,\kappa) r_{\times} \frac{(W_s-W_{\times})^{n+2}}{{\Omega_K(r_{\times})}}$
 \\
\hline\hline
\end{tabular}
\end{table*}
\begin{figure*}
\includegraphics[width=8.91cm]{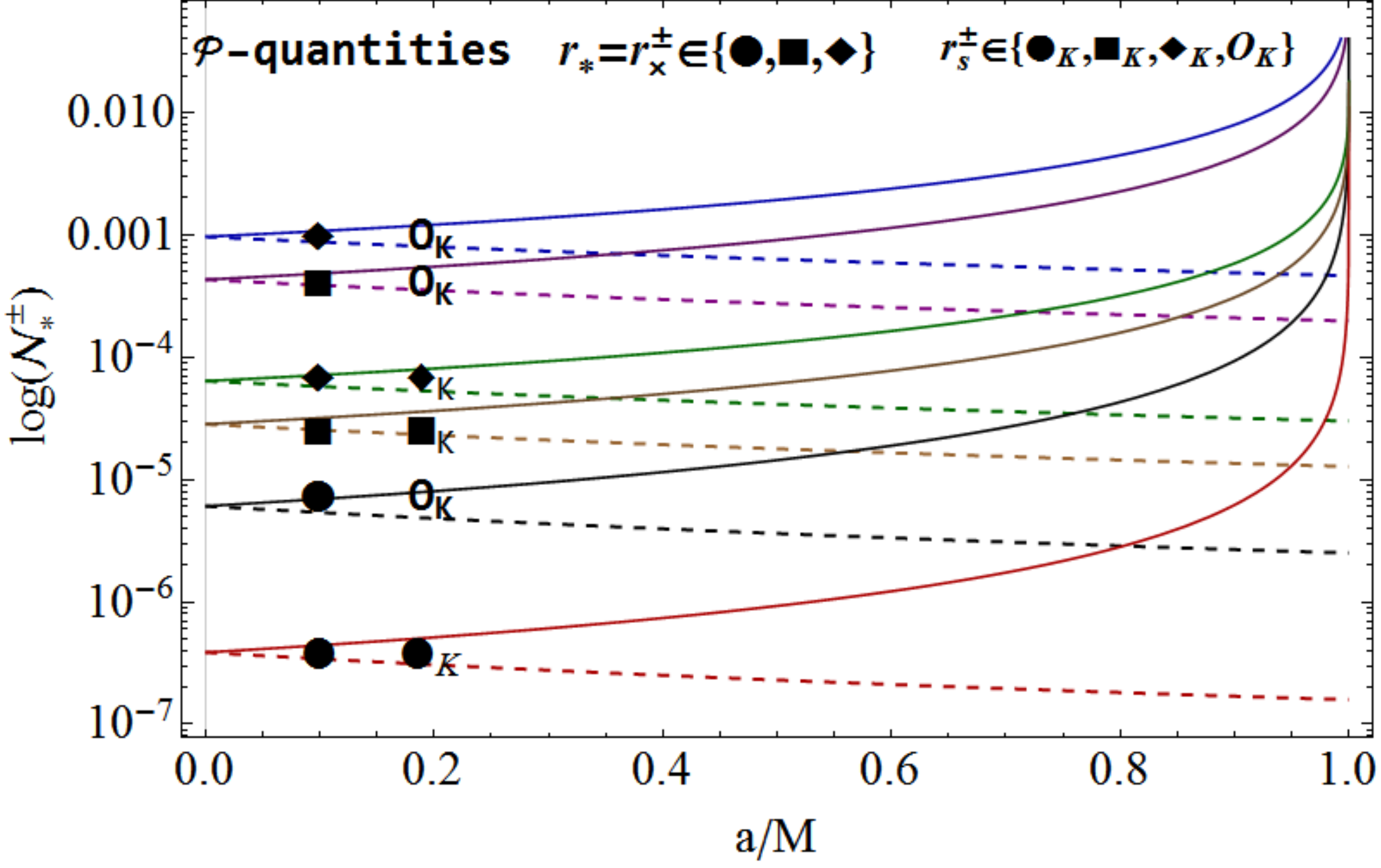}
\includegraphics[width=8.91cm]{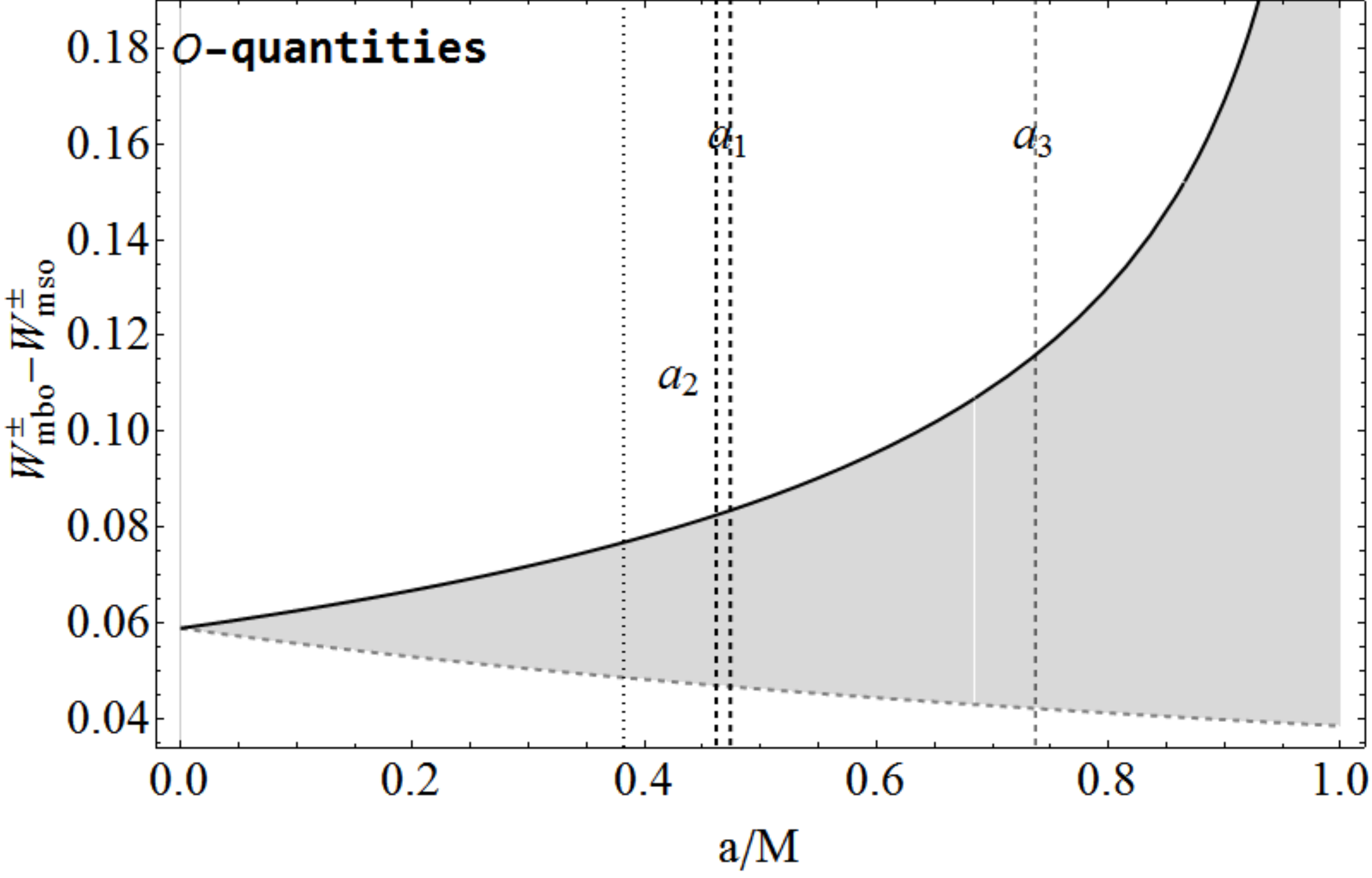}
\includegraphics[width=8.91cm]{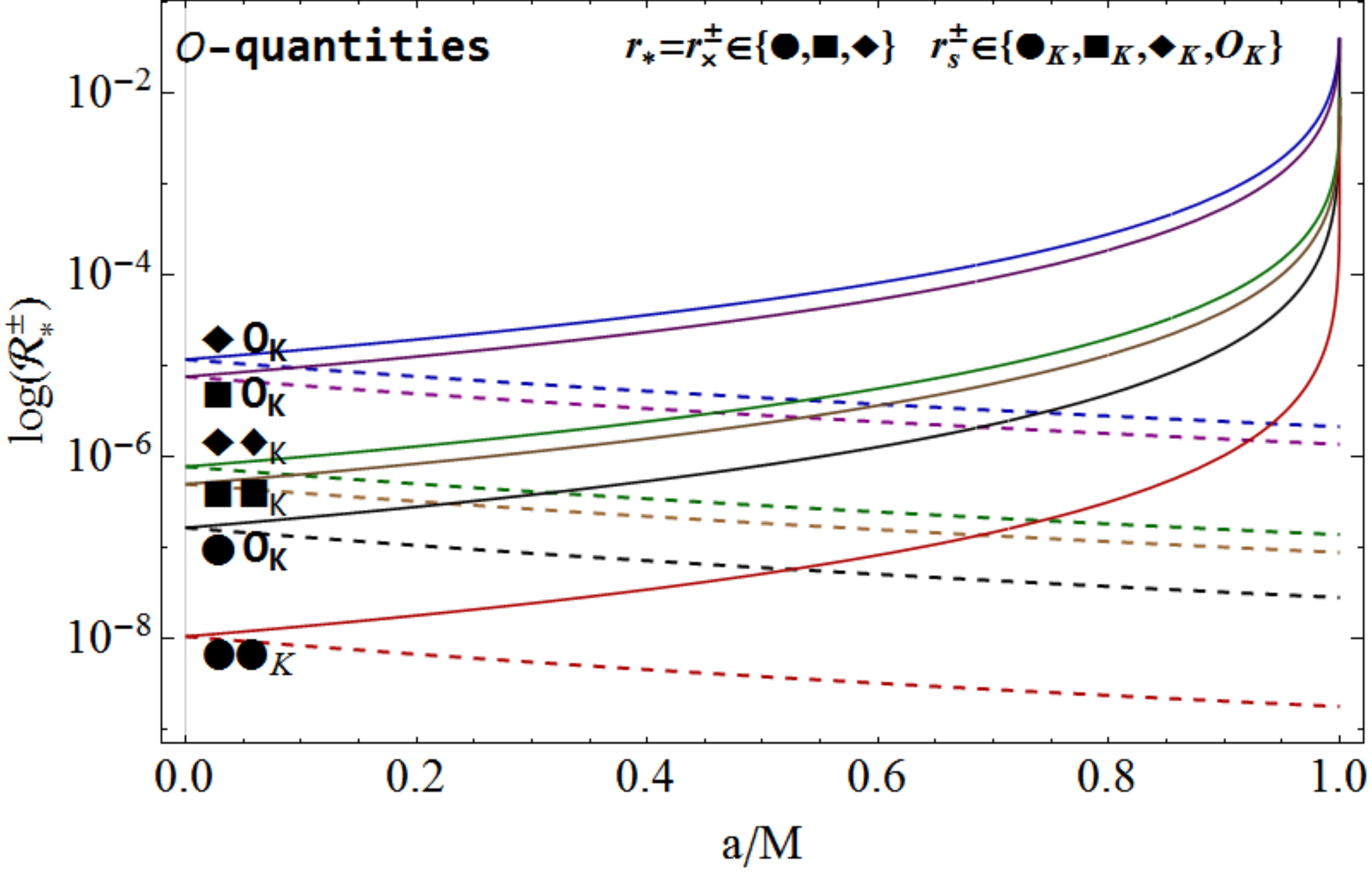}
\includegraphics[width=8.91cm]{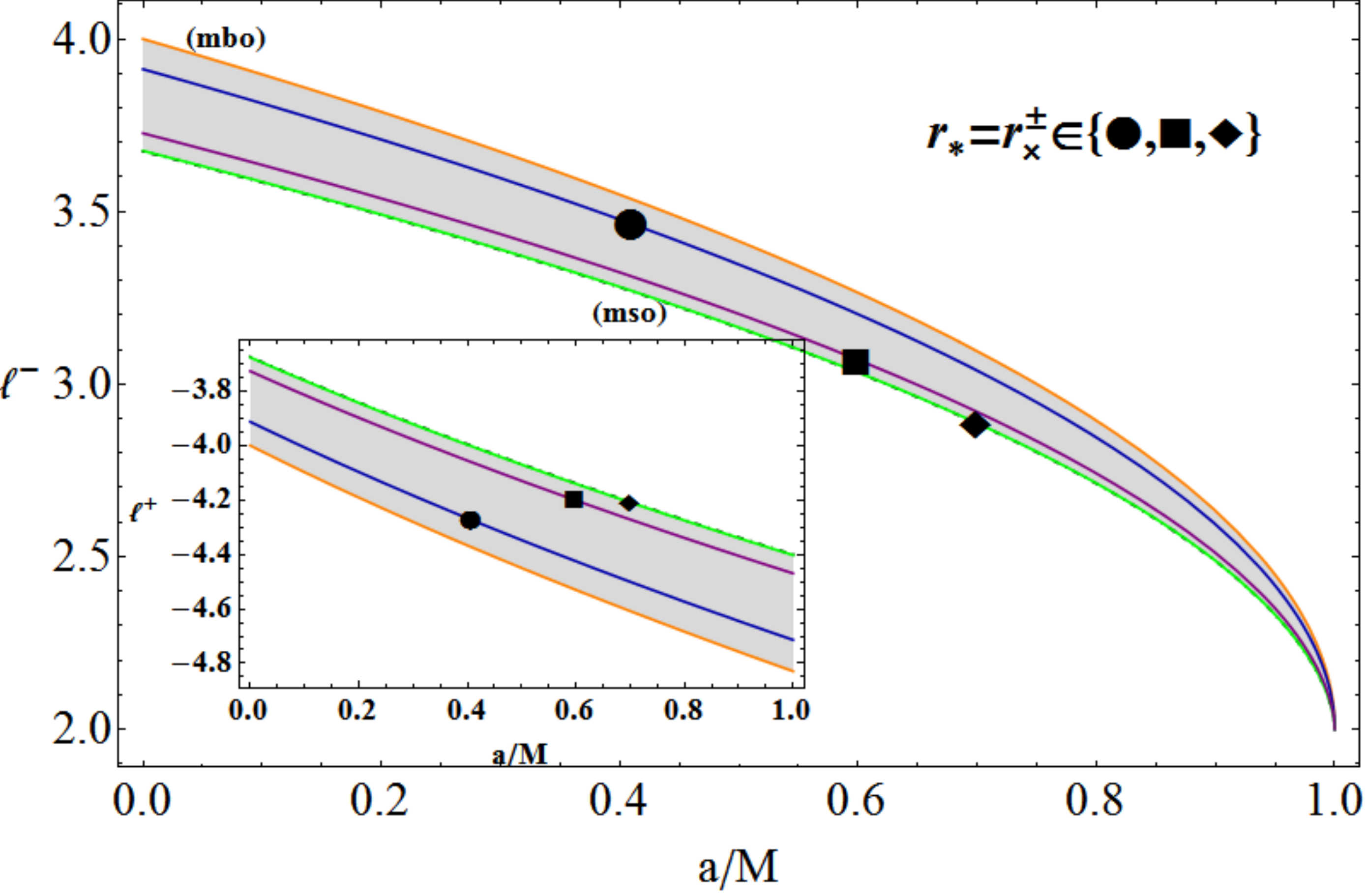}
\includegraphics[width=8.91cm]{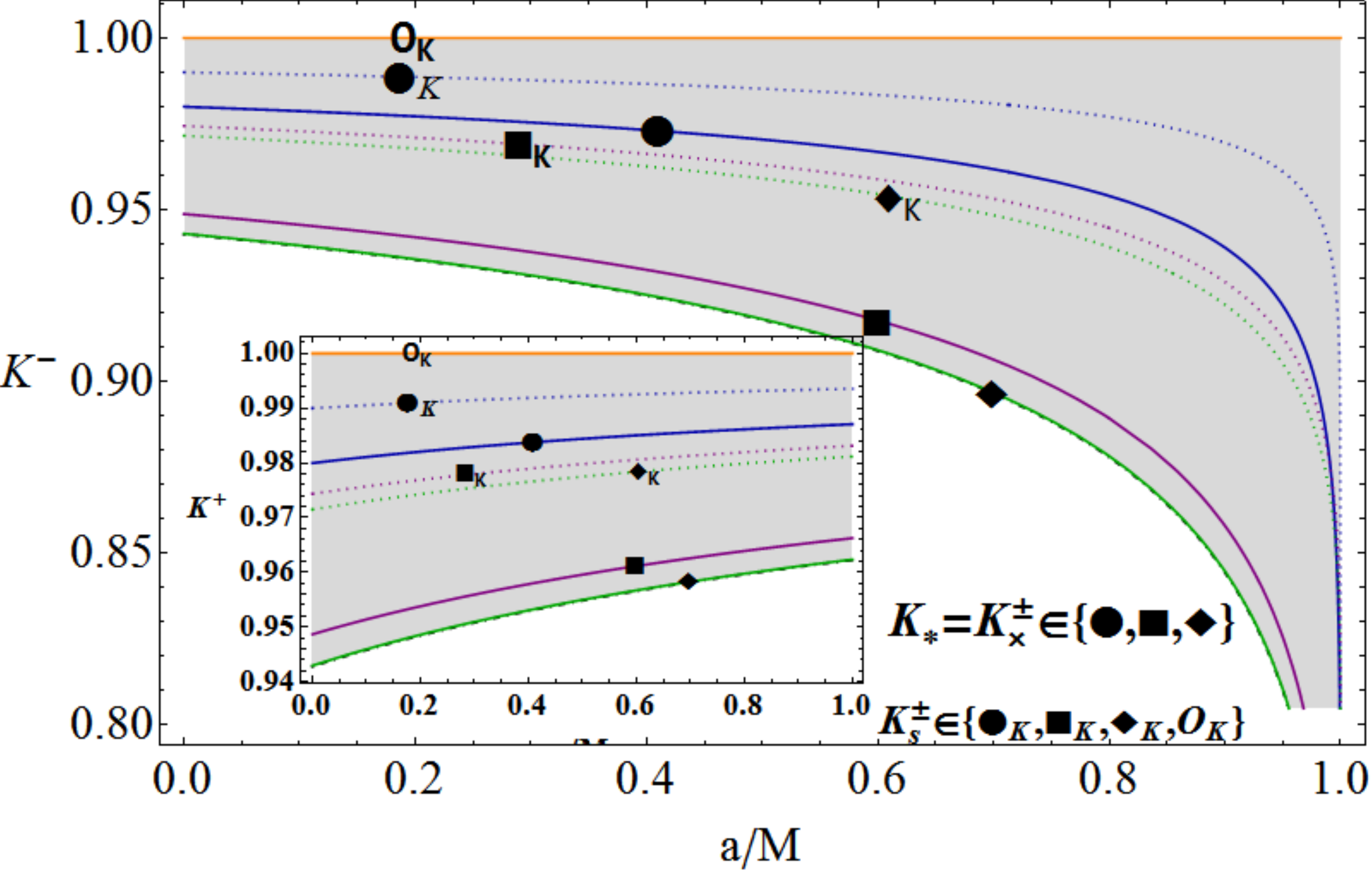}
\includegraphics[width=8.91cm]{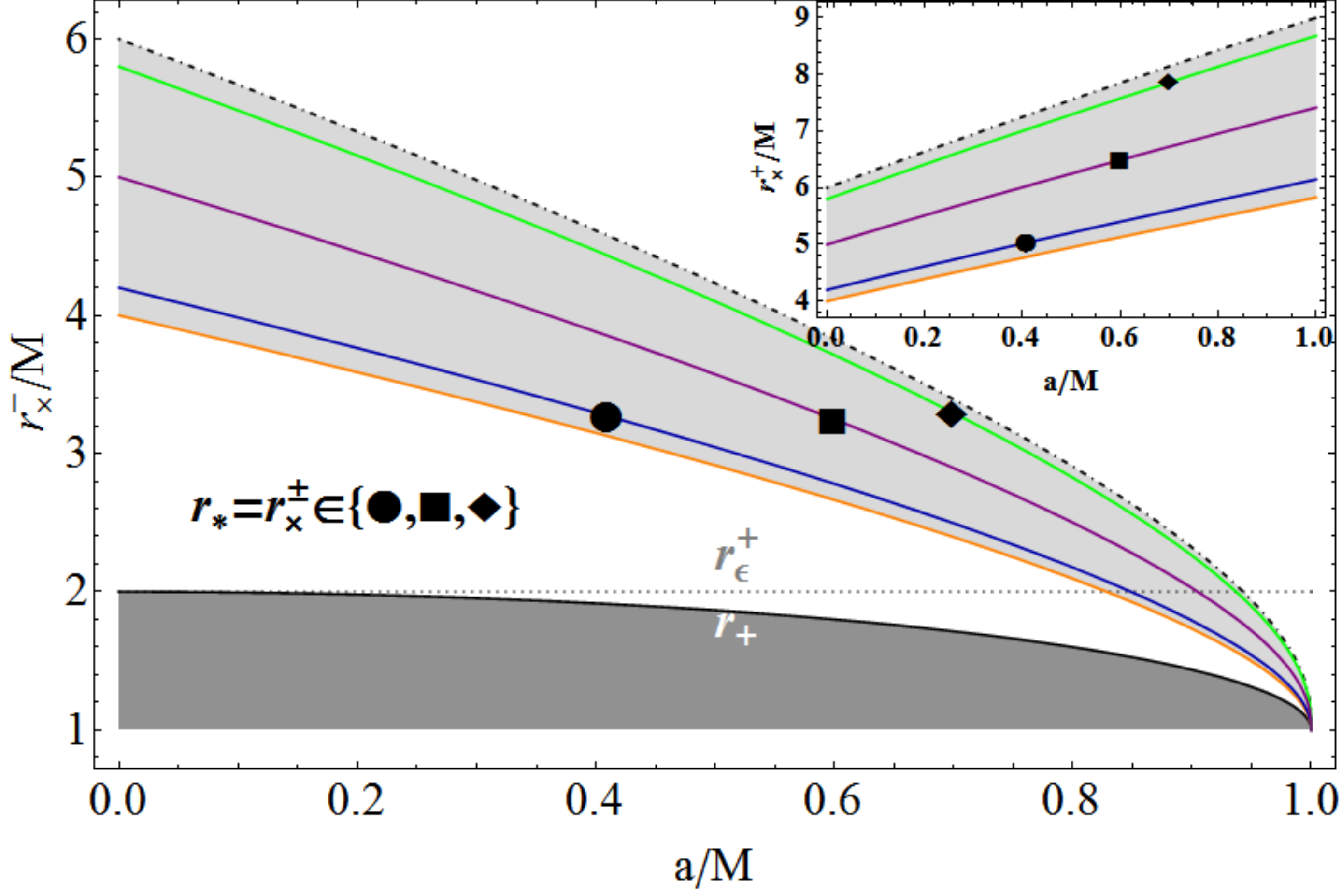}
\caption{Evaluation of $\mathcal{P}$- and $\mathcal{O}$-quantities of Table\il(\ref{Table:Q-POs})  for corotating and counterrotating tori, versus Black hole dimensionless spin $a/M$.     Plots of $\mathcal{N}_{*}^{\pm}\equiv{r_*} (W^{\pm}(r_{s})-W^{\pm}_{*})^\kappa(\Omega_K(r^{\pm}_*))^{-1} $ (\emph{upper left panel}) for $\mathcal{P}$-quantities analysis; the maximum difference $(W^{\pm}(r_{mbo})-W^{\pm}(r_{mso}))$     (\emph{upper right panel})  (Eqs\il(\ref{Eq:provi-Ang-Lesb},\ref{Eq:soft-chogra})) and  $\mathcal{R}_{*}^{\pm}\equiv(W^{\pm}(r_{s})-W^{\pm}_{*})^\kappa$ (\emph{center left panel}),  for $\mathcal{O}$-quantities analysis  as functions of \textbf{SMBH}  dimensionless spin $a/M$, for corotating ([\textbf{-}]--continuum curves) and counterrrotating  ([\textbf{+}]--dashed curves)   tori at  different $r_*=r_{\times}^{\pm}\in\{\bullet,\blacksquare,\blacklozenge\}$ and $r_{s}\in\{\bullet_K,\blacksquare_K,\blacklozenge_K,\mathrm{O_K}\} $ where  $\varpi=n+1$, with  $\gamma=1/n+1$ is the polytropic index. 
 Radii $(r_*,r_s)$ and  the correspondent  angular momentum  $\ell$ and $K$ parameters  are shown with $\{\bullet,\blacksquare,\blacklozenge,\bullet_K,\blacksquare_K,\blacklozenge_K,\mathrm{O_K}\}$.  $\Omega_K$ is  the Keplerian angular velocity, $r_{\times}$ is the accreting tori cusp (inner edge of accreting torus), $r_s$ is related to thickness $\mathbf{{\widehat{\mathbf{h}}}}_s$ of the accreting  matter flow  as in Figure\il(\ref{Fig:Hologra}).  $r_{mbo}$ is the marginally bounded orbit. $\Omega_K^{+}$  has been considered for the counterrotating fluids. The maximum location of inner edge is  $r_{\times}\lessapprox r_{mso}$.  }\label{Fig:Manyother}
\end{figure*}
We note that    each of  these quantities  can be express in  the  general  form
 $\mathcal{O}(r_\times,r_s,n)$ of Table\il(\ref{Table:Q-POs}), where $q(n,\kappa)$ and $d(n)$ are different functions of the polytropic index  $\gamma=1+1/n$ and polytropic constant $\kappa$ or as   $\mathcal{P}$ form, for example  the mass flow rate through the cusp (mass loss, accretion rates)  $\dot{M}_{\times}$,  and the cusp luminosity $\mathcal{L}_{\times}$ (and the accretion efficiency $\eta$),
 measuring the
rate of the thermal-energy    carried at the  cusp.  The relativistic frequency $\Omega$  reduces  to Keplerian  values $\Omega_K$ at the edges of the accretion torus, where  the pressure forces   are vanishing ({The limiting  case of  polytropic index  $\gamma=0$  would correspond to the case of zero pressure represented  by the dust  of test  particles)}). Note that this also justifies  the approximations for collision energy evaluation of   Fig.\il(\ref{Fig:AliceB}).
Parameters  $(\kappa,n)$ within the constraints  $q(n,\kappa)=\bar{q}=$constant,  fix  a  polytropic-family. The   $\mathcal{O}(r_\times,r_s,n)$ depends on the location of the inner edge of the accreting torus, constant for  $K$-modes, but variable   with the \textbf{RAD} $\ell$-modes,  and on  theradius   $r_s$ which is the related to the thickness, $\overleftarrow{\mathbf{h}}_{s}$, of the matter flow\footnote{Note that in some sense this quantity is still dependent on a generalized definition of  $K$-parameter.
It is worth specifying here this aspect, as it  is  also briefly mentioned  in  Sec.\il(\ref{Eq:CM-ene}):
if  $\ell\in \textbf{L1}$ then $K \in[K_{\min},K_{\max}]$, where $K_{\min}>K_{mso}$ and  $K_{\max}<1$.
$K_{\max}$ corresponds to the accreting point $r_{\times}$, while $r_s$ is directly associated to the accreting flux thickness.
Clearly  $r_s$ can be write in terms of a $K$ parameter value $K_s\in[K_{\max},1[$--Figs\il(\ref{Fig:Manyother}).} (see also Fig.\il(\ref{Fig:Hologra})).
 Then, as the cusp approaches the limiting radius  $r_{mbo}$,  the potential  $W_\times\approx0$,  which is also the limiting asymptotic value for very large  $r$ as well as for the emergence of the proto-jets for $\ell\in \textbf{L2}$.

 $\mathcal{P}$- and $\mathcal{O}$-quantities of Table\il(\ref{Table:Q-POs})  are evaluated for corotating and counterrotating tori in Figs\il(\ref{Fig:Manyother}),  for  dimensionless  \textbf{SMBH}  spin $a/M\in [0,1]$.
    To simplify our discussion, we have considered for the analysis of the two   classes of the functions the following  reduced quantities:
\bea\label{Eq:provi-Ang-Lesb}
&&
\text{\textbf{[$\mathcal{O}$-quantities]}:}  \mathcal{R}_{*}^{\pm}\equiv(W^{\pm}(r_{s})-W^{\pm}_{*})^\varpi\\
&&\nonumber \text{ and } (W^{\pm}(r_{mbo})-W^{\pm}(r_{mso}))
\\&&\label{Eq:soft-chogra}
\text{\textbf{[$\mathcal{P}$-quantities]}:  }     \mathcal{N}_{*}^{\pm}\equiv\frac{{r_*} (W^{\pm}(r_{s})-W^{\pm}_{*})^\varpi}{\Omega_K(r^{\pm}_*)}\\
&&\nonumber \mbox{where}\quad r_s=r_s(a)\in]r_{\times}(a),1[,
\eea
 serving as an immediate and simple way to understand the trends of these functions of the \textbf{BH} spin, the fluid  angular momentum and the cusp location. In fact, this analysis is especially focused on the characterization of the $\mathcal{P}$- and $\mathcal{O}$-quantities for an accreting couple  $\cc^-_\times<\cc^+_\times$,
as functions of \textbf{SMBH}  dimensionless spin $a/M$,  at  different  locations of the inner edges, $r_*=r_{\times}^{\pm}$ and the radius  $r_{s}$, which is   related to the  thickness $\mathbf{{\overleftarrow{\mathbf{h}}}}_s$ of the accreting  matter flow  as in Figure\il(\ref{Fig:Hologra}).  Then $\varpi=n+1$, with  $\gamma=1/n+1$  being the polytropic index. $\Omega_K$ is  the Keplerian angular velocity, where $\Omega_K^{+}$  has been considered for the counterrotating fluids. The maximum location of the accreting torus   the inner edge is  $r_\times\lessapprox r_{mso}$.
Note that  in Eq.\il(\ref{Eq:provi-Ang-Lesb})the maximum difference $(W^{\pm}(r_{mbo})-W^{\pm}(r_{mso}))$ is provided in relation to the analysis of the $\mathcal{O}$ and $\mathcal{R}$ quantities, this function of the $a/M$ is also plotted in Figs\il(\ref{Fig:Manyother}). This difference clearly    features the limiting case where $r_s\lessapprox 1$ and $r_{\times}\gtrapprox r_{mso}$,  providing the maximum (limiting) distance  $r_{\times}-r_s$, and the limiting maximum flow thickness $\overleftarrow{\mathbf{h}}_s$ case. On the other hand,  Figs\il(\ref{Fig:Hologra}) show  also  lines  $(\exp[W^{\pm}(r_{mbo})]-\exp[W^{\pm}(r_{\times}))]=$constant  for different values of \textbf{SMBH} spin $a\in[0,M]$ and  the inner edge of accreting torus  $r_{\times}\in]r_{mbo},r_{mso}[$.  Note that
$\exp[W]=K$.
The couple of parameters  $\{r_s(a),r_{\times}(a)\}$  in Eq.\il(\ref{Eq:soft-chogra}) has been fixed in the examples of Figs\il(\ref{Fig:Manyother}), to simplify the  comparison of the $\mathcal{O}^{\pm}$ and $\mathcal{P}^{\pm}$ quantities in the  corotating and counterrotating tori,  and to characterize the dependence of these quantities on the \textbf{SMBH} spin-to mass ratio $a/M$. In fact, within this choice, all  the quantities $\mathcal{O}$ and $\mathcal{P}$ or, equivalently,  $\mathcal{N}$ and $\mathcal{R}$ of Eq.\il(\ref{Eq:soft-chogra}), become functions of the  spin $a/M$ only; on the other hand, at fixed $a/M$,  we compare these quantities for different radii  $r_s$ and $r_{\times}$, related by   fixed relations through the  different laws $r_s(a)$ and $r_{\times}(a)$, as expressed in Figs\il(\ref{Fig:Manyother}).

It is clear that the $\mathcal{O}$ and $\mathcal{P}$ quantities have  similar dependence on the triplet of parameters $\mathbf{p}^{(3)}_{\digamma}\equiv(a/M; r_*,r_s)$.  We consider then the doublet $\mathbf{p}^{(2)^{\pm}}_{\digamma}\equiv(r_*^{\pm},r_s^{\pm})$.
There is always $\left.\partial_{a^*} G^{\pm}\right|_{\mathbf{p}_{\digamma}^{(2)^{\pm}}}\lessgtr0$ respectively for counterrotating and corotating tori, and $\left.G^+(a)\right|_{p_{\digamma}^{(2)^{+}}}\leq\left. G^-(a)\right|_{\mathbf{p}_{\digamma}^{(2)^{-}}}$, where
$ G^{\pm}\in\{\mathcal{Q}^{\pm},\mathcal{P}^{\pm}\}$, and  $G^+=G^-$    for $a=0$ only; here $a^*\equiv a/M$.
On the other hand, the spread  between corotating and counterrotating curves increases with the \textbf{BH} spin, i.e., $\partial_{a^*} (\left. G^{-}\right|_{\mathbf{p}_{\digamma}^{(2)^{-}}}-\left.G^+\right|_{\mathbf{p}_{\digamma}^{(2)^{+}}})>0$. Note that we  adopted the parameters  $\mathbf{p}^{(2)^{+}}_{\digamma}\equiv(r_*^{+},r_s^{+})$ and $\mathbf{p}^{(2)^{-}}_{\digamma}\equiv(r_*^{-},r_s^{-})$ at fixed  $a/M$, this means that we are comparing $\ell$counterrotating  tori  related through these parameters as indicated in  Figs\il(\ref{Fig:Manyother}).
Interestingly then, there is in general  $\partial_{a^*} (\left.G^-\right|_{\bar{\mathbf{p}}^{(2)^{-}}_{\digamma}}-
\left.G^-\right|_{{\breve{\mathbf{p}}}^{(2)^{-}}_{\digamma}})<0$,  where $\left.G^-\right|_{\bar{\mathbf{p}}^{(2)^{-}}_{\digamma}}\geq\left.G^-\right|_{{\breve{\mathbf{p}}}^{(2)^{-}}_{\digamma}}$,
 while $\left.G^-\right|_{\bar{\mathbf{p}}^{(2)^{-}}_{\digamma}}=\left.G^-\right|_{{\breve{\mathbf{p}}}^{(2)^{-}}_{\digamma}}$,
 asymptotically for $a=M$, and $(\bar{\mathbf{\mathbf{p}}}^{(2)^{\pm}}_{\digamma}, \breve{\mathbf{p}}^{(2)^{\pm}}_{\digamma})$ are two general couples of parameter, therefore associated to two different curves in Figs\il(\ref{Fig:Manyother}).
As $(\bar{\mathbf{\mathbf{p}}}^{(2)^{-}}_{\digamma}, \breve{\mathbf{p}}^{(2)^{-}}_{\digamma})$ are functions of $a/M$ themselves as shown in Figs\il(\ref{Fig:Manyother}),   this  result  may be interpreted  as  the combined effect of the  increasing $\mathbf{SMBH}$ spin, and   the approaching of $(r_s, r_{\times})$ to the central attractor, which means an  increase of the dragging effects.
Viceversa, in the counterrotating tori, there is   $\partial_{a^*} (\left.G^+\right|_{\bar{\mathbf{p}}^{(2)^{+}}_{\digamma}}-\left.G^+\right|_{{\breve{\mathbf{p}}}^{(2)^{+}}_{\digamma}})\approx0$, where $\left.G^+\right|_{\bar{\mathbf{p}}^{(2)^{+}}_{\digamma}}\geq0\left.G^+\right|_{{\breve{\mathbf{p}}}^{(2)^{+}}_{\digamma}})$, which may be read saying that the $\mathcal{O}$ and $\mathcal{P}$ quantities for the  counterrotating case   are less dependent on the  \textbf{BH} spin-up process then the corotating ones. Note  that in the corotating tor case there is a sharp increase with the \textbf{SMBH} spin $a/M$ as $a\gtrsim 0.9M$, and for these value of the spin of the central attractor, some  tori   can be formed in the ergoregion, or be partially contained  in the ergoregion--Sec.\il(\ref{Sec:coroterf}).
Then, as the accretion rate ($\mathcal{P}$-quantity) is larger for corotating then for counterrotating tori (in the adopted parametrization), and increases with the \textbf{SMBH} dimensionless spin for corotating fluids, while decreases for counterrotating ones, this implies that the corotating tori may represent  very  good  candidates to explain the \textbf{SMBH} mass according to the process considered in\cite{apite1,apite2,apite3,Li:2012ts,Oka2017,Kawa,Allen:2006mh},   rather then the counterrotating ones\footnote{{  However comparing the  counterrotating and corotating accretion we need to analyse more carefully the accretion flow.
 The counterrotating case is in fact a special case of accreting flux
 where a reversal of flow rotation  close  to the \textbf{BH}  (near the ergosurface) is expected.
 In this sense the spin of a black hole has a special role
 on the determination of   the accretion flow  as  well as of the accretion rate see  \cite{Middleton:2014cha,DeVilliers,Nelson:1997nj,Nixon(2011),Roedig:2013lqa}.
In the evaluation of the quantities  in  Table\il(\ref{Table:Q-POs})  we  based our  analysis on the assessment of the flow thickness which in our model  depends   on the cusp location, the  specific angular momentum and the parameter $K$, and  the quantities  $\mathcal{P}$ depend  on the fluid relativistic angular frequency at the cusp.
We consider accretion flow from the outer counterrotating torus of  a  $\pp^-<\cc_{\times}^+$ couple, the counterrotating   flow impacts  on the inner torus  which   is generally close to the outer torus cusp (the maximum limiting  tori separation is $\bar{\lambda}\approx 8M$ in the geometry of near extreme Kerr \textbf{BH} with $a=M$).}}.
 \subsubsection{Structure of polytropic tori}\label{SeC:poly}
In this section we develop some general considerations on the equation of states and the  polytropic  \textbf{RAD} tori, which has been used   in Sec.\il(\ref{Sec:more-l}) for the study of $\mathcal{P}$ and $\mathcal{O}$ quantities. 
Geometrically thick tori consider the hypothesis of a barotropic law $p=p(\varrho)$, this is also used  in more general  situation where the  magnetic field contribution is included in the  torus force balance \cite{abrafra,PuMon13}.
The entropy distribution depends on the initial conditions of the system and on
the details of the dissipative processes,
 and it is constant along the flow.
Focusing on polytropic tori regulated by the equation of state: $p=\kappa \varrho^{\gamma}$, where  $\kappa>0$ is a constant   and $\gamma$ is the polytropic index, in \cite{PuMonBe12}
it has been shown that for the Schwarzschild geometry $(a=0)$ there is a specific classification of eligible geometric polytropics  and a specific class of polytropics is characterized by a discrete  range of values for the index $\gamma$--see also \cite{Raine}.
 Nevertheless, most of the considerations traced out for the static attractors, $a=0$,  holds also for a more generic  effective potential function, in particular   for a Kerr \textbf{SMBH} where $a\neq0$-- ultimately
 a similar classification of the polytropics for $a=0$ was proved to be  valid also in the rotating case of  the  Kerr geometry--\cite{pugtot}.  Here we recall some of those considerations in the \textbf{RAD} framework,  in particular in the case of $\ell$counterrotating couples.
Using  Eq.\il(\ref{Eq:scond-d}), we can write  the density $\varrho$ as function  $\gamma$. Therefore, we can   propose a general classification for the tori  ($\cc, \cc_{\times}$), as for proto-jets $\oo_{\times}$, assuming a particular representation of the density function.
However, concentrating our attention   on the \textbf{RAD} components  $\cc$ and $\cc_{\times}$ for which  $K<1$, there is :
\bea\nonumber
&&
\varrho_{\gamma}\equiv
\kappa^{1/(\gamma-1)}\bar{\varrho}_{\gamma}\quad\mbox{and}\quad
\bar{\varrho}_{\gamma}\equiv\left[\frac{1}{\kappa}\left(V_{eff}^{-\frac{\gamma -1}{\gamma
}}-1\right)\right]^{\frac{1}{(\gamma-1)}}\;\mbox{for}\;\gamma\neq1 \\
&&\label{peter}  \mbox{thus}\quad
\varrho_{\gamma}\equiv \mathds{C}^{1/(-1+\gamma )},\quad  \mathds{C}\equiv (V_{eff}^{-2})^{\frac{\gamma-1}{2 \gamma }}-1.
\eea
Then $\mathds{C}$ is actually a function of  $K\in ]K_{\min},K_{\max}]$, where $K_{\max}<1$, regulating whether the torus is quiescent or in accretion. Then we concentrate on the case  $K<1$ with the condition  $\varrho>0$; this case is verified, according to Eq.\il(\ref{peter}), when $\gamma>1$
\footnote{The boundary conditions have been fixed   so that the tori inner and outer edges are  $[r_{in},
, r_{out}]\subset]r_+, \infty]$; this may be read as the extensive range  of existence for $V_{eff}(\ell)$. The case  $\gamma=1$ is considered apart where
$
\varrho_k\equiv V_{eff}^{-(1+k)/{ k}}/({1+k})$ and
at the boundaries
\(
\varrho^{out}_{k}=\varrho^{in}_{k}\left({V^{out}_{eff}/V^{in}_{eff}}\right)^{-({1+k})/{k}},
\)--see the \cite{pugtot}.}
see also the ranges\footnote{
For  any $\kappa>0$ and $\gamma>1$, there is the density  $\bar{\varrho}_{\gamma}=0$   in the limit of  large $r$ (i.e. $r\gg r_{cent}$) where  $V_{eff}\approx1$. We should note that equally this is also the case   of the boundary configuration for the  open $\oo_\times$ Boyer surfaces,   where the minimum pressure (and density) point is located in    $r_{\jj}\geq r_{mbo}^{\pm}$ correspondent to $K_{J}\geq 1$, an overflow of matter in funnels of material occurs \cite{open}.} in Eq.\il(\ref{Eq:def-partialeK}).  In the case of   very small tori, i.e. with small elongation $\lambda$ on the equatorial plane,  the  hydrostatic pressure component  is less relevant    for the force balance, and the torus approaches the  geodesic limit of a one-dimensional (stable) ring of dust {(test particle approximation as the  pressure forces vanish)}.
On the other hand, keeping the minimal condition  $\bar{\varrho}_{\gamma}>0$
for the   $(\cc,\cc_\times)$   tori, there is  $\partial_k\bar{\varrho}_{\gamma}<0$, for any polytropic index.
In the limit of large $ \kappa$, in each point of the torus  in equilibrium or in accretion,    the pressure  and the density decrease  to zero (with the torus volume).
 The pressure $p$, associated to the solution in Eq.\il(\ref{peter}),  depends on $k^{\frac{1}{1-\gamma }}$. It decreases with $\kappa$ more slowly  then $\varrho$.

We conclude this section with some additional considerations on the role of the density function $\varrho_\gamma$ and the  $K$-parameter in the  \textbf{RAD} framework. We consider \textbf{RAD} potential
$\left.V_{eff}^{\mathbf{C}^2}\right|_{K}$  in Eq.\il(\ref{Eq:def-partialeK}), for a couple of tori,  where, for a fixed distribution of fluid specific angular momentum, $\{\ell_i\}_{i=1}^n$, we  can set the fixed parameters $(K_i,K_o)$, defining the  boundary conditions in  $\left.V_{eff}^{\mathbf{C}^2}\right|_{K}$ and used to describe the \textbf{RAD} $K_{\ell}$-modes.
The function  $\varrho_\gamma$, for each \textbf{RAD} component, is actually function of $K$. To fix the  \textbf{RAD}, it is important to consider the ratio $K_{i}/K_{o}$ for two adjacent tori--\cite{ringed,dsystem}.
 Using this approach, we   finally can obtain a series of constraints related directly to the density functions $\varrho_\gamma$, rather then on the  $K$ parameters.
 In fact, as mentioned in Sec.\il(\ref{Sec:Kerr-2-Disk}), the surfaces of constant density coincide with the constant pressure and potential surfaces. Taking  explicitly into consideration Eq.\il(\ref{peter}),
one has for a  torus that  $K= \left[(\varrho_{\gamma})^{\gamma -1}+1\right]^{\gamma /(1-\gamma)}$.

Then the  relation  $K_i+\epsilon K_o=\chi$=costant implies for the tori $\cc_i<\cc_o$, that
\be
\varrho^i_\gamma=\left[\left(\chi -\epsilon  \left[(\varrho^o_\gamma)^{\gamma -1}+1\right]^{-\frac{\gamma }{\gamma -1}}\right)^{\frac{1}{\gamma }-1}-1\right]^{\frac{1}{\gamma -1}},
\ee
with $\epsilon=$constant.

Theory of accreting tori adopted here implies on the other hand that each $\varrho^i_\gamma$ and $\varrho^o_\gamma$ are constant. 
Finally, to simplify our arguments, we have assumed that the polytropic indices (but not necessarily the polytropic constant $\kappa$)  are equal for each torus. This hypothesis may be especially relevant for   the  $\ell$corotating couple.
Note that  we can then directly impose several  constraints  for the density function. Some simple examples,  including special (composite) density profiles are, for example, the case
 $\varrho_{[+]}^{-}= \varrho_\gamma^i-\varrho_\gamma^o=\varrho_\Phi=$constant,  where \[K_i= \left(\left[\varrho_\Phi-\epsilon \left({K_o}^{-\frac{\gamma -1}{\gamma }}-1\right)^{\frac{1}{\gamma -1}}\right]^{\gamma -1}+1\right)^{-\frac{\gamma }{\gamma -1}}.\]

These solutions create special tori surfaces from the condition on the  constant pressure. Moreover,  within the limits  considered  before, these constraints  can found application also in the collision analysis, to infer the final states (es. final merger tori) from these constraints\footnote{Other notable cases might be  founded by the constraints
 $\varrho_{[\times]}= \varrho_\gamma^i \varrho_\gamma^o=$constant,
 $\varrho_{\{\times\}}= \varrho_\gamma(K^i K^o)=$constant
 or  $\varrho_{\{+\}}^{\pm}= \varrho_\gamma(K^i\pm K^o)=$constant. It is possible to show that not all these profiles are related to quiescent of accreting  toroids.}.
 %

\section{General discussion on interacting tori and  energetic of associated processes}\label{Sec:grosss}
This first evaluation of the  energetic related to the tori collision processes, performed by considering colliding  particles  associated with the   \textbf{RAD} tori,  certainly suggests a promising  and interesting aspect of the agglomerate  energetics. Here we add some  clarifications about these processes more specific  in the   \textbf{RAD} framework collision they are associated to.
Situations showed in Fig.\il(\ref{Fig:AliceB}) deal basically  with two very distinct phenomena depicting the tori collisions, the first  $\mathbf{-j-}$ concerns the only couples $\cc_\times^-<\cc_{\times}^+$ or $\cc^-<\cc_{\times}^+$ (as there must be an inner torus screening between the central \textbf{BH} and the  accreting outer torus), and arises due to the    matter overflow  from the  cusp of the outer accreting torus, which  impacted onto the inner corotating torus.  The second process  $\mathbf{-jj-}$   instead  does not necessary involve  accretion from the outer torus, and the main process features an impact from the quiescent or accreting  outer torus   on the inner  one or, more precisely, a contact between two  adjacent tori of the  couple, as such they may involve $\ell$corotating or $\ell$counterrotating tori, thought the conditions for such collisions to occur turn to be constrained especially by the \textbf{BHs} dimensionless spin defining  some special \textbf{BHs} classes.
We will add some further comments on this collision mechanics below.

Focusing on the $\mathbf{-j-}$ case, the initial {state} of the inner torus is in fact not   relevant for analysis  of this very initial phase of collision and the assessment of the  conditions for a collision  after accretion from the outer torus. In the case where the outer torus is in accretion, collision is driven by the dynamics of the external torus, the inner torus acts as a screening  mass between the accreting torus  and the central black hole, similar situations are for example considered in \cite{Marchesi,Gilli:2006zi,Marchesi:2017did,Masini:2016yhl,Storchi-Bergmann}.  Clearly the initial state of  the inner torus is relevant in the second phase of the collision mechanisms, when we are forced to explore the situation under a wider  perspective  to consider the final state of the \textbf{RAD}.
In the $\mathbf{-j-}$ mechanism, the spacing region $\bar{\lambda}_{i,o}$,  defined as in Fig.\il(\ref{Fig:Quanumd}),  plays an  essential role together with the initial constrains on the outer torus. Both aspects are thoroughly discussed in Sec.\il(\ref{Sec:theory-weell}).

Nevertheless, the evaluation of the energetics of collision processes in the \textbf{RAD}, either for  $\mathbf{-j-}$ or $\mathbf{-jj-}$  mechanism, cannot avoid a more general, global approach,  by considering  collision as a ``global'' phenomenon involving matter flow from the outer torus and a large part of the inner torus, rather then focusing  on each particle of the flow.
On the other hand, the amount of matter  overflow from each accreting torus, may   also be affected  by  the accreting torus (global) oscillations;   we do not consider here this  aspect,  but we refer for example  to --\cite{abrafra}.

We now concentrate on the $\mathbf{-j-}$ case, particularly on the  $\ell$counterrotating  $\cc_{\times}^-<\cc_{\times}^+$ as seen in Fig.\il(\ref{Fig:AliceB}), which pictures  the outer torus as  source of accreting matter falling onto the inner one.
More specifically, regarding the $(\mathbf{-j-})$ mechanism, we should consider the outer torus mass accretion rate $\dot{M}$ and matter flux, as well as the flow thickness,  defined in Table\il(\ref{Table:Q-POs})  and carefully studied in Sec.\il(\ref{Sec:more-l}). These quantities  are functions of the \textbf{BHs} spin and range of variation of the the  fluid specific angular momentum. Therefore, in this investigation  we  find  the  maximum  and minimum  values  of these quantities, for given initial  conditions on the outer  (counterrotating) torus. These aspects are all considered in Sec.\il(\ref{Sec:more-l}), where we  also refer to a direct and explicit comparison between the corotating and counterrotating tori. Some additional notes about the free-particle assumption for matter leaving the cusp are also included, this hypothesis is translated into the use of Keplerian   velocity for the particle.
We are not concerned here with the conditions, or the \textbf{RAD} evolutive path, leading to a possible couple $\cc^-_\times<\cc_{\times}^+$ which are  investigated   in \cite{dsystem} and considered also in Sec.\il(\ref{Sec:Kerr-2-Disk}),  nevertheless  the initial conditions on the density mass  and angular momentum are necessary   for the evaluations of  all the quantities of Table\il(\ref{Table:Q-POs}).
 For the evaluation of the mass  accretion rate, cusp luminosity and other relevant quantities of accretion tori,
we assumed also  the inner and outer tori are filled with polytropic fluids, and  in Sec.\il(\ref{SeC:poly}) we revisit this hypothesis, the polytropic equation of state  allowing to rewrite in an  explicit way the  density   $\rho$  in terms of  the $K$-parameter.
The tori morphology, the   thickness and elongations of each  component of the couple is constrained in Sec.\il(\ref{Sec:theory-weell}),  the morphology of the aggregate components    are important to asses  the possibility of tori collision,  the state of tori (if quiescent or in accretion) and in general the \textbf{RAD} instability. This  information provides also some hints on the density  of the fluid, as the $K$  parameter is directly linked to the torus density.
We shall see that some rather small tori may be described as well by the  model adopted in the investigation for each \textbf{RAD} component, especially for the $\ell$counterrotating  couples $\cc_{\times}^-<\cc_{\times}^+$, according to specific classes of \textbf{BHs} attractors --Sec.\il(\ref{Sec:more-l}). Actually  small \textbf{RAD} tori of this kind of  couple  are expected to be especially  characteristic of slow rotating and low mass attractor.
Construction of accretion models  have  narrowed many of the relevant features of geometrically thick  disks and particularly of  the so called  Polish doughnuts which are very close to the models adopted for the \textbf{RAD} components.  There is a certain variability in  featuring such tori, depending essentially on the  central attractors  (as  it is the case also for the \textbf{RAD} model), and on the force balance condition regulating the \textbf{RAD} dynamics. Generally, these  tori are opaque,  having  extreme optical depths.    Expected to be  cold, such tori have    temperature   less or much  less  than the virial temperature. They are  radiation pressure supported and cooled by advection with low viscosity. Very importantly for the \textbf{RAD} energetic  is that geometrically thick tori are characterized by high accretion rates, with super-Eddington or highly super-Eddington  luminosity (with  very small accretion efficiency). We refer to Sec.\il(\ref{Sec:more-l}) for evaluation of related quantities in dependence of the parameters $(\ell, K)$ and the \textbf{BH} dimensionless spin.  We know the entropy is constant along the flow, the enthalpy flux has been investigated in Sec.\il(\ref{Sec:more-l}).
 We remind nevertheless  that super-Eddington accretion, thought often associated to  strong matter outflows,
  does not \emph{necessarily} imply this, and this fact   turns to be very  significant
in the modeling the processes of the   \textbf{BHs} seeds growing  towards the \textbf{SMBH} sizes to justify the masses of these extreme  compact objects
at high redshifts. A further explanation on the other hand  involves  different
accretions stages of the \textbf{SMBHs} accretion tori life, and this particular mechanism is in fact seen here as one of most promising situations  in  which \textbf{RAD} structure can be  directly  involved through several accretion stages  induced by \textbf{RAD} instability.
Given the importance of this aspect for the energetic of  the \textbf{RAD} processes, and for  the  mechanisms of accretion of medium-small masses \textbf{BHs} towards  \textbf{SMBH} at high redshift,  we have dedicated Section\il(\ref{Sec:more-l}) to the study of these quantities, focusing nevertheless on stressing the distinction between corotating and counterrotating  tori.
The initial conditions, prior to the collision, i.e., the range of values for the specific angular momentum $\ell$ of the two tori and the parameter  $K$,    have been dealt with in dedicated works.   The assessment of the initial  $(i)$ parameters $(\ell_i(i),\ell_o(i))$ are then  important for the evaluation of the energetic of the process and the range of density values for the fluid and the tori elongation on the equatorial plane. While not  specificized otherwise, we shall indicate $(i)$ for initial and $(f)$ for a final state of a process.
The $\ell$counterrotating couples can have different evolutive paths according to initial data con angular momentum, with respect to  the \textbf{BH} spin, and to  the fact that the angular momentum may change  magnitude, remaining in \textbf{L1}, or  it can possibly  increase up to values in \textbf{L2} or \textbf{L3}. Nevertheless, these possibilities turned to be significantly  constrained by the  \textbf{SMBH} dimensionless spin\footnote{A state of a \textbf{RAD} component is fixed when the parameters $(K,\ell)$ and the tori  morphology are established. The possibility that  a \textbf{RAD} couple may also return to a previous state of the evolution path was named \emph{state loop} in the couple  evolutionary graph in \cite{dsystem}. Considering the results traced in  \cite{dsystem}, we  briefly summarize the general constraints  on the fluid specific angular momentum and the \textbf{BH} dimensionless spin for any state of the  couple $\pp^-<\pp^+$, made by  an inner corotating torus and outer counterrotating torus. These constraints were then discussed together with the constraints  on the $K$-parameters--see also \cite{ringed}.
First, let us introduce the   \textbf{BHs} spins
$a_{\iota}=0.3137M$ and  $a_{0}=0.3726M$, the last spin has also been introduced in Sec.\il(\ref{Eq:CM-ene}) see also Fig.\il(\ref{Fig:AliceB}). These spins, rather close in values, refer to  slowly spinning   \textbf{SMBHs}  with respect to the limiting spins  considered  in Eq.\il(\ref{Eq:spins}). In \cite{dsystem}, we introduced the concept of correlation which here  is intended for occurrence of the possibility of    tori  interaction,  or a tori collision.
The situation is as follows:
Couple  $\cc^-<\cc_{\times}^+$ is clearly in correlation
 while for
the  situation of  $\cc_{\times}^-<\cc^+$, with an inner torus in accretion and  an outer  quiescent torus,
  only   sufficiently slow \textbf{BHs}, $a< a_{0}$, can be characterized by the couple correlation.
   If the inner torus is in a proto-jet unstable phase,  collision never occurs, unless
  the outer torus is not in accretion, or a proto-jet for attractors with $a<a_{\iota}$.
	On the other hand, the situation when an  outer torus is a proto-jet with an inner (corotating) torus in accretion  be only be observed in the spacetimes with   $a>a_{0}$, and there is no tori  collision.
	 In the  state with both  proto-jets,  collision is possible only for  $a<a_{\iota}$;
	if the outer configuration is a proto-jet and the inner torus  is in equilibrium,  then there is a collision. Viceversa, if the outer torus is in equilibrium, and the inner structure is a proto-jet no correlation occurs.}  \cite{dsystem}.

If the outer torus is counterrotating,  we may assume
the falling material delivers negative angular momentum, increasing the mass of the inner accreting torus (possibly affecting $K_i$ parameter) and, simultaneously, lowering its angular momentum; this is based on the  simplest assumption that $\delta\ell_i\equiv \ell_i(f)-\ell_i(i)\leq0$, where in the final $(f)$ state, $\ell_i(f)$ is   taken almost constant.
The lowering of the angular momentum magnitude  of the inner torus in principle would imply a increase  of the cusp location (and of $K_{\max}$) and a decrease of the  torus center as well, i.e., $r_{\times}^-(f)>r_{\times}^-(i)$ and $r_{cent}^-(f)<r_{cent}^-(i)$--see Figs\il(\ref{Fig:Truema}). But this aspect comes into competition,
as we will discuss below, with the simultaneous mass increase, which may correspond to a change in $K$ (this is actually a function of matter density, but also of the torus specific angular momentum).   Moreover, under suitable conditions  it could also  turn to  block the accretion of the inner torus.
If  the cusp moves inwards, the increase of the torus mass  $m^-$,  due to the new injection of matter from the outer torus occurs, and  the simultaneous decrease of specific  angular momentum  may also lead to an decrease of the accretion rate. However, Sec.\il(\ref{Sec:more-l}), and especially  the  analysis  in Figs\il(\ref{Fig:Manyother}), presents  particularly the behavior of accretion rate with respect to the cusp location and the specific angular momentum.

This dynamics  on the other hand concerns also the second tori  collision process,  $-\mathbf{jj}-$,   occurring  when  $\bar{\lambda}_{i,o}=0$, and the collision is due to two possible evolutive paths of the colliding couple or a combination of the two: in the first,  $-\mathbf{jj_o}-$, collision is caused by the  outer torus which, in  accretion or quiescence, stretches onto the equatorial plane,  $r_{in}^o(f)<r_{in}^o(i)$; this change can happen for a change in the angular momentum magnitude (thus  $r_{cent}^o(f)<r_{cent}^o(i)$), or a net increase of the  $K$-parameter at $\ell$ approximative constant   (thus  $r_{cent}^o(f)\approx r_{cent}^o(i)$ and $r_{in}^o(f)<r_{in}^o(i)$, while $r_{out}^o(f)>r_{out}^o(i)$ and the  torus elongation  $\lambda^o(f)>\lambda^o(i)$) or a combination of the two--\cite{dsystem}.
  The second $-\mathbf{jj_i}-$ is due to   the inner torus which  is growing $\ell^-(f)\approx\ell^-(i)$ or stretching towards the accretion  at approximative constant  $K_i$, or viceversa an increases of $K^i\approx K_{\max}^i$ at $\ell^i$ almost constant (with also $r_{out}^i(f)>r_{in}^i(i)$) implying arising of an accretion phase for the inner torus. Processes $-\mathbf{jj_i}-$ and $-\mathbf{jj_o}-$ may be   therefore described also in the framework of $\ell$-modes or $K_{\ell}$-modes respectively. They result  in an ultimate stretching in the equatorial plane where $r_{out}^i$ moves outwards, thought not necessarily $r_{cent}^i$ grows, or viceversa, with an increases of $\ell^i$, and a shift outward of $r_{cent}^i$, leading to $\bar{\lambda_{i,o}}=0$.
A general discussion of the evolutive processes of the couple can be found in  \cite{dsystem}
while   we shall draw here some qualitative considerations, more specifically  on the conditions \emph{after} collision. Thought we do not  approach directly the modeling of the collision phenomena between the fluid of the two tori  requiring  a  different approach  being a task far  from the one considered here, we note   that, for this purpose,  it is in fact essential to fix the initial conditions on the colliding couples, as we have done in the evaluation of the test particle collision.  A colliding couple $\cc_{\times}^-<\cc_{\times}^+$  can lead to a state after the collision made by one torus or two final tori i.e. \textbf{-I-} a \textbf{RAD} made up by  one torus, $\pp^\pm$, with final state described by parameters $(\ell(f), K(f))$  or  $\mathbf{-II-}$  a couple  $\pp^-<\pp^+$. Note that  the tori state  in \textbf{-I-}  and \textbf{-II-}  has not  been specified as it  could be in accretion or quiescent.  However, assuming the background  spacetime remains unaltered during the process, i.e, the accretion matter  does not change the spin or mass of the central attractor neither the torus self-gravity leads to a feedback reaction  these situations are in reality very much constrained by $\ell$ and  a further analysis bounds the  relative values of  $(K^i, K^o)$ in the final states.
Then,  as considered in \cite{dsystem} the second evolutionary path, $\mathbf{-II-}$, could be also a transient stage leading finally  to  one torus, $\mathbf{-I-}$,  or generally being part of a series of further  evolutionary phases, where the couple is subject to different state transitions.
In the case  $\mathbf{-I-}$, a one torus solution can be found when, either because of in accretion (decreasing its angular momentum magnitude at almost constant $K$), or for growing gaining a greater value of  $K$, the outer torus may eventually merge with the inner resulting in  a  single torus, destroying the \textbf{RAD}  and  carrying the system to the commonly considered accretion environment when a  central \textbf{SMBH}  is surrounded by one  orbiting  accretion  torus.
It is clear that these considerations enter also in the discussion of the processes of the \textbf{BHs} accretion tori formation.

$\mathbf{-II-}$ In  the case where the final state  is made by a couple of  tori, we  concentrate on the situation for the inner torus with parameters  $(\ell(f), K(f))$,  discussing briefly  the situation for the outer torus also.
 Two main processes are in competition :\textbf{(a)} on one hand there, is an  increase of matter provided by the outer torus such that  $K(f)$  is the new parameter value for the inner torus and $m(f)(K(f))>m(i)(K(i))$ where $m$ is the torus total mass.  One evolutive path can feature an outer torus which is  losing mass and momentum approaching  the accretion  (the torus being smaller), while  matter impacting on the inner torus may be not sufficient  for accretion to occur, i.e., to lead to a situation were  $K_i(f)<K_i^{max}(\ell_i(f))$. This inequality has to be evaluated together with    $\ell(f)$  and $\ell(i)$.
 However, it should be also considered that this situation does not necessary occur, because in fact in the evaluation of the role of the  $K(f)$  parameter actually  what is most relevant is the value of  $K(f)$ related to the limits $[K_{mso},K_{\sup}[$ (or more precisely $[K_{\min},K_{\sup}[$), where  $K_{\sup}=K_{\sup}(\ell)$ is a function of the specific angular momentum at fixed \textbf{BH} spin.
 Therefore  the problem of change in  the inner torus  rotation law  has to be addressed  prior   the evaluation of mass increase.
 Nevertheless,   the  initial and final state of each torus can be studied in  fact as an  $\ell$corotating  couple (the states of the single torus before and after the collision do not change the direction of rotation with respect to the black hole, changing only in  magnitude of the specific angular momentum). An increased  $K(f)>K(i)$ for  $\ell(f)>\ell(i)$ does not lead necessarily to an increase of the torus elongation and the establishing of the accretion (precise limits and methods are in \cite{ringed}), while there is  $r_{center}(f)>r_{center}(i)$  for the tori centers and $r_{\times}(f)<r_{\times}(i)$ for the tori accreting points, thus, if the inner torus acquires enough mass from the outer torus such that $K(i)(\ell(i))+\Delta_K=K(f)(\ell(f))(r_{\max})$ and $\ell(f)\in \textbf{L1}$,  then there is $\lambda(f)>\lambda(i)$ and the accretion of the final state of the inner torus can lead to collision with the outer torus. This holds also  if the outer torus loses its angular momentum, $|\ell^+|$,   entering a phase where $\lambda(f)<\lambda(i)$ and there is  $r_{cent}(f)<r_{cent}(i)$ for the tori centers. However, in this dynamical process  constrains on $K$ will remains decisive.
It is clear that the specific angular momentum after collision is the  first significant parameter to fix the state, as $K$   depends on the limits on $\ell$ and  $K({f})$, and on the difference  $\Delta K(f)\equiv K_{\sup}(f)-K(f)=\Delta K=K_{\sup}(i)-K(i)$.  Accreting matter brings  also specific angular momentum, then   assuming that the outer torus is small,
 we  may assume    $\ell(i)\approx \ell(f)$ while  $K(f)\leq K(i)+\Delta K_{i,f} (r_s^o,r_{\times}^o)$
where   $\Delta K_{i,f} (r_s^o,r_{\times}^o)$ is a function of the quantity of matter accreting from the outer torus on the inner.
Actually, the relation between the specific angular momentum $(\ell_o^+,\ell_i^-)$ is properly fixed, the maximum elongation of  the outer and inner tori, their thickness, their spacing,  and ranges of specific angular momentum, are constrained by the central attractor dimensionless spin.
 Quantity $\Delta K_{i,f} (r_s^o,r_{\times}^o)$ is dependent also on the cusp location (which may vary),  and location of $r_s$; this radius enters   through the quantity $W(r_s):\,W(r_\times)<W(r_s)<1$, and it is related to the thickness $\mathbf{\overleftarrow{\mathbf{h}}}_s$   of the accreting flow, analyzed in Sec.\il(\ref{Sec:more-l}), see   also Fig.\il(\ref{Fig:Hologra}). Therefore, all these quantities will depend on the difference
$\Delta W=W(r_s)-W(r_\times)$.
 Note that generally the outer torus may be  far more larger then the inner one as   the  \textbf{BH} spin increases.

Since  these processes   can involve a   substantial part of the inner torus,  it is  necessary to discuss the situation for the inner  torus of the couple to trace some conclusions on the \textbf{RAD}  final state  after collision.  On the other hand, in the \textbf{RAD}  framework, tori dynamics cannot generally  be effectively considered  for tori being disentangled. It is possible to consider  the evolution of each torus disentangled from the others only in some periods of the \textbf{RADs} evolution.
The possibility that tori of the aggregate do not collide, evolving independently from each other is expected to the  especially  relevant for super-massive Kerr \textbf{BHs} and Kerr \textbf{SMBHs} with near extreme dimensionless spin.

In this regard, we conclude this section by  discussing a further relevant aspect to be consider in the energetic involving a \textbf{SMBH} interacting with orbiting tori.
 These objects can  be subjected to
the runaway instability, a phenomenon, here directly affecting only the \textbf{RAD} inner  accreting torus, but indirectly the entire couple.  Therefore, the couples where runaway instability may arise  are $\cc_\times^-<\pp$, or $\cc_{\times}^+<\cc^\pm$,  as these  may involve either corotating or counterrotating torus.
The runaway instability is  due to the interaction of the  \textbf{BH}  with the accreting torus via accretion of matter flowing  into the \textbf{BHs}.
  The basic mechanism features an interactive dynamical model involving both the attractor and  the accreting torus:
 basically  because of the change in the mass \textbf{BH} parameter, the accreting torus cusp can  move deeply  towards the torus, i.e. the tori accreting points are related by  $r_{\times}(f)>r_{\times}(i)$ where $i$ and $f$ are for initial state  prior  accretion  and $f$ is for the torus  state when runaway occurs. Note that  we do not explicit the role of the torus center (implying explicitly  a torus  angular momentum change). These effects can  be specially relevant in the \textbf{SMBH} accretion processes.
 Increasing the mass lost rate $\dot{M}$
  due to  the cusp will bend inside the torus.
 The black hole parameters  and  specifically \textbf{BH} mass increase, depending on the  mass accretion rate and the matter impacting  on  the \textbf{BH} event horizon. Consequently,  the spacetime structure
 is modified, and  this also implies  decreasing the spin-to-mass ratio. In terms of  \textbf{RAD} dynamics, there can  be a shift from an   attractor class to another \cite{multy,dsystem}.  Consequently  the
accretion tori and then  the \textbf{RAD}   never reaches a  steady state. Actually, a similar  mechanisms  of \textbf{BH}-accreting torus interaction can   arise  also after a  spin-shift, and
more generally, this feedback process  from the \textbf{BH} should be seen together with any other phenomena involving  any change of the \textbf{BH} parameters,  spin and mass (and eventually even symmetries).
Nevertheless, from the point of \textbf{RAD} morphology structure,  the cusp may also move inwardly
towards the black hole,   decreasing the mass transfer, and therefore stabilizing
the entire  process (see for the dependence of the accretion  rate on the cusp location and other parameters the analysis of Sec.\il(\ref{Sec:more-l})). Viceversa, there is the occurrence of the runaway instability when  the cusp   moves outwards,
penetrating the torus, resulting finally in an increases of mass transfer rate.
Note that the accreting torus could even   be completely destroyed by such instability.
From methodological point of view,
the study of this  situation can be  carried out by considering  stationary
models \cite{Font02}. This can be interesting for the applications  of this approach to \textbf{RAD} environments and this method  in fact is also applied  in the analysis of the \textbf{RAD} possible evolutive paths discussed here: in our language the inner torus $\cc_{\times}^i$ releases a part of  its  mass,  with a total loss of $\mathfrak{F}_K K(i)$, and angular
momentum $\mp \mathfrak{F}_{\ell} \ell^{\pm}$  (where $\mathfrak{F}_K\ll 1$ and $\mathfrak{F}_{\ell}\ll 1$)  to the central
attractor which swallows it,  the final stage of this first part of the process is then concluded and  a new calculation can
consider  further state with a new background, which is  changed after change of the  \textbf{BHs}, the new position of the cusp is then evaluated. Generally, the study of this phenomenon  can be  performed in a non-dynamical framework.
 The evolution of
the central black hole is read as a sequence of exact  black
holes solution, each picture of the sequence being labeled with a   different  mass, whose increase is function of the  mass
accretion rate.  Consequently, following the dependence of the mass accretion rate (see Sec.\il(\ref{Sec:more-l})), the  runaway instability is regulated  by
several  parameters, from the \textbf{BHs} dimensionless spin, to  the cusp location, the flow thickness and  ultimately by the  structure of the innermost part of the torus.
Pictures of  the systems during stages of its dynamics was studied, this method was carefully analyzed and discussed for geometrically thick torus around Kerr \textbf{BHs}    in \cite{pugtot}. Here we apply this approach to the  characterization of the \textbf{RAD} dynamics.
The  relevance of  the runaway instability for the \textbf{RAD} dynamics   should also
consider  the process  time-scales (which in general  is considered to be short, depending  also on different parameters and especially the attractor)  compared with the time-scale of the dynamical processes of the accreting torus model, and in the \textbf{RAD} framework, of the \textbf{RAD} dynamics.
In a system made by a \textbf{RAD} couple and its \textbf{BH} attractor  as in Fig.\il(\ref{Fig:AliceB}), while the accretion from the inner corotating torus may establish the runaway instability, this can be  accompanied by a further process of torus-torus interaction, which would bring, according to the mechanisms considered above, also a change in the cusps, and if a runaway of the inner  torus is established, then also the  outer torus is affected.
   Similarities between  \textbf{BH}-inner torus runaway  process,   and  the mechanism of torus-torus interaction appear. The consequences of tori collision may analogously lead the  evolution of both the elements of the mechanism,   going through a positive or negative feedback reaction. One of these stabilizes the torus and the  couple process, the other turning in a more complex situation where evolutive loops may also occur,  giving rise also to more accretion cycles--a mechanism supposed for example in \cite{dsystem}.
   This is particularly interesting in the case of double accretion, or accretion from outer counterrotating torus, where
   the  presence of an  screening quiescent inner corotating torus blocks the runaway from the outer torus accretion.
In this respects we could talk about torus-torus
and  torus-\textbf{BH} runaway instability. A torus-torus runaway may be seen as analogue problem with respect to  the \textbf{BH} runaway instability, not concerning, however, the background spacetime.  The runaway instability has also been considered playing a part in the  \textbf{GRB} production, and more generally in the energy  extraction from the central \textbf{BH} engine,
 released by the accretion mechanism itself together with the extraction of  the rotational energy of the  rotating black hole  via the Blandford-Znajek mechanism.
 All these aspects could be combined in the  \textbf{RAD} scenario  with the  dynamics of the interacting tori.
\section{Conclusion  and future perspectives}\label{Sec:conl}
The physics of accretion
around super-massive black holes has been extensively studied, both on a theoretical ground and by means of numerical simulations bringing  knowledge of both the  accretion  disk dynamics,  inherently   disk formation and disk instabilities. These studies have often dealt with issues which are   partially  still unresolved, as a  full description of  the  accretion mechanism  or even jet launching. A complete physical picture of all these phenomena around  strong  attractors  remains to be framed in a unique setting, since  there are strong signals that there  should  be  some unique  interpretive framework to deal  with the pieces of the accretion puzzle.
Here we concentrated our attention on the possibility that more than one accretion configuration, a couple of  tori, may be formed around a central super-massive Kerr \textbf{BH}.
Similar structures are  considered as screening bubbles of dust or other materials in studies of   X-ray \textbf{BH} emission. {Analogue  effects produced by generic  cloud  might arise from  configurations  considered here, taking care to consider axial symmetries and coincidence of the orbital plane with the equatorial plane (the specific toroidal model considered here as \textbf{RAD} component is also optically thick)}.
Up to now    some ad hoc  mass distribution   additional to the main accretion disks  was considered, avoiding  a discussion of structure, morphology and more general constraints--see \cite{Nixon:2013qfa,Dogan:2015ida,Bonnerot:2015ara,Aly:2015vqa} where  titled and strongly misaligned disks are considered.
{In a broader perspective of analysis one might think that  some aspects
   on the \textbf{AGN} activities,  connected to the formation of  such materials (and especially  optically thick material), may fit to an adaptation to the \textbf{RAD} context. Cloud accretion has been also claimed as a mechanism, together with  star  tidal disruption, to enhance the \textbf{SMBH} accretion rate--see  for example \cite{Bonnell}. The other  possible  case, represented by a series of small and random accretions, overlaps with the predicted  \textbf{RAD} evolution phases     due to its  inner ringed  structure.
Screening  effects of  X--ray emission  have been instead studied for example  in \cite{1992ApJ...400..163B,Marchesi,Gilli:2006zi,Marchesi:2017did,Masini:2016yhl,DeGraf:2014hna,Storchi-Bergmann}. Such investigation  also  suggests that   optical and   X-ray emission profile obscuration  may arise from
 different phenomena   caused by  several  materials including dust surrounding the
 inner part of the galactic nuclei,  randomly distributed around the central \textbf{BH}. Depending on  the gas  density, the light emitted  could  be absorbed in the optic and in the X-ray electromagnetic band, distinguishing  \textbf{AGN}  as obscured, not obscured, or much
obscured (or \emph{Compton thick})-see also recent analysis \cite{WA}.
 Assuming   a \textbf{RAD} scenario one could  analyse  this  obscuration. The observed tori are constrained  as  the inner torus of the \textbf{RAD}, or the inter--torus  (quiescent and corotating), located between two tori where the outer  (counterrotating) one is accreting towards the central attractor. Although certainly more detailed studies focused on this aspect are necessary, a screening torus proposed here   could not be excluded in our opinion as a concurrent cause of the obscuration  effects.
}

This  work  bridges this gap  with the  aim to provide a reliable  general set of constraints on the ringed disks  and  emerging instability.
This analysis led also  to     three   sideline  results: \textbf{\textsl{1.}} a very first evaluation of the energy release  after tori collision from two colliding tori of the same \textbf{RAD}, according  to two different collision mechanisms discussed in Sec.\il(\ref{Eq:CM-ene}),  proving  that energy release in such phenomena may be really huge depending  on the  attractor rotation. \textbf{\textsl{2.}}
  Second,  the results  trace a link between the attractor characteristics   and the  orbiting tori helping, on one side, to  determine the model to be adopted  and to distinguish the  central \textbf{BH} with its associated \textbf{RAD}--Sec.\il(\ref{Sec:Kerr-2-Disk}).  \textbf{\textsl{3.}}
  {Third, we provided an extensive comparative study of properties of the counterrotating accretion tori  orbiting around a Kerr \textbf{SMBH} with respect to the corotating tori. A substantial part of development and especially of analysis of the phenomenology  associated with  \textbf{RAD} features the presence of   $\ell$counterrotating couple. Although the hypothesis of having counterrotating tori has been considered in the literature, most studies  are    predominantly oriented towards the analysis  of the corotating accreting tori interacting with their central spinning attractor.
    On the other hand, the \textbf{RAD} investigation made it necessary to explicitly address the possibility of the formation of counterrotating extended toroidal matter configurations especially around \textbf{\textbf{SMBHs}}. We set constraints on the formation and accretion of such tori, This analysis has also led to a careful assessment of the accretion rate, cusp luminosity and accretion efficiency in dependence of the \textbf{BH} spin and the cusp locations for the two classes of orbiting fluids. To our knowledge, this systematic study has been for  the first time directly  addressed  in  literature.}

   In this analysis we have considered every  central Kerr black hole attractor:  from very slow, up to the Schwarzschild limit,  to the limit of extreme Kerr \textbf{BH}. We realized  a systematic investigation of the parameter space for the  tori,  with particular attention to the case of  tori in  accretion and tori collision, providing a classification of couples  according to the dimensionless spin of the central Kerr \textbf{BH}.
We demonstrate  that only specific couples of accretion tori  can orbit    around a central Kerr \textbf{BH}--Sec.\il(\ref{Sec:const}).  The dimensionless spin of the Kerr black hole strongly constrains the possible couple of orbiting  tori  in number,  location and relative range of variation for specific fluid angular momentum and density parameter.
There are strong indications that (see \cite{ringed,dsystem,open}) \textbf{RADs} are capable to   feature   a ``co-evolution''  of \textbf{BH} system and galactic life of the \textbf{BHs},
implementing a   \textbf{BH}-\textbf{RADs} correlation  paradigma, relating specially featured  \textbf{RADs}  evolutions to one  special class of Kerr \textbf{BHs}, according to their dimensionless spin--\cite{multy}.

\medskip

The systems  investigated here  offer    several   methodological  and observational challenges.  Describing a set of virtually separated  tori orbiting one attractor  as an entire configuration, requires a certain number of assumptions.
A first aim of this paper was to establish the conditions for the existence and instability of  the couple of tori,   considering also the possibility of tori collision   at a certain stage of the couple evolution.
To do this, we adopted both an  analytical  and numerical approach,
considering an effective potential function for the tori couples  and by considering a direct integration for certain particular representative cases,  guided in the parameter choice  by the analysis of geometric  properties  of the Kerr spacetime.
The general relativistic hydrodynamic  equations we solved  are coupled with proper boundary conditions, enabling  to take account  of the presence of two tori and eventually emergence of their  collision regime. This onset
 was modified  in a simple way  in order
to preserve   the  general relativistic hydrodynamic approach  for each  individual torus which is commonly  rendered through  an effective potential function.  We casted  therefore in Sec.\il(\ref{Sec:Kerr-2-Disk}) the set of  the Euler  equations  using  a composite effective potential for the couple.

We believe our study  show  there is a strong motivation to discuss the results of this approach in comparison with more complex scenarios where the contribution of the radiation, viscous and magnetic factors  having a predominant role at least in some phases of the evolution of the formed structures. For each torus, this implies also the need to define the specific nature and \emph{history} of the \textbf{BH} attractor and of each \textbf{RAD} component.

The theory of  the  perfect fluid relativistic torus orbiting
a Kerr \textbf{BH}  has  been developed since the first fundamental works in many different applications.
For the first time to our knowledge, it is however applied  here and in \cite{ringed,open} to the description of a  couple of tori.
The results reported in this work  are certainly  a first step towards the understanding of the dynamics of a tori couple and therefore we expect
several  different  extensions of the model  and its improvements.
Particularly, we
plan to extend the investigation  to the  scenarios   where  different rotational laws are considering for the accretion tori.
In addition, it would be  interesting to investigate, using
more elaborated  approaches, further situations  where, for example, a change of the \textbf{BH} parameters  is expected   due to accretion, or where  the tori self-gravity is considered.
A possible mechanism expected to be relevant in the case of a thick torus  is for example the runaway instability \cite{Abra83},  raised eventually from accretion  of   the inner \textbf{RAD} torus (in the $\ell$corotating or $\ell$counterrotating tori of the kind $\cc_{\times}^+<\cc^-$ or $\cc_{\times}^-<\cc_{\times}^+$.) As consequence of the   mass loss through the cusp of the inner  tori,  the \textbf{BH} parameters  would be affected by  the accreting material,  therefore the spacetime   geometry modified  and so the instability of the accreting tori occurs, changing its cusp $r_{\times}$ location.
 It is worth pointing out that in our model prescription   of  such  runaway mechanism   may eventually trigger a drying-feeding process in which the  inner torus would be characterized by  several stages  of instabilities.
 We should  note that  each torus in this special \textbf{RAD} has high or very high accretion rates; we expect in  processes analogue to  feeding-drying,  this can be give rise to a sort of  ``clumpy'' episodic accretion  process.
As  mentioned above, a further challenge acknowledges  the  existence of very different \textbf{RADs} tori histories which may be distinguished in  three periods of \textbf{BH}-ringed accretion disk  life: the first (\textbf{I}) featuring tori formations, the second (\textbf{II}) facing the accretion of one or two tori onto the central \textbf{BH} and the third (\textbf{III})   the eventually emerging of tori due to collisions.
Picturing these situations is clearly a complex task.
Each of these periods obviously counts  on different contributions to the balance of forces for the torus, hence the need to deal with the problem with different tools. We were motivated to provide general applicable results   with the least number of assumptions that would make this a very specific model restricted to a particular attractor and  particular torus. We substantially reduced the parameter  space of our model, providing ranges of variation of the variables and parameters which may fit also to  some extent for other   models, providing also  attractor classes on the bases of the tori features  indicating the attractor which we should chase to find evidences for.
Dynamics and tori instability are generally determined by the balance of
gravitational, magnetic, centrifugal  and hydrostatic pressure, together with dissipative effects, as viscosity and resistivity, eventually torus self gravity and radiation pressure.
Each of these ingredients is relevant in one of the disk period and mainly determine the disk model to be adopted.
On the other hand, the disk model is tightly bounded to the attractor, thus the choice of a geometrically thick accretion torus around a \textbf{SMBH} where the curvature contribution to the force balance is  predominant, is usually well founded.
Setting our analysis in this model, we are able to provide general and strict limits  especially on periods  \textbf{II} and \textbf{III}--(a more careful analysis of the evolution of these systems is discussed  in \cite{open,dsystem}). By neglecting here the period of formation of the single tori that can be attributed to the influence or combination of different phenomena, we concentrate on the second period and the emergence of the collision.
A second   aim
of this paper, addressed in Sec.\il(\ref{SeC:coll}) was in fact to investigate the emerging of tori  collision and the related  mechanisms.
In all of these catastrophic events, collision  energy is expected to be released.
An evaluation of CM-energy
   in the test particles approximation is provided. Although it    suffers of various  issues (e.g. fine-tuning problems \cite{Berti:2009bk}), this approach  is effectively   widely used as a   useful first approximation. In the future  investigations,  the role   of other factors  as  radiation,   magnetic fields, and temperature could be included.
As shown in Fig.\il(\ref{Fig:AliceB}),  collisional efficiency of interacting matter of $\ell$counterrotating tori increases with increasing \textbf{BH} spin being very high for near-extreme \textbf{BHs}.
 Results on collision thus  suggest
that   collisional double tori effects  may represent a  new and unexpected sources of radiation
that might be observed through the  phenomenology associated with both the stable and unstable  configurations, giving clear indications that considering only part of the expected contribution a large release energy is expected from the collision. The phenomenology associated with these toroidal complex structures may be indeed  very wide. This new  complex  scenario in facts lead  to reinterpret the phenomena analyzed so far in the single-torus framework.
 The radially oscillating tori of the couple could be related to the high-frequency quasi periodic oscillations  (\textbf{QPOs}) observed in non-thermal X-ray emission from compact objects, keeping  fingerprint of the discrete radial profile of the couple structure.
More generally instability  of such configurations, we expect, may reveal  crucial significance for the high energy astrophysics related  especially to accretion onto supermassive \textbf{BHs}, and the extremely energetic phenomena occurring  in quasars and \textbf{AGNs} that could be observable by the planned X-ray observatory  \texttt{ATHENA}\footnote{\url{http://the-athena-x-ray-observatory.eu/}}.

\medskip

Further  important aspect,  investing the methodological point of view, concerns
the applicability in different contexts from the stationary hydrodynamic model set-up adopted  here for the single torus.
If this first investigation on tori collisions is to be understood as a proposal where  introduction of additional ingredients should be taken into account by specific models for the single torus,
we expect that in the balance of different contributions, the presence of other magnetic or viscosity effects  cannot qualitatively change the  constraints enlightened here.
To  be certain of adaptability and applicability of the results plotted here, we need to consider first the  adaptability to other models  of the results concerning  the limit of one-torus model.
 The hydrodynamical model considered here is widely applied in many contexts and surely showing a remarkably good fitting on the more complex models  as well discussed  for example in \cite{Lei:2008ui} when it comes to the analysis of morphology presented here.
These results can then be used as initial data and comparative model in  any numerical analysis of more complex situations,  sharing  the same symmetries, as it is generally adopted in many general relativistic hydrodynamic (GRHD) or  and general relativistic magnetohydrodynamic (GRMHD)  approaches  for the single accretion disk case.
In the current analysis of dynamical one-torus system of   both   GRHD  and  GRMHD set-up, the geometrically  thick tori considered in this work for each \textbf{RAD} component,  are  commonly adopted as initial configurations for the numerical  analysis--\cite{Fragile:2007dk,DeVilliers,Porth:2016rfi}.
   This is a  good hint of the reliability in wider contexts  represented by different accretion disks supported by more complex  accretion.
There is no expectation that   there would be a qualitative overcome of
constraints provided here, especially for period \textbf{II} which are reasonably not \emph{qualitatively} affected by   other contributions. The \emph{morphological}  features of the equilibrium and locations are certainly  predominantly determined   in these studies by the centrifugal and gravitational components (see for example the very much debated definition of inner edges of accreting disks, which  is commonly accepted running  in the range $]r_{mbo},r_{mso}]$ according to the specific tori model).
Thus, this work  is also  intended  to be a guideline for constraints on numerical dynamical analysis.

\medskip

A further aspect influencing the model development is the \textbf{RAD} formation. These coupled  tori are expected to form as a result of   the  interaction of the  central attractor    with the  environment in {\textbf{AGNs}}, where   corotating  and counterrotating accretion stages are
mixed.
A major, suggestive possible host-environment for \textbf{BH-RADs} are \textbf{AGNs}  \textbf{SMBHs}.
There are evidences  suggesting what  these \textbf{RADs}  structures may play a major role in Galaxy dynamics and particularly in \textbf{AGNs}.
Several studies are in  support of  the existence  of \textbf{SMBHs}  characterized by  multi-accretion episodes  during their life-time in Galaxy cores.
Consequently \textbf{SMBHs} life may report  traces  of  its  host Galaxy dynamics as a diversified feeding   of a \textbf{SMBH}. These processes may involve
repeated galaxy mergers
or also interacting
binary \textbf{BH}, X-ray binaries or \textbf{SMBHs} binary systems.
As a consequence of these activities, matter around  attractor could give an equilibrium configuration as
counterrotating  and    misaligned tori \cite{Aly:2015vqa,Dogan:2015ida}.
Sequences of orbiting toroidal structures  with strongly   different features as,  for example, different rotation orientations with respect to the central Kerr \textbf{BH} are argued to be produced from
chaotical, discontinuous   accretion episodes   where corotating and counterrotating accretion stages are likely to  be mixed \cite{Dyda:2014pia,Aligetal(2013),Carmona-Loaiza:2015fqa,Lovelace:1996kx,Gafton:2015jja}.
  In this environment where also the galactic magnetic field can eventually play an important role in the formation and rotation orientation of each tori, strongly  misaligned tori with respect to the central \textbf{SMBH} spin  may appear  \cite{Nixon:2013qfa,Dogan:2015ida,Bonnerot:2015ara,Aly:2015vqa}.
Further phenomena seen  as possible candidate for \textbf{RADs} formations may be \textbf{BHs} kick-out in \textbf{BH}-populated Galaxy core--see for example \cite{mnras-pa}.
The turbulent attractor life would be source for  very  different  accreting matter in some kinds of binary systems, or may involve the self gravity of a unique original accreting  disk  which  finally splits into several rings.

\medskip

From  observational view point,
 we believe our results may   be  of significance for the high energy  astrophysics. In fact  the presence of such structures is capable to substantially modify    the    single disk  scenario, which has been effectively taken  so far  as the interpretative common  ground.  Explanation of some of most intriguing and unveiled issues of \textbf{BHs} physics interacting with matter may be reset in this new framework where single tori paradigma would be then  just seen as a limit or special case related to an  evolutive  phase of non isolated  \textbf{BHs} life. The presence of inner tori may also enter as a new unexpected ingredient in the accretion-jet puzzle, as proposed also in   \cite{S11etal,KS10,Schee:2008fc} and \cite{open}.
These is a huge  amount of possibilities to be investigated as  tori interactions or  oscillations can be associated to a variety of phenomena with relevant energy release, as we have also partially faced here.
Accretion or collision constitute possible scenario for the entire ringed  disk instability  eventually leading to interesting observational effects.  Tori in ringed disk   may  collide and merge,  or, eventually
the accreting matter from the outer torus of the couple can  impact on the  inner torus, or    the outer torus may be inactive  with  an active inner torus  accreting  onto the \textbf{BH}, or both tori  may be active.
 These multi-configurations  may be at the root of phenomena eventually  detectable  by  X-ray detectors   as  the
shape of X-ray emission spectra,  the X-ray obscuration and absorption
by one of the torus, in the extremely energetic  radiative phenomena in quasar  and  \textbf{AGNs}. Signatures of \textbf{RADs} may be found in the emission shape lines as
peaked profiles of radially stratified emission\cite{KS10,S11etal,Schee:2008fc}.
 The radially oscillating tori of the ringed disk could be related to the high-frequency quasi periodic oscillations observed in non-thermal X-ray emission from compact objects (QPOs), a still obscure feature of the  X-ray astronomy related to the inner parts of the disk.
Relatively indistinct excesses of the relativistically broadened emission-line components were predicted in difference works, arising in a well-confined radial distance in the accretion structure originating by a series of episodic accretion events.
Our analysis first shows that  occurrence of these situations  is strictly constrained by the black hole spin.
  This aspect has  also important implications  on the possible observational  effects    providing  a perspective on the  phenomena  emerging from  their dynamics, isolating those situations where actually  these configurations may be chased.

\begin{acknowledgement}
D. P. acknowledges support from the Junior GACR grant of the Czech Science Foundation No:16-03564Y.
Z. S. acknowledges  the Albert Einstein Centre for Gravitation and Astrophysics supported by grant No.
 14-37086G.
 \end{acknowledgement}

\appendix

\section{Corotating tori in the ergoregion: \textbf{RAD} tori separation and  constraints}
\label{Sec:coroterf}
\begin{figure}[h!]
\includegraphics[width=1\columnwidth]{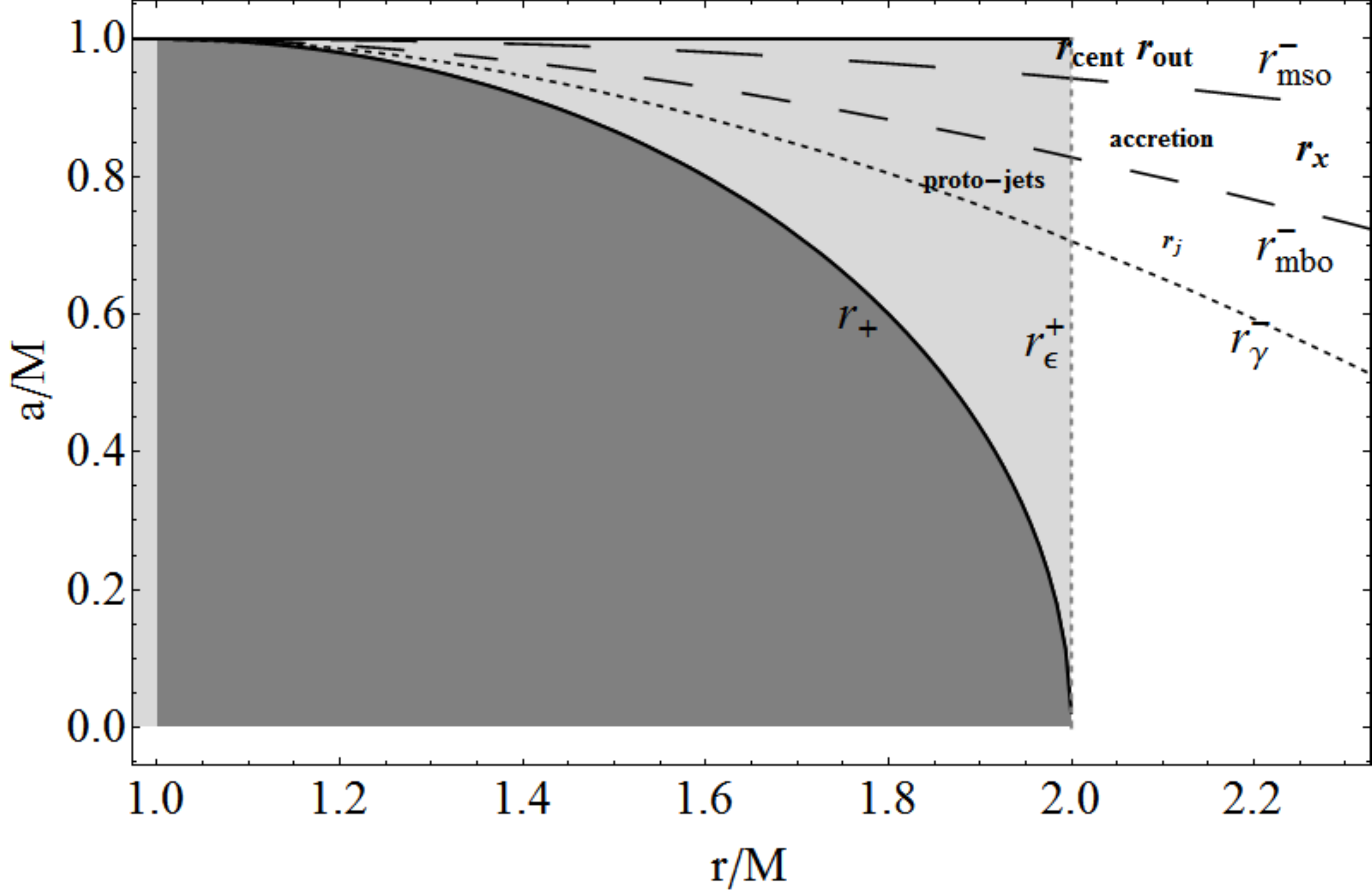}
\caption{The outer ergoregion $\Sigma_{\epsilon}^+\equiv]r_+,2M]$ (light-gray region) of the Kerr  geometry. Gray region is $r<r_+$, $r_+$ is the \textbf{BH} horizon. $r_{\epsilon}^+=2M$ is the outer ergosurface. Location of inner edge of quiescent tori and tori center, the region of inner edge of accreting tori $r_{\times}$ and proto-jets are shown.   }\label{Fig:TwomodestateErg}
\end{figure}
The possibility that corotating  toroidal extended matter configurations  could be  formed in the outer ergoregion (``all-contained'' accretion tori), $\Sigma_{\epsilon}^+$, and even  to cross the outer ergosurface, $r_{\epsilon}^+$,  has been   investigated in \cite{pugtot}.
This situation was then     explored  in the \textbf{RAD} context, particularly in relation to   the \textbf{RAD} tori separation in corotating  tori sub-sequences of \textbf{RAD} components (which are possible to form in $\Sigma_{\epsilon}^+$), and the counterrotating ones, confined in the outer region of the boundary $r_{\epsilon}^+$, in dependence on the  Kerr \textbf{BH} spin--\cite{ringed,open,long}.   Fundamental differences  between slow rotating  \textbf{SMBHs} and faster spinning \textbf{SMBHs} were observed. The tori separation is a particularly relevant  phenomenon  for faster spinning \textbf{BHs}.  For these spacetimes, quiescent and accreting tori solution of the Euler equation (\ref{E:1a0}), located entirely  in $\Sigma_{\epsilon}^+$ are possible, proving that, as in  Fig.\il(\ref{Fig:TwomodestateErg}), different classes of  faster spinning \textbf{SMBHs}  provide different  topological  constraints (equilibrium conditions) for the tori.
In \cite{open,long} in particular, this aspect was addressed in relation to the role of ergoregion in the emission and the structure  of the proto-jets, whereas a  more attentive  look to the attractors classification   was considered in  \cite{multy}.
To complete the discussion on $n=2$ \textbf{RAD},   we briefly mention  the possibility of the  ergosurface penetration and all-contained accretion tori
(or ``dragged surfaces''). The tori location with respect to the central \textbf{BH},  with specification on their equilibrium states, are represented in  Fig.\il(\ref{Fig:TwomodestateErg}),  where the critical points of the hydrostatic pressure are also shown.
 Clearly, this analysis concerns  the case of couple with an inner corotating torus,  in accretion or not, i.e. all the couples $\pp^-<\pp^{\pm}$ are considered according to the constraints and restrictions discussed  in  Sec.\il(\ref{Sec:Kerr-2-Disk}). The inner configuration containing or crossing the ergosurface  (in the sense illustrated in  \cite{pugtot}) can be a ``screening torus'', as considered in several analysis of the X-ray emission\cite{Marchesi,Gilli:2006zi,Marchesi:2017did,Masini:2016yhl,Storchi-Bergmann}.
 Finally, these tori should have  a special relevance also in the evaluation of collisional effects, considered here in  Fig.\il(\ref{Fig:AliceB}), occurring in  $\Sigma_{\epsilon}^+$   with possible applications to  the energy extraction processes assumed to occur from the Lense-.Thirring effect in the $\Sigma_{\epsilon}^+$. The role of the dragging effects has been also considered in Sec.\il(\ref{Sec:more-l}) in the evaluation of the mass-accretion rates for corotating and counterrotating tori in dependence of the \textbf{BH} spin.


\begin{thebibliography}{99}


\bibitem
{WW-AAVV}
S. A. Walker,  et al  
\mnras, 468, 2 2506-2516 (2017)



\bibitem
{Ricci.:2017wmr}
  C. Ricci, {\it et al.},
  \mnras, { 468}, 1273 (2017)



\bibitem
{Tadhunter:2017qji}
  C. Tadhunter, R. Spence, et al 
  arXiv:1702.02573 [astro-ph.GA] (2017)


\bibitem
{Regan:2017vre}
 J. Regan, E. Visbal et al. 
 \nat Astronomy,  1,  0075 (2017)

\bibitem
{edge-1}
  S. McAlpine,  et sl 
\mnras,  468, (3), 3395-3407 (2017)



\bibitem
{Tucci:2016tyc}
 M. Tucci \& M.  Volonteri,
  \aap, { 600},  A64(2017)



\bibitem
{Xie:2017jbz}
 F.~G. Xie\&  F. Yuan,
  \apj, { 836}, 1,  104 (2017)

\bibitem
{Collinson:2016rmg}
 J.~S. Collinson  et al
  \mnras,\  { 465}, 1,  358 (2017)




\bibitem
{Parker:2017wnh}
 M.~L.  Parker { et al.},
  Nature, { 543},  83 (2017)

\bibitem
{Zhang:2015eka}
 J. Zhang, Z.W. Xue, J.J. He, E.W.  Liang \& S.N.  Zhang,
  \apj, { 807}, 1,  51 (2015)


{\bibitem{1992ApJ...400..163B} G. Bao\&  Z. Stuchlik, \apj, 400, 163 (1992).
}


 \bibitem
 {Marchesi}
 S. Marchesi, et al.,
\apj, 830, 100, 20(2016)

\bibitem
{Gilli:2006zi}
 R. Gilli, A. Comastri,  and G.  Hasinger,
  \aap, { 463}, 79 (2007)

\bibitem
{Marchesi:2017did}
 S.  Marchesi, et al
  \apj, { 836},  1,  116 (2017)


\bibitem
{Masini:2016yhl}
A.  Masini,  {\it et al.},
 \aap,  { 589}, A59 (2016)


 \bibitem
 {DeGraf:2014hna}
  C. DeGraf, A. Dekel, J. Gabor, and F. Bournaud,
  \mnras,\  { 466},  1462 (2017)

  \bibitem
  {Storchi-Bergmann}
T. Storchi-Bergmann  et al.
\apj,  835,  2 (2017)




\bibitem
{2011PhRvD..84h4002K} Kov{\'a}{\v r}, J., Slan{\'y}, P., Stuchl{\'{\i}}k, Z., et al.\ 2011, \prd, 84, 084002
\bibitem
{2013CQGra..30b5010K} J. Kov{\'a}{\v r}, O. Kop{\'a}{\v c}ek, V. Karas, \& Y. Kojima,\ 2013, Classical and Quantum Gravity, 30, 025010
\bibitem
{2014PhRvD..90d4029K} J. Kov{\'a}{\v r}, P. Slan{\'y},, C. Cremaschini, et al.\ 2014, \prd, 90, 044029
    \bibitem
    {2016PhRvD..93l4055K}  J. Kov{\'a}{\v r}, P. Slan{\'y}, C. Cremaschini , et al.\ 2016, \prd, 93, 124055
{\bibitem{Slany:2013rml}
  P.~Slany, J.~Kovar, Z.~Stuchlik and V.~Karas,
  Astrophys.\ J.\ Suppl.\  {\bf 205},  3 (2013).
\bibitem{Kovar:2011uh}
  J.~Kovar, P.~Slany, Z.~Stuchlik, V.~Karas, C.~Cremaschini and J.~C.~Miller,
  Phys.\ Rev.\ D {\bf 84}, 084002 (2011).
\bibitem{Schroven:2018agz}  K.~Schroven, A.~Trova, E.~Hackmann and C.~L\"ammerzahl,
  Phys.\ Rev.\ D {\bf 98} no.2,  023017 (2018).
  \bibitem{Trova:2018bsf}
  A.~Trova, K.~Schroven, E.~Hackmann, V.~Karas, J.~Kovar and P.~Slany,
  Phys.\ Rev.\ D {\bf 97}, no.10,  104019  (2018).
}
 \bibitem
 {pugtot}
 D. Pugliese, \& G.  Montani,
  \prd, 91, 083011 (2015)
%
 \bibitem
 {ringed}
 D. Pugliese\& Z.  Stuchlík,
\apjs,  { 221}, 2,  25 (2015)

  \bibitem
  {open}
  D. Pugliese\& Z.  Stuchlík,
 \apjs,  { 223}, 2,  27(2016)

 \bibitem
 {dsystem}
   D. Pugliese\& Z.  Stuchlík,
 \apjs,  { 229},  2, 40(2017)


 \bibitem
 {multy} Pugliese, D., \& Stuchl{\'{\i}}k, Z.\ 2018, Classical and Quantum Gravity, 35, 185008

\bibitem
{Proto.Jets} Pugliese, D., \& Stuchl{\'{\i}}k, Z.\ 2018, Classical and Quantum Gravity, 35, 105005



\bibitem
{long}
   D. Pugliese\& Z.  Stuchlík,
  JHEAp {\bf 17} 1 (2018)

  \bibitem
  {Pugliese:2018zlx}
  D.~Pugliese and G.~Montani,
  arXiv:1802.07505 [astro-ph.HE].

\bibitem
{S11etal}
V. Sochora, V.  Karas, J. Svoboda, \& M. Dovciak,
\mnras, 418, 276--283 (2011)



\bibitem
{KS10}
V. Karas, \&V. Sochora,	
\apj,  725,  2,  1507--1515 (2010)


\bibitem
{Schee:2008fc}
J.   Schee, \&  Z. Stuchl{\'{\i}}k,
  Gen.\ Rel.\ Grav.,  41, 1795 (2009)
\bibitem
{Schee:2013asiposs}
J.  Schee,  \& Z. Stuchl{\'{\i}}k,
  JCAP,   2013 (2013)


\bibitem
{NixonKing(2012b)}
C. Nixon,    A. King,  D. Price,  J. Frank,
\apj, {757}, L24 (2012)

\bibitem
{AKM03}
M. Ansorg,
A. Kleinwachter, \&  R.  Meinel, \mnras, {339}, 515-523 (2003)



\bibitem
{Aligetal(2013)}	
 C. Alig,  M. Schartmann,  A. Burkert, K. Dolag,
\apj,  771,  2, 119 (2013)




\bibitem
{Lovelace:1996kx}
  R.~V.~E.   Lovelace\& T. Chou,
  \apj,   { 468},  L25 (1996)


\bibitem
{Gafton:2015jja}
  E.  Gafton, E. Tejeda, et al 
  \mnras,  { 449}, 1,  771 (2015)



\bibitem
{natures}
 	J. M.  Miller, et al
\nat, {526}, 542--545 (2015)


\bibitem
{Blanchard:2017zfe}
P.~K.  Blanchard,  {\it et al.},
  arXiv:1703.07816 [astro-ph.HE] (2017)

\bibitem
{Carmona-Loaiza:2015fqa}
J.M. Carmona-Loaiza, M. Colpi,  M.  Dotti\& R. Valdarnini,
 \mnras,  { 453},  1608 (2015)


\bibitem
{Dyda:2014pia}
S.  Dyda,   R.V.E. Lovelace, et al 
  \mnras,  { 446}, 613 (2015)


\bibitem
{Volonteri:2002vz}
 M. Volonteri, F.  Haardt, \& P. Madau,
  \apj, { 582},  559 (2003)



\bibitem
{Jaroszynski(1980)}
 M. Jaroszynski,    M. A. Abramowicz, B. Paczynski, Acta Astronm., {30},  1 (1980)




\bibitem
{Pac-Wii}
 B.   Paczy{\'n}ski,
 {Acta Astron.}, {{30}}, 4 (1980)



\bibitem
{cc}
 B.  Paczy{\'n}ski, \& P.  Wiita,
\aap, {{88}}, 23 (1980)


\bibitem
{Koz-Jar-Abr:1978:ASTRA:}
 M. Koz{\l}owski, M. Jaroszy{\'n}ski,  M.~A.   Abramowicz,
\aap, {{63}}, 209 (1998)



\bibitem
{abrafra}
 M.~A.  Abramowicz \&  P.~C. Fragile,
   Living Rev. Relativity, {16}, 1 (2013)



\bibitem
{Shafee} R. Shafee,J. C, McKinney,  et al.
\apj, {687}, L25 (2008)



\bibitem
{Fragile:2007dk}
 P. C.  Fragile,  O. M. Blaes, P. Anninois, \&  J. D. Salmonson,
\apj, {668}, 417--429 (2007)


\bibitem
{DeVilliers}
J-P. De Villiers, \&  J. F. Hawley,
 \apj, {577}, 866 (2002)




 \bibitem
 {arXiv:0910.3184}
 Z. Stuchl{\'{\i}}k, P.  Slan{\'y}, \& J.  Kovar,
    Class. Quantum Gravity, {{26}}, 215013 (2009)


\bibitem
{Porth:2016rfi}
   O. Porth, et al.
  arXiv:1611.09720 [GRqc] (2016)

\bibitem
{F-D-02}
J. A.  Font, \& F.  Daigne,
\apj, {581}, L23--L26 (2002)



\bibitem
{Hawley1990}
J. F.  Hawley,
   \apj, {356}, 580 (1990)
%
\bibitem
{Hawley1987}
J. F.  Hawley,
   \mnras, {225}, 677 (1987)



%




\bibitem
{Hawley1991}
  J. F.    Hawley,
\apj, {381}, 496 (1991)



 \bibitem
 {Hawley1984}
 J. F. Hawley, L. L. Smarr, J. R.  Wilson,
  %
\apj, {277}, 296 (1984)

\bibitem
{Fon03}
 J. A.  Font,
Living Rev.\ Relat.,  {{6}},  4 (2003)


\bibitem
{Lei:2008ui}
  Q.  Lei,  M. A. Abramowicz,  P. C. Fragile, et al.
\aap, {498}, 471 (2008)

\bibitem
{Boy:1965:PCPS:}
 R.~H. Boyer,
Proc. Camb. Phil. Soc., {61}, 527 (1965)

\bibitem
{Raine}
J. Frank,
A. King,
D. Raine,
 \emph{{Accretion Power in Astrophysics}}, (Cambridge University Press, Cambridge 2002)



 \bibitem
 {PuMonBe12}
 D. Pugliese, G.  Montani,\& M.~G.~  Bernardini,
  \mnras,  { 428}, 952 (2012)



  \bibitem
  {PuMon13}
  D.  Pugliese, \& G.  Montani,
  Europhys.\ Lett., 101, 19001 (2013)




\bibitem
{2011}
  H.  Kucakova,  P.  Slany,    Z. Stuchl{\'{\i}}k,
JCAP, {{01}}, 033(2011)


\bibitem
{Rez-Zan-Fon:2003:ASTRA:}
 Rezzolla, L., Zanotti, O., Font,  J.~A. 2003
\aap, {412}, 603



\bibitem
{Stuchlik:2012zza}
   Z. Stuchl{\'{\i}}k,   Slan{\'y}, P.,  Torok, G., Abramowic,  M. A. 2005,
  \prd, {{71}}, 2



 \bibitem
 {Sla-Stu:2005:CLAQG:}
Slan{\'y},  P. \&   Z. Stuchl{\'{\i}}k 2005
\newblock   Class. Quantum Gravity, {22}, 3623


 \bibitem
 {astro-ph/0605094}
  Z. Stuchl{\'{\i}}k\& Slan{\'y}, P. 2006,
  AIP Conf.\ Proc. { {861}}, 770





  \bibitem
  {Stu-Sla-Hle:2000:ASTRA:}
  Z. Stuchl{\'{\i}}k,   Slan{\'y} P.,  Hled{\'{\i}}k  S. 2000,
\aap, {{363}}, 425


\bibitem
{Stu-Kov:2008:INTJMD:}
 Z. Stuchl{\'{\i}}k, \&   J. Kov\'{a}\v{r},  2008,
 { Int. J. Mod Phys {D}}, {{17}}


 \bibitem
 {M.A.Abramowicz}  M. A. Abramowicz,
 Acta. Astron., {21}, 81(1971)



\bibitem
{Chakrabarti0}S. K.  Chakrabarti,    \mnras, {245}, 747 (1990)


\bibitem
{Chakrabarti}S. K. Chakrabarti,     \mnras, {250}, 7 (1991)


\bibitem
{Abramowicz:2008bk}
M. A.  Abramowicz,
   arXiv:astro-ph/0812.3924 (2008)

\bibitem
{Stuchlik:2013esa}
Z.  Stuchlik,  A.  Kotrlova, and G.  Torok,
 \aap, { 552}, A10 (2013)


\bibitem
{PQN}
    D. Pugliese, H. Quevedo, \&  R. Ruffini,
  \prd, { 83}, 024021 (2011)

   \bibitem
   {PQK}
   D. Pugliese, H. Quevedo, \&  R. Ruffini,
  \prd, { 84}, 044030 (2011)

  \bibitem
  {PQKN}
  D. Pugliese, H. Quevedo, \&  R. Ruffini,
  \prd, {88},  2,  024042 (2013)









 \bibitem
 {Pugliese:2011aa}
 D. Pugliese\& J. A. V. Kroon,
  Gen.\ Rel.\ Grav.,  { 44}, 2785 (2012)




\bibitem
{Krolik:2002ae}
J. H.  Krolik\&  J. F.  Hawley,
  \apj,  {{573}}, 754 (2002)

\bibitem
{BMP98}
 B. C. Bromley,
 W. A. Miller,  V.  I.   Pariev,
\nat, {391}, 54, 756 (1998)


\bibitem
{Agol:1999dn}
E. Agol\&  J. Krolik,
  \apj, { {528}}, 161 (2000)

 \bibitem
 {Afshordi:2002gi}
N. Afshordi\& B. Paczynski,
  Astrophys.\ J.\  { 592}, 354 (2003)




\bibitem
{ergon}
  D. Pugliese, \& H. Quevedo,
 Eur.\ Phys.\ J.\ C, { 75}, 5,  234 (2015)

 \bibitem
 {Stu:conto}
Z.  Stuchlik, \& S. Hledík, K.Truparová
Class.\ Quant.\ Grav., \ 28,  15,  155017  (2011).
	






\bibitem
{Harada:2010yv}
T.  Harada, \&  M. Kimura,
  \prd, { 83}, 024002 (2011)
%


\bibitem
{Tursunov:2013zha}
  A.  Tursunov, M. Kolos, A.  Abdujabbarov, B. Ahmed, Z. Stuchlik,
  \prd, { 88}, 124001 (2013)




 \bibitem
 {1980BAICz..31..129S} Z. Stuchlik, Bulletin of the Astronomical Institutes of Czechoslovakia, 31, 129 (1980)
\bibitem
{Blaschke:2016uyo}
  M.~Blaschke and Z.~Stuchlik,
  Phys.\ Rev.\ D {\bf 94} 8,  086006 (2016).



\bibitem
{Stuchlik:2013yca}
Z.  Stuchlik, \&  J.  Schee,
  Class.\ Quant.\ Grav., \  { 30}, 075012 (2013)





\bibitem
{Stu-Sche:2012:CLAQG}
Z.  Stuchl{\'{\i}}k, \&  J.  Schee,
Class. Quant. Grav., {29},  065002 (2012)



\bibitem
{Stuchlik:2017rir}
  Z. Stuchl{\'{\i}}k, M. Blaschke  and  J. Schee,
  Phys.\ Rev.\ D { 96}, 10,  104050 (2017)


\bibitem
{apite1}M. Volonteri,  M. Sikora, J.-P. Lasota, \apj, 667, 704 (2007)

 \bibitem
 {apite2}M. Volonteri, \apj, 663, L5 (2007)
\bibitem
{apite3}M.Volonteri, A\&AR, 18, 279 (2010)


\bibitem
{Li:2012ts}
 L.~X. Li.,
  \mnras,  { 424}, 1461 (2012).



\bibitem
{Oka2017} T. Oka,   S. Tsujimoto, et al.
Nature Astronomy-Letter, (2017)


\bibitem
{Kawa}N. Kawakatu,  K.  Ohsuga,
\mnras,  417,  4,  2562-2570 (2011)


\bibitem
{Allen:2006mh}
 S. W.  Allen, et al 
  \mnras, 1, { 372},  21 (2006)


\bibitem
{Japan}M. A. Abramowicz,
Astronomical Society of Japan,  37, 4, 1985,  727-734


\bibitem
{Middleton:2014cha}
 M.  Middleton,  J.  Miller-Jones and R. Fender,
  \mnras, 439,2,  1740 (2014)



\bibitem
{Nelson:1997nj}
R.~W.   Nelson {\it et al.},
  \apj, { 488}, L117 (1997)



\bibitem
{Nixon(2011)}
 C. J. Nixon, P. J. Cossins,  A. R. King, J. E. Pringle,
\mnras,  412,  3,  1591-1598 (2011)



\bibitem
{Roedig:2013lqa}
 C. Roedig and A. Sesana,
  \mnras, { 439}, 4,  3476 (2014).



\bibitem
{Font02}
  J. A. Font, \&  F. Daigne,
 \mnras,  { 334}, 383 (2002)




\bibitem
{Nixon:2013qfa}
 C. Nixon, A.  King, \& D. Price, D.,
  \mnras,  { 434}, 1946 (2013).


 \bibitem
 {Dogan:2015ida}
 S. Dogan, C. Nixon, A. King,  D.~J. Price,
 \mnras, { 449},  2,  1251 (2015)




\bibitem
{Bonnerot:2015ara}
 C. Bonnerot,E.~M. Rossi, G.  Lodato \& D.~J. Price,
\mnras,\  {455}, 2,  2253 (2016)




{\bibitem
{Bonnell} I.~A. Bonnell, \&  W.~K.~M. Rice,\  Science, 321, 1060 (2008).
}
{\bibitem{WA}K. Zubovas, A. King,
 arXiv:1901.02224 [astro-ph.GA],  (2008)
}

\bibitem
{Aly:2015vqa}
H. Aly,   W. Dehnen, C. Nixon, \& A. King,
  \mnras,  { 449}, 1,  65 (2015).



\bibitem
{Abra83}
 M. A. Abramowicz, M.  Calvani, \& L. Nobili,
\nat 302, 597--599 (1983).

\bibitem
{Berti:2009bk}
 E.  Berti, V. Cardoso, L. Gualtieri et al.
  \prl, { 103}, 239001 (2009).

 \bibitem
 {mnras-pa}
 P. G. Jonker,  M. A. P. Torres,  A. C.  Fabian et al.
\mnras, 407, (1), 645-650 (2010).



 \end{thebibliography}
\end{document}